\documentclass[aps,pre,floats,floatfix,superscriptaddress,nofootinbib,twocolumn]{revtex4}
\usepackage{bbm}
\usepackage{amsmath,amssymb,amsthm}
\usepackage{xcolor}
\usepackage{graphicx}
\usepackage{makecell}
\usepackage{url}
\usepackage{algorithm}
\usepackage{algorithmic}
\usepackage[colorlinks=true,allcolors=blue]{hyperref}

\begin{document}

\title{Link prediction with hyperbolic geometry}

\author{Maksim Kitsak}
\affiliation{\footnotesize Faculty of Electrical Engineering, Delft University of Technology, Mathematics and Computer Science, 2600 GA Delft, The Netherlands\looseness=-1}
\affiliation{\footnotesize Network Science Institute, Northeastern University, 177 Huntington avenue, Boston, Massachusetts 02115, USA\looseness=-1}
\author{Ivan Voitalov}
\affiliation{\footnotesize Network Science Institute, Northeastern University, 177 Huntington avenue, Boston, Massachusetts 02115, USA\looseness=-1}
\affiliation{\footnotesize Department of Physics, Northeastern University, 110 Forsyth Street, 111 Dana Research Center, Boston, Massachusetts 02115, USA\looseness=-1}
\author{Dmitri Krioukov}
\affiliation{\footnotesize Network Science Institute, Northeastern University, 177 Huntington avenue, Boston, Massachusetts 02115, USA\looseness=-1}
\affiliation{\footnotesize Department of Physics, Department of Mathematics, Department of Electrical and Computer Engineering,
Northeastern University, 110 Forsyth Street, 111 Dana Research Center, Boston, Massachusetts 02115, USA\looseness=-1}

\begin{abstract}
Link prediction is a paradigmatic problem in network science with a variety of applications. In latent space network models this problem boils down to ranking pairs of nodes in the order of increasing latent distances between them. The network model with hyperbolic latent spaces has a number of attractive properties suggesting it must be a powerful tool to predict links, but the past work in this direction reported mixed results. Here we perform a systematic investigation of the utility of latent hyperbolic geometry for link prediction in networks. We first show that some measures of link prediction accuracy are extremely sensitive with respect to inaccuracies in the inference of latent hyperbolic coordinates of nodes. This observation leads us to the development of a hyperbolic network embedding method, the \textsc{hyperlink} embedder, which we show maximizes the accuracy of such inference, compared to existing hyperbolic embedding methods. Applying this method to synthetic and real networks, we then find that when it comes to predicting obvious missing links hyperbolic link prediction---for short, \textsc{hyperlink}---is rarely the best but often competitive, compared to a multitude of other methods. However, \textsc{hyperlink} appears to be at its best, maximizing its competitive power, when the task is to predict less obvious missing links that are really hard to predict. These links include missing links in incomplete networks with large fractions of missing links, missing links between nodes that do not have any common neighbors, and missing links between dissimilar nodes at large latent distances. Overall these results suggest that the harder a specific link prediction task the more seriously one should consider using hyperbolic geometry.
\end{abstract}

\maketitle

\section{Introduction}
\label{sec:intro}

Link prediction is a paradigmatic example of forecasting network dynamics~\cite{Peng2015,lu2011link,Menon2011,peixoto2018reconstructing}, with diverse applications including the reconstruction of networks based on partial data~\cite{Marchette2008,guimera2009missing,Kim2011} and prediction of future social ties~\cite{Peng2015,Adamic2003friends,Newman2016}, protein interactions~\cite{VonMering2002a,Yu2008,kovacs2019network}, and user ratings in recommender systems~\cite{zhou2007bipartite,lu2012recommender,bobadilla2013recommender,schafer1999recommender}.

Latent space network models~\cite{gilbert1961random,mcfarland1973social, McPherson2001, Krioukov2010hyperbolic,newman2015generalized} offer an intuitive and simple approach to link prediction. In these models, network nodes are points in a latent space, while connections are established with probabilities that decrease with latent distances between nodes. Latent distances model similarity between nodes, and the main idea behind these models is to model homophily: more similar nodes are more likely to be linked. Link prediction then reduces to ranking unconnected node pairs in the order of increasing latent distances between them: the closer the two unlinked nodes in the latent space, the higher the probability of a missing link~\cite{brew2010latent,zhu2016scalable,peixoto2018reconstructing,garcia-perez2020precision}.

Among many latent space models considered in literature, only the one that assumes that the latent space is hyperbolic reproduces sparsity, self-similarity, scale-free degree distribution, strong clustering, the small-world property, and community structure~\cite{Serrano2008,Krioukov2010hyperbolic,Papadopoulos2012popularity,zuev2015emergence}. All these properties are often observed in many real networks~\cite{Lazega2006social,NEWMAN2010,barabasi2016network}, and hyperbolic geometry captures them all. In addition, the hyperbolic network model is likely to be the simplest or parsimonious with respect to these properties, as in some of its limiting regimes it has been proven to be statistically unbiased, satisfying the maximum entropy principle~\cite{VanderHoorn2018sparse,Krioukov2016Clustering}.

Given the combination of these attractive properties, one could naturally expect that the hyperbolic latent space model must be a powerful tool in link prediction. Yet the previous studies on this subject reported mixed results~\cite{Serrano2012uncovering,Papadopoulos2015network1,Papadopoulos2015network,Alessandro2018leveraging,Muscoloni2018minimum,garcia-perez2020precision}.

Here we perform systematic investigation of the efficiency of link prediction using latent hyperbolic geometry. We organize the presentation of the results as follows.

In Sec.~\ref{sec:methods} we recall the definitions of the hyperbolic latent space network model, which for short we call random hyperbolic graphs (RHGs), and outline the basic idea behind link prediction based on this model. We also recall the definitions of the main measures of link prediction accuracy---AUC (area under receiver-operating characteristic), AUPR (area under precision-recall curve), and Precision---and discuss  what these measures actually measure: while AUPR cares mostly about most obvious easy-to-predict missing links, AUC puts more weight on less obvious and harder-to-predict missing links between more dissimilar nodes, albeit with the cost of not caring that much about false positives.

Our main results are then given in Secs.~\ref{sec:model_true_coords} and~\ref{sec:inferred_coords}. In Sec.~\ref{sec:model_true_coords} we calculate analytically the AUC and AUPR on RHGs with \emph{known} hyperbolic coordinates of all nodes. That is, the same coordinates are used both to generate RHGs and to predict missing links in them, an ideal situation yielding the upper bound for the link prediction accuracy using hyperbolic geometry. To understand the robustness of link prediction in the case where coordinates are inferred (Sec.~\ref{sec:inferred_coords}), so that they are not equal exactly to the true coordinates, we add uniform noise to the true coordinates, and analyze the AUC, AUPR, and Precision as functions of the noise amplitude to find that: (1) AUC is not that sensitive to noise, but (2) AUPR and Precision decrease quickly as noise grows. The latter result implies that the AUPR and Precision scores of link prediction using hyperbolic geometry in real networks can be high only if node coordinates are inferred with sufficiently high accuracy. This is because the most likely missing links candidates are those between similar nodes at small hyperbolic distances, which are most sensitive to coordinate inaccuracies.

To predict missing links in networks with unknown coordinates one first needs to infer these coordinates. Motivated by the results in Sec.~\ref{sec:model_true_coords} calling for high-accuracy coordinate inference, and given that no existing hyperbolic coordinate inference algorithm is sufficiently accurate, in Sec.~\ref{sec:inferred_coords} we develop an alternative one, which we call the \textsc{hyperlink} embedder, the focus of which is on high precision in coordinate inference. We  present its overview in Sec.~\ref{sec:inferred_coords}, while all the details are delegated to Appendix~\ref{sec:hyper_inference}, where we also compare it to some existing inference algorithms to show that its accuracy is indeed higher. A software package implementing the \textsc{hyperlink} embedder is hosted by the Bitbucket repository~\cite{codeHLembedder}.

We then apply the \textsc{hyperlink} embedder to a collection of RHGs with ``forgotten'' coordinates, and to real networks, calling the overall link prediction procedure the \textsc{hyperlink} method, and comparing it to a representative collection of other link prediction methods.

Section~\ref{sec:discussion} contains both high-level (Tables~\ref{table:result1}~and~\ref{table:result2}) and more detailed summaries of all the results. The results are definitely not that the \textsc{hyperlink} or any other method is a clear winner in all the considered scenarios according to all the considered link prediction accuracy measures. We discuss what methods are strong in what scenarios. The \textsc{hyperlink} appears to be the strongest in the most difficult link prediction tasks. That is, the more challenging a particular link prediction task/scenario, the better off is the \textsc{hyperlink} compared to other methods. We conclude the paper with an outline of open problems at the end of Sec.~\ref{sec:discussion}.

These results emphasize that the \textsc{hyperlink} is definitely not the universally best link prediction method, which simply cannot exist as was recently shown in~\cite{valles2018consistencies,Ghasemian2019evaluating,ghasemian2019stacking,garcia-perez2020precision}. That is, there can exist no \emph{one size fits all} solution for the link prediction problem. Different methods are good at predicting different types of links. Therefore, as far as a particular link prediction method is concerned, the best one can do is to document what particular link prediction scenarios the method is good at; that is, what types of links the method is good at predicting, which is exactly the subject of this paper.

\section{Methods}
\label{sec:methods}
We begin the exposition by discussing the latent geometric link prediction framework and the null model that we utilize to predict missing links.

\subsection{Link prediction with latent geometry}
Link prediction with hyperbolic geometry is a two-step procedure. First, one needs to infer node coordinates in the hyperbolic space and calculate hyperbolic distances between node pairs. This coordinate inference procedure is often referred to as network mapping or embedding. The second step of the procedure is to identify most likely missing link candidates.  This subsection focuses on the second step of this procedure. The technical details of the null geometric model and the network mapping algorithm constituting the first step, are provided in Secs.~\ref{sec:hrg}~and~\ref{sec:hlnutshell} and Appendix~(\ref{sec:hyper_inference}). We refer to the network mapping algorithm and the entire hyperbolic link prediction framework as the \textsc{hyperlink} embedder and the \textsc{hyperlink}, respectively.

The latent geometric link prediction framework is applicable to all latent geometric models, where connections are established independently with decreasing connection probability function $p(x)$. Intuitively, the smaller the latent distance between two nodes, the higher the probability of a link between them. Then, if two nodes located close to each other in the latent space are not connected, it is likely that there is a missing link between them.

Specifically, consider a latent geometric model where nodes are assigned positions $\{\mathbf{x}_i\}$ in a certain latent space $\mathcal{M}$, and every node pair $\{ij\}$ is connected with probability $p_{ij} = p\left(x_{ij}\right)$, where  $x_{ij} = d\left(\mathbf{x}_i, \mathbf{x}_j\right)$ is the latent distance between the nodes, and
$p:\mathbb{R}_{+}\to [0,1]$ is the decreasing connection probability function specified by the  model. After all connections are established, some  links are removed with probabilities $1-q_{ij}$. These pairs of nodes are referred to as missing links.

Any unconnected node pair $\{ij\}$ in the resulting network is either not connected in the network formation process, or connected in the network formation and later removed with probability $1-q_{ij}$.  Therefore, the probability for an unconnected pair of nodes $\{ij\}$ separated by $x_{ij}$ to be a missing link, is
\begin{equation}
\tilde{p}\left(x_{ij}\right) = \frac{p\left(x_{ij}\right)\left(1-q_{ij}\right)}{ 1 -p\left(x_{ij}\right) + p\left(x_{ij}\right)\left(1-q_{ij}\right)}.
\label{eq:general_missing_links}
\end{equation}

In the particular case of a decreasing connection probability function $p(x)$ and the random link removal process, $q_{ij} = q$
\begin{equation}
\tilde{p}\left(x_{ij}\right) = \frac{\left(1-q\right)p\left(x_{ij}\right)}{ 1 - q p\left(x_{ij}\right)}
\label{eq:missing_links}
\end{equation}
is the decreasing function of $x_{ij}$ for any $q>0$. Thus, the most probable candidates for missing links are indeed unconnected node pairs located at small latent distances, as stated, and the latent geometric link prediction algorithm only needs to rank unconnected node pairs in the increasing order of latent distance between them.

It is important to note, however, that this approach is only guaranteed to work in the case the links are removed uniformly at random. In the general case, missing link probabilities in Eq.~(\ref{eq:general_missing_links}) depend both on latent distances $\{x_{ij}\}$ and missing link rates $\{1-q_{ij}\}$ and further information on the nature of $\{q_{ij}\}$ is needed to rank missing link candidates properly.

\subsection{Random hyperbolic graphs}
\label{sec:hrg}

While the latent geometric framework described above is applicable to all latent space models, in our paper we use the RHG as a null model for link prediction.

RHGs have been extensively studied in the literature~\cite{Krioukov2009,Krioukov2010hyperbolic,Boguna2010sustaining,Papadopoulos2015network1,Papadopoulos2015network,Kitsak2017latent,Aldecoa2015,Garcia2018multiscale} and have been shown to reproduce common properties of many real networks including heterogeneous distributions of node degrees, strong clustering, as well as community structure~\cite{Krioukov2010hyperbolic,zuev2015emergence,muscoloni2018nonuniform}.

The latent space of the RHG model is the two-dimensional hyperbolic disk of constant negative curvature $K=-1$ and radius $R$. The hyperbolic distance $x$ between any two points in the hyperbolic disk is given by the hyperbolic law of cosines:
\begin{equation}
\cosh x = \cosh r \cosh r' - \sinh r \sinh r' \cos \Delta \theta,
\label{eq:hypercos}
\end{equation}
where $(r, \theta)$ and $(r', \theta')$ are the hyperbolic coordinates of the two points within the disk and $\Delta \theta = \pi - |\pi - |\theta - \theta'||$ is the angle between them.

The RHG has three parameters --- hyperbolic disk radius $R>0$, temperature $T \in [0,1)$ and node density parameter $\alpha > 1/2$ --- and is defined as follows:
\begin{enumerate}
\item Draw node coordinates $\{r_i,\theta_i\}$, $i=1,2,\ldots,N,$ from probability density functions:
\begin{eqnarray}
\theta_i \leftarrow \rho(\theta) &=& 1/(2\pi),~\theta_{i} \in [0, 2\pi], \\
r_i \leftarrow \rho(r) &=& \frac{\sinh (\alpha r)}{ \cosh (\alpha R) - 1},~r_{i}\in[0,R]
\label{eq:rho_r}
\end{eqnarray}

\item Compute distances $\{x_{ij} \}$ between all node pairs using  Eq.~(\ref{eq:hypercos}).

\item Connect node pairs with probability
\begin{equation}
p\left(x_{ij}\right) = \frac{1}{1 + e^{\frac{x_{ij} - R}{2 T}}}.
\label{eq:conn}
\end{equation}
\end{enumerate}

We summarize basic RHG properties in Appendices~\ref{sec:hrg_prop}: parameter $\alpha$ controls the exponent $\gamma=2\alpha+1$ of the power-law degree distribution, while clustering is a decreasing function of temperature $T$ approaching zero in the $N\to\infty$ limit as $T\to1$. In this limit, clustering is zero for any $T\geq1$.

\subsection{HYPERLINK embedder in a nutshell}
\label{sec:hlnutshell}
To infer hyperbolic coordinates of nodes in a given network with random links removed, we aim to find the set of node coordinates $\{\mathbf{x}_{i}\} \equiv \{(r_i,\theta_i)\}$, $i=1,2,\ldots,N$, that maximize the posterior probability $\mathcal{L}\left( \{\mathbf{x}_{i}\} | a_{ij}, \mathcal{P}, q \right)$, which is the probability density function of coordinates $\{\mathbf{x}_{i}\}$ in an RHG with adjacency matrix~$a_{ij}$, parameters $\mathcal{P}= \{\alpha, T, R\}$, and link removal probability $1-q$.
By the Bayes' rule this probability is
\begin{equation}
\mathcal{L}\left( \{\mathbf{x}_{i}\} | a_{ij}, \mathcal{P}, q \right) = \frac{\mathcal{L}\left(  a_{ij}| \{\mathbf{x}_{i}\}, \mathcal{P}, q \right) {\rm Prob} (\mathbf{x}_i)}{\mathcal{L}\left( a_{ij}| \mathcal{P}, q \right)},
\end{equation}
where $\mathcal{L}\left(  a_{ij}| \{\mathbf{x}_{i}\}, \mathcal{P}, q \right)$ is the likelihood that network $a_{ij}$ is generated as an RHG with subsequent random link removal with probability $1-q$, ${\rm Prob} (\mathbf{x}_i)$ is the prior probability of node coordinates generated by the RHG, and $\mathcal{L}\left(  a_{ij}|  \mathcal{P}, q \right)$ is the probability that the network has been generated as the RHG with random link removal.
Since node pairs are connected independently, this likelihood is

\begin{equation}
\mathcal{L}\left(  a_{ij}| \{\mathbf{x}_{i}\}, \mathcal{P}, q \right) = \prod_{i < j} \left[\tilde{p}\left(x_{ij}\right)\right]^{a_{ij}}  \left[1 -\tilde{p}\left(x_{ij}\right)\right]^{1 - a_{ij}},
\end{equation}
where $\tilde{p}\left(x_{ij}\right)$ is the effective connection probability in the RHG generation process with subsequent random link removal:
\begin{equation}
\tilde{p}\left(x\right) \equiv q p(x),
\end{equation}
where $p(x)$ is the RHG connection probability function in Eq.~(\ref{eq:conn}). Finally, in RHGs node coordinates $\{\mathbf{x}_i\} \equiv \{r_i, \theta_i\}$, and the prior probability is given by
\begin{equation}
{\rm Prob} (\mathbf{x}_i) = \frac{1}{\left(2 \pi\right)^{N}} \prod_{i=1}^{N} \rho(r_{i}),
\end{equation}
where $\rho(r_i)$ is as in Eq.~(\ref{eq:rho_r}).

The \textsc{hyperlink} embedder aims to find node coordinates $\hat{\mathbf{x}}_i$ that maximize the likelihood $\mathcal{L}\left( \{\mathbf{x}_{i}\} | a_{ij}, \mathcal{P}, q \right)$, or equivalently its logarithm
\begin{equation}
\begin{aligned}
&\ln \mathcal{L}\left( \{\mathbf{x}_{i}\} | a_{ij}, \mathcal{P}, q \right) =  K + \sum^{N}_{i=1} \ln \rho(r_i) + \\
 +& \sum_{i < j} \left[a_{ij} \ln \tilde{p}\left(x_{ij}\right) + \left(1 - a_{ij}\right) \ln \left(1 - \tilde{p}\left(x_{ij}\right)\right) \right],
\end{aligned}
\end{equation}
where constant $K$ absorbs all terms independent of $ \{\mathbf{x}_{i}\}$.

Similar to other maximum-likelihood estimation (MLE) based embedders~\cite{Boguna2010sustaining,Papadopoulos2015network1,Papadopoulos2015network,Perez2019mercator}, node coordinates $\hat{\mathbf{x}}_i$ are computed iteratively: starting with initial random coordinate configuration, the \textsc{hyperlink} embedder updates node coordinates at each iteration step to increase $\ln \mathcal{L}\left( \{\mathbf{x}_{i}\} | a_{ij} \right)$ and stops when we arrive to a stable configuration. One feature of the \textsc{hyperlink} embedder which is different from other MLE-based embedders is that at each iteration step $\ell$ the embedder adds synthetic noise of variable magnitude $a(\ell)$ to angular node coordinates:
\begin{eqnarray}
\hat{\theta}_{i} &\leftarrow& \hat{\theta}_{i} + a(\ell) X_{i},
\end{eqnarray}
where $X_i$ is a random number drawn from the uniform distribution on the circle $[0,2\pi]$.
These coordinate perturbations allow the \textsc{hyperlink} embedder to avoid getting trapped for long time in local maxima of the log-likelihood function and to find (nearly) optimal solutions much faster, thus increasing the coordinate inference accuracy given the same amount of computational resources (see Appendix~\ref{sec:hyper_inference} for details).

\subsection{Link prediction accuracy}
\label{sec:link_accuracy}
We evaluate the accuracy of the \textsc{hyperlink} as well as other link prediction methods through random link removal experiments. To this end, we first remove existing links uniformly at random with probability $1-q$ from the network of interest $G$.  We refer to the remaining network as the \emph{pruned} network and denote it by $\tilde{G}$. We refer to removed links as missing links and denote them by $\Omega_{R}$. The set of remaining links in $\tilde{G}$ is referred to as $\Omega_{E}$.

To test the link prediction method of interest we compute likelihood scores for all unconnected node pairs in $\tilde{G}$, $\overline{\Omega}_{E}$, which include both missing links $\Omega_{R}$ and true nonlinks $\Omega_{N}$, so that  $\overline{\Omega}_{E} = \Omega_{R} \cup \Omega_{N}$. We then rely on these scores to rank unconnected node pairs in the decreasing order of missing link likelihood and refer to them as missing link candidates. We denote the fraction $\lambda \in [0,1]$ of most likely missing link candidates as set $\Omega_{M}(\lambda)$. In the case $\lambda = 0$, $\Omega_{M}(\lambda)$ is the empty set, while in the $\lambda = 1$ case  $\Omega_{M}(\lambda)=\Omega_{R} \cup \Omega_{N} = \overline{\Omega}_{E}$.

In the case the exact number of missing links is known, the most direct way to assess link prediction accuracy is to consider the same number of the most likely missing link candidates and evaluate its intersection with the set of missing links.  This metric is known as Precision and is formally defined as
\begin{equation}
{\rm Precision} =   \frac{|  \Omega_{R} \cap \Omega_{M}(\lambda^{*}) |} {|\Omega_{R}|},
\end{equation}
where fraction $\lambda^{*} = 1-q$ is chosen such that $| \Omega_{M}(\lambda^{*})| = | \Omega_{R}|$. The Precision score is bounded by $0$ and $1$ with the upper bound corresponding to the ideal link predictor ranking  all missing links in $\Omega_{R}$ higher than nonlinks in $\Omega_{N}$.

\begin{figure}
\includegraphics[width=3in]{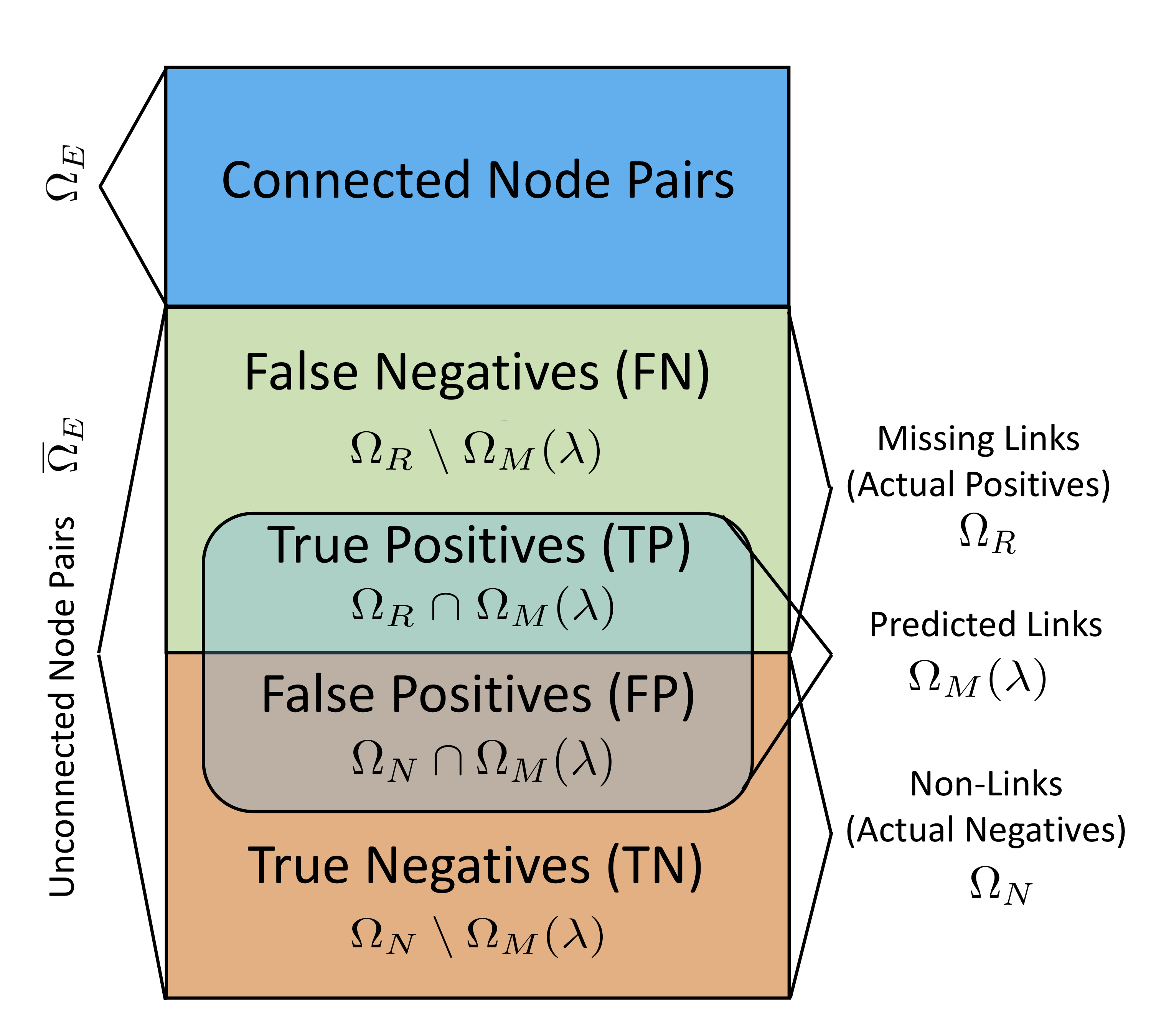}
\includegraphics[width=3in]{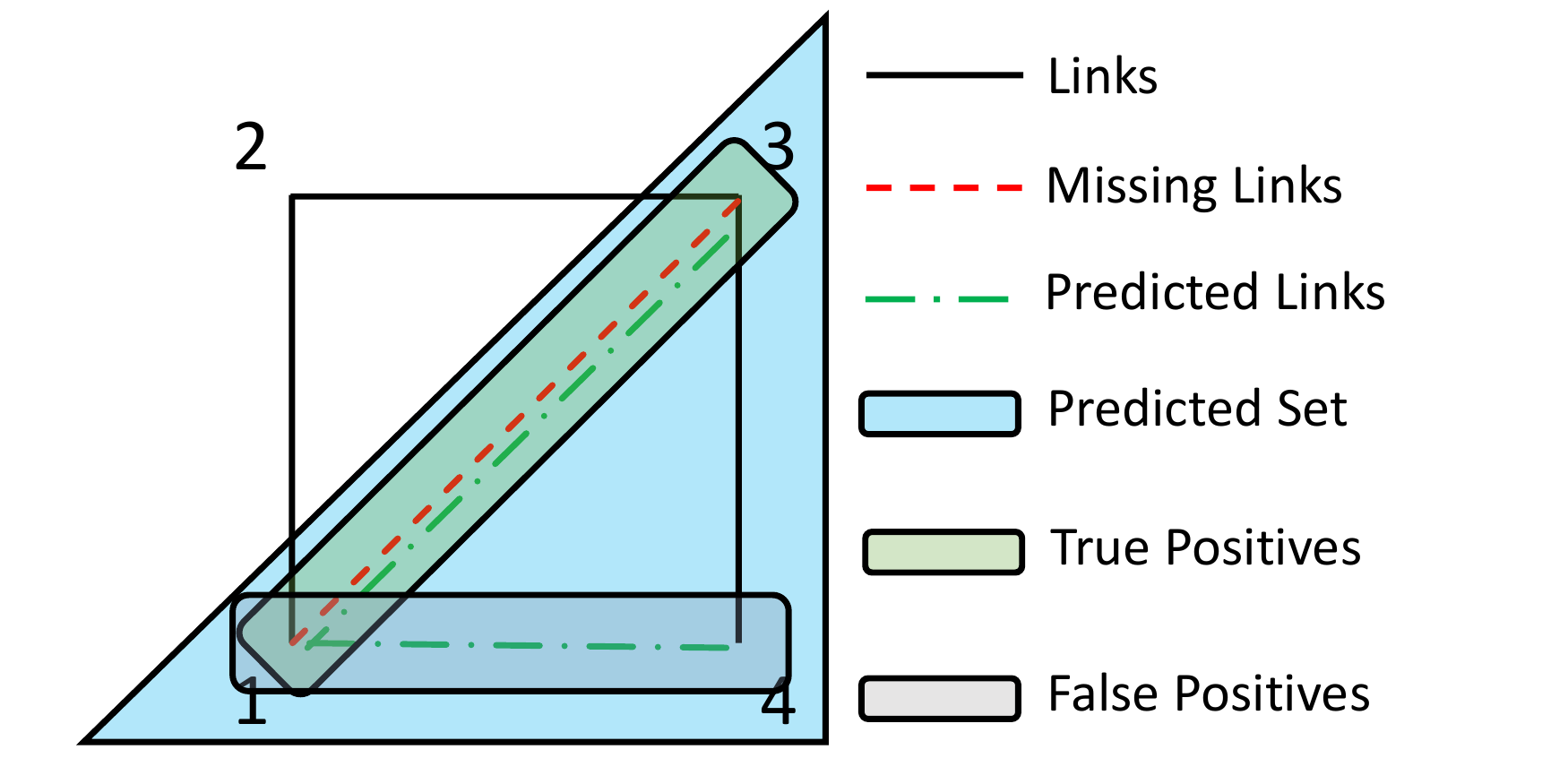}
\caption{\footnotesize Confusion matrix and a toy example of link prediction. Top: Confusion matrix for link prediction. Bottom: Toy link prediction example. Existing links are shown with solid black lines. Missing links, $\Omega_{R} = \{13\}$, are shown with red dotted lines, while predicted missing links, $\Omega(\lambda) = \{13, 14\}$ are shown with green dashed lines. In this example the sizes of the confusion matrix sets are ${\rm TP} = 1$, ${\rm FP} = 1$, ${\rm FN} = 0$, and ${\rm TN} = 1$.}
\label{fig:confusion_mat}
\end{figure}

In practical circumstances, however, the exact number of missing links is often unknown. Further, depending on the application, one might be interested to minimize the number of false positives in the prediction set, possibly by the expense of false negatives, or vice versa, minimize the number of false negatives by the expense of false positives. One example of the former case where one is interested to minimize the number of false positives, i.e., good citizens misclassified as criminals, is the criminal justice system. This example is in contrast to cancer screening, where the number of false negatives, or not-identified cancer cases, should be minimized. In both cases one is interested to explore the performance of the link predictor for a range of $\Omega_{M}(\lambda)$ sizes.

A number of link prediction metrics have been developed to this end with the receiver operating characteristic (ROC) and the precision-recall (PR) being the most popular.

To formally introduce ROC and PR curves we first define the confusion matrix. The latter consists of four values --- the numbers of \emph{true positives} (TP), \emph{false positives} (FP), \emph{false negatives} (FN), and \emph{true negatives} (TN), Fig.~\ref{fig:confusion_mat} --- and is extensively used in statistical classification problems. Link prediction is not a genuine classification problem since one is only interested to predict links and not their absence. Nonlink node pairs are predicted implicitly as unconnected node pairs that are not part of~$\Omega_{R}$.

In the context of link prediction, the number of true positives is the number of correctly identified missing links from $\Omega_{M}(\lambda)$, Eq.~(\ref{eq:TP}). The number of false negatives is the remaining number of missing links that are not part of the $\Omega_{M}(\lambda)$, Eq.~(\ref{eq:FN}).  The number of false positives is the number of missing link candidates in $\Omega_{M}(\lambda)$ that are not correctly identified, Eq.~(\ref{eq:FP}). Finally, the number of true negatives is the number of unconnected node pairs that are neither true positives nor false positives nor false negatives [see Eq.~(\ref{eq:TN}) and Fig.~\ref{fig:confusion_mat}]:
\begin{eqnarray}
\label{eq:TP}
{\rm TP}\left( \lambda \right) &=& | \Omega_{R}  \cap \Omega_{M}(\lambda)  |, \\
\label{eq:FN}
{\rm FN}\left( \lambda \right) &=&| \Omega_{R} \setminus \Omega_{M}(\lambda) |,\\
\label{eq:FP}
{\rm FP}\left( \lambda \right) &=& |\Omega_{N} \cap \Omega_{M}(\lambda) |,\\
\label{eq:TN}
{\rm TN}\left( \lambda \right) &=& | \Omega_{N}\setminus \Omega_{M}(\lambda) |.
\end{eqnarray}

Since network sizes vary, it is common to normalize confusion matrix elements, obtaining true positive, false positive, false negative, and true negative rates, formally defined as
\begin{eqnarray}
 {\rm tpr}(\lambda) &\equiv& \frac{{\rm TP}\left( \lambda \right) }{|\Omega_{R}|},\\
 {\rm fnr}(\lambda) &\equiv& \frac{{\rm FN}\left( \lambda \right) }{|\Omega_{R}|},\\
 {\rm fpr}(\lambda) &\equiv& \frac{{\rm FP}\left( \lambda \right) }{|\Omega_{N}|},\\
 {\rm tnr}(\lambda) &\equiv& \frac{{\rm TN}\left( \lambda \right) }{|\Omega_{N}|}.
\end{eqnarray}
\begin{figure}
\centering
\includegraphics[width=3.5in]{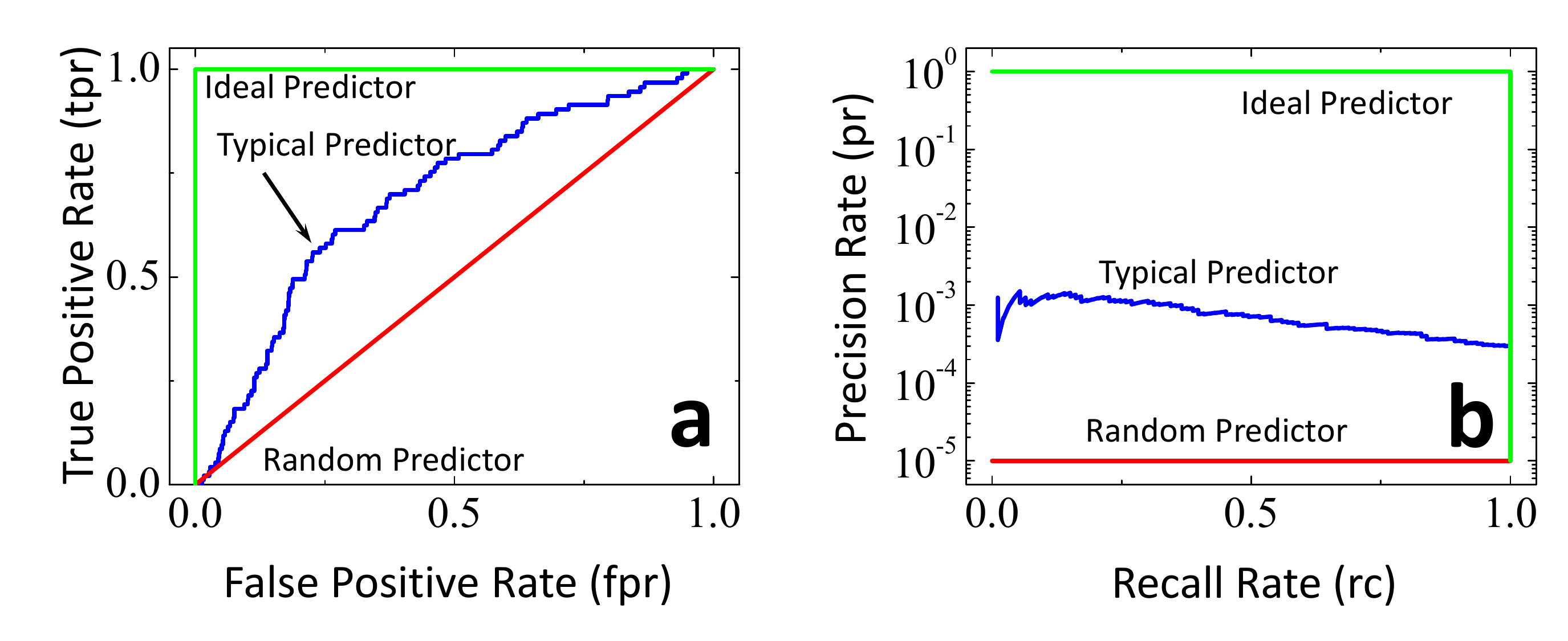}
\caption{\footnotesize Sketches of typical, {\bf a}, ROC and, {\bf b}, PR curves.}
\label{fig:typical_roc_pr}
\end{figure}

An ROC statistics or curve is defined as the parametric plot of the true positive rate ${\rm tpr}(\lambda)$  as a function of the false positive rate ${\rm fpr}(\lambda)$ obtained by varying the fraction of considered link candidates $\lambda \in [0,1]$. The ideal predictor is expected to rank all node pairs corresponding to missing links, $\Omega_{R}$, higher than nonlinks, $\Omega_{N}$, resulting in unit true positive rate and zero false positive rate for  $\lambda = 1-q$, ${\rm tpr}(1-q) = 1$, ${\rm fpr}(1-q) = 0$. The corresponding ROC curve of the ideal predictor is thus a rectangle going through the upper left corner $(0,1)$ of the ROC space. A fully random link predictor, on the other hand, will guess missing links at random from $\overline{\Omega}_{E}$ and is expected to yield equal true positive and false positive rates,  ${\rm tpr}(\lambda) = {\rm fpr}(\lambda)$ for all $\lambda$ values, resulting in the diagonal ROC curve, Fig.~\ref{fig:typical_roc_pr}{\bf a}.

The standard way to quantify ROC-based prediction accuracy is through the AUC:
\begin{equation}
{\rm AUC} = \int_{0}^{1}{\rm tpr}(\lambda) {\rm fpr}'(\lambda) {\rm d} \lambda .
\end{equation}
AUC values vary in between $0$ and $1$ with ${\rm AUC} = 0.5$ corresponding to a fully random predictor and ${\rm AUC} = 1.0$ corresponding to the perfect predictor.

The AUC score can be interpreted as the probability that a randomly chosen missing link is assigned a higher link prediction score than a randomly chosen unconnected node pair. ROC curves are easy to read and interpret, which is arguably the basic reason  behind their popularity.

At the same time, there is a growing consensus that ROC curves and corresponding AUC scores are insensitive in class imbalance problems, where the size of the positives is disproportional to that of the negatives~\cite{Davis2006relationship}. Link prediction in sparse networks is one example of class imbalance. Here the number of missing links is of the order of $N$ and is significantly smaller than the number of nonlinks, which is of the order of $N^{2}$. Intuitively, in this situation the ${\rm tpr}(\lambda)$ rate grows much faster than the false positive rate since the latter is normalized by  $|\Omega_{N}|$ and, as a result, most ROC curves tend to be substantially above the random baseline, yielding AUC scores close to $1.0$, regardless of the link prediction method.

An alternative to the ROC curve is the PR characteristic, defined as the parametric plot of the precision rate ${\rm pr}(\lambda)$ as a function of the recall rate  ${\rm rc}(\lambda)$ obtained by varying $\lambda \in [0,1]$, where the two rates are defined by
\begin{eqnarray}
 {\rm pr}(\lambda) &\equiv& \frac{{\rm TP}\left( \lambda \right) }{|\Omega_{M}(\lambda) |},\\
 {\rm rc}(\lambda) &\equiv& \frac{{\rm TP}\left( \lambda \right) }{|\Omega_{R} |}= {\rm tpr}(\lambda).
 \label{eq:rc3}
\end{eqnarray}
That is, the recall rate is identical to the true positive rate, while the precision rate differs from the latter by a different normalization --- to the number of predicted links versus the number of removed links.

In the case of an ideal predictor, the precision rate is maximized, ${\rm pr}(\lambda) =1.0$ for $\lambda \leq 1-q$,  while the recall is growing from ${\rm rc}(0) = 0$ to ${\rm rc}(1-q) = 1$, resulting in the rectangular PR curve going through the upper right corner $(1,1)$ of the  PR space. A fully random predictor, on the other hand, maintains constant precision rate equal to the ratio of the number of true missing links to the total number of unconnected node pairs, ${\rm pr}^{rand}(\lambda) = \frac{|\Omega_R|} {|\Omega_R|+|\Omega_N|}$ for all $\lambda$ values, Fig.~\ref{fig:typical_roc_pr}{\bf b}. The standard metric quantifying PR-based prediction accuracy is the AUPR:
\begin{equation}
{\rm AUPR} = \int_{0}^{1}{\rm pr}(\lambda) {\rm rc}'(\lambda) {\rm d} \lambda.
\end{equation}
AUPR values vary between $\frac{|\Omega_R|} {|\Omega_R|+|\Omega_N|}$ and $1$ with the unit score corresponding to the ideal predictor. In the case of sparse networks $\Omega_{R} \ll \Omega_{N}$, leading to ${\rm AUPR}\ll 1$ in the case of a random predictor.  Unlike ROC curves, PR characteristics do not directly depend on the number of true negatives and, as a result, do not suffer from the class imbalance problem in case of sparse networks.

\subsection{AUC versus AUPR}

While both AUC and AUPR quantify link prediction accuracy, they tend to weigh missing link candidates differently. AUPR scores tends to emphasize highly ranked missing links candidates, i.e., those corresponding to small $\lambda$ values. AUC scores, on the other hand, put more weight on missing links candidates corresponding to larger $\lambda$ values.

Indeed, AUPR averages precision rate ${\rm pr(\lambda)}$ over the recall rate ${\rm rc(\lambda)}$. Since the recall rate is given by ${\rm rc} (\lambda) = \frac{|\Omega_R \cap \Omega_{M}(\lambda)|}{|\Omega_R|}$, Eq.~(\ref{eq:rc3}),  good link predictors tend to reach ${\rm rc(\lambda)} = 1 $ values when the size of missing link candidates set $\Omega_{M}(\lambda)$ becomes comparable to that of $\Omega_{R}$:  $|\Omega_{M}(\lambda)| \approx |\Omega_{R}| \ll |\Omega_{N}|$.  The latter inequality holds in the case of sparse networks, where the number of links is much smaller than the number of nonlinks. Thus, $|\Omega_{M}(\lambda)| \ll |\Omega_{N}|$, which corresponds to $\lambda \ll 1$ values.
Thus, AUPR link prediction scores are dominated by small $\lambda$ fractions, i.e., by the most likely and, typically, most obvious missing link candidates in $\Omega_{M}$.

AUC scores, on the other hand, average true positive rate ${\rm tpr}(\lambda)$ over false positive rate ${\rm fpr}(\lambda)$. The latter takes large values when $|\Omega_{M}(\lambda)|$ becomes comparable to $|\Omega_{N}|$, i.e. for $\lambda$ values close to $1$. AUC scores, thus, are emphasizing not only easy-to-predict links at small $\lambda$ values but also harder to predict links in $\Omega_{M}$ at intermediate and large $\lambda$ values.

In summary, AUC and AUPR scores complement each other by weighing missing link candidates in $\Omega_{M}$ differently. Thus, in our paper we compute both metrics to obtain a comprehensive view on the utility of hyperbolic geometry in link prediction. In addition to AUPR and AUC scores, we also compute Precision scores, which are the scores to use if the number of missing links is known exactly, although such knowledge is rarely the case in practice.

\section{Link Prediction with Known Coordinates}
\label{sec:model_true_coords}
\begin{figure}
\includegraphics[width=2.5in]{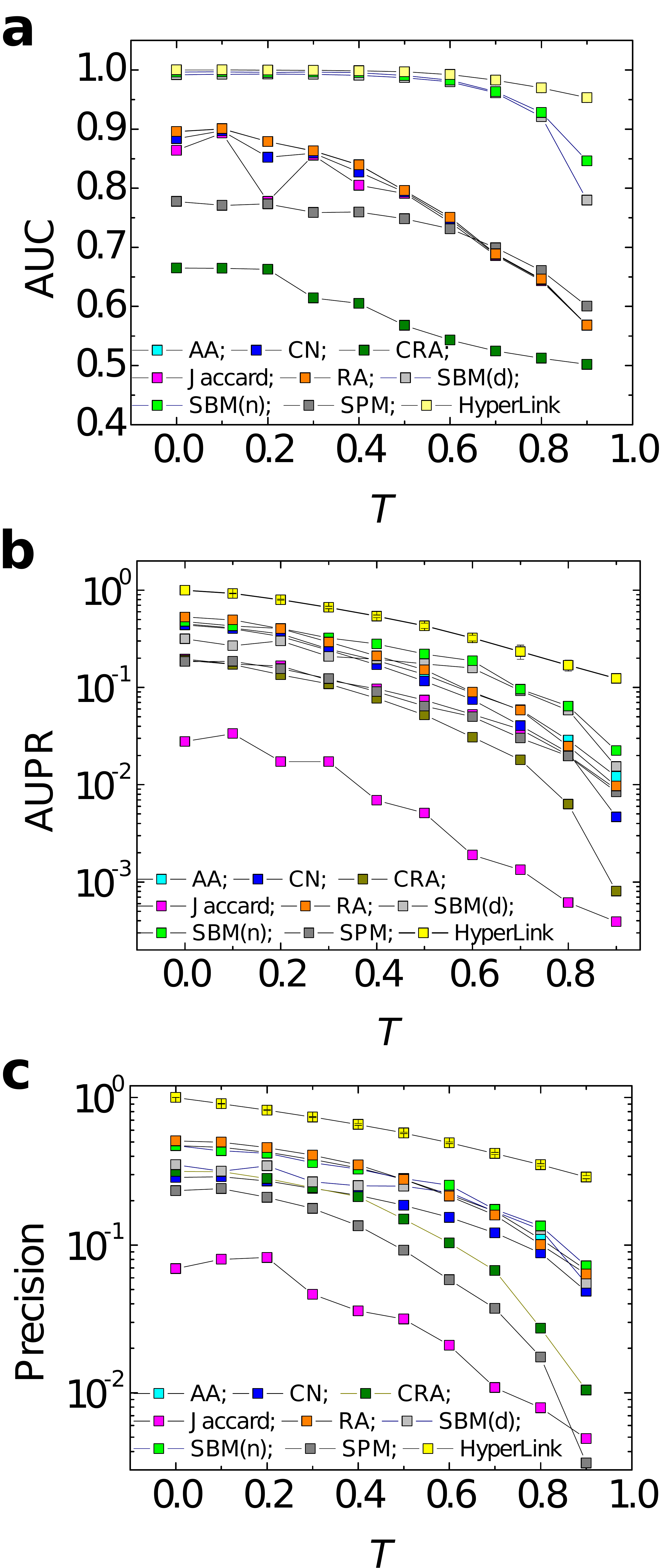}
\caption{\footnotesize {\bf  Link prediction on RHGs with known coordinates.} In all experiments we remove links uniformly at random with probability $1-q=0.5$. Then missing links are predicted using hyperbolic distances between unconnected node pairs. Link prediction accuracy is quantified using, {\bf a} AUC, {\bf b} AUPR, and {\bf c} Precision scores  plotted as a function of RHG temperature $T$. All results correspond to RHGs  with $N=10^{4}$ nodes, $\gamma=2.5$, and $\overline{k}=10$. The \textsc{hyperlink} link prediction scores are compared to those of  AA, CN, CRA, Jaccard, RA, SBM(d,n), and SPM methods (see Appendix~\ref{sec:lp}).
}
\label{fig:11}
\end{figure}

Before investigating link prediction accuracy in real networks, we conduct link prediction experiments with RHGs with known coordinates. In doing so we pursue several goals. The RHGs provide the upper bound for link prediction accuracy of the \textsc{hyperlink} if the same node coordinates are used both for the graph construction and for link prediction~\cite{garcia-perez2020precision}, so that we want to quantify this upper bound. Second, we want to measure link prediction accuracies of other methods, listed in Appendix~\ref{sec:lp}, and compare them to that of the \textsc{hyperlink}. Establishing these results provides a baseline for interpreting link prediction results on real networks. To achieve these goals, we first calculate analytically the AUC and AUPR in RHGs with known coordinates and with coordinates disturbed by noise of varying magnitude. The latter result allows us to quantify in a controlled environment the level of coordinate inaccuracy beyond which the \textsc{hyperlink} becomes essentially impuissant.

We start with the analysis of \textsc{hyperlink} accuracy in the case of randomly missing links in RHGs. After the generation of an RHG we visit each of its links and remove it with probability $1-q$, arriving at a pruned network. We then rank missing link candidates using distances between all unconnected node pairs calculated with coordinates from which the network was originally generated.

As seen in Fig.~\ref{fig:11}, the predictive power of the \textsc{hyperlink} is maximized as $T \to 0$ and decreases as $T$ increases. This result is expected. In the $T\to 0$ limit the RHG is deterministic since the connection probability in Eq.~(\ref{eq:conn}) becomes the Heaviside step function, $p(x) \to \Theta(R-x)$. As a result, all node pairs with $x< R$ are connected and other node pairs are not. Then, an unconnected pair of nodes at distance $ x < R $ is guaranteed to be a true positive and all unconnected pairs at $x \geq R$ are true negatives. As $T$ increases, connections are allowed at distances $x>R$ with increasing probability and, as a result, underlying geometry plays a smaller role in the formation of links, explaining the decreasing link prediction accuracy as a function of $T$, as quantified by all scores in Fig.~\ref{fig:11}.

Even though all scores, AUC, AUPR and Precision, are decreasing functions of $T$, they behave differently. AUC scores remain constant in the  $T\in \left(0,\frac{1}{2}\right)$  interval and then exhibit a slow decay to ${\rm AUC = 0.95}$ at $T=0.9$. AUPR and Precision scores, on the other hand, decrease rapidly in the entire testing interval of $T \in [0,0.9]$ from ${\rm AUPR} = 1$ (${\rm Precision} = 1$) at $T=0$ to ${\rm AUPR} = 0.34$ (${\rm AUPR} = 0.29$)  at $T=0.9$.

 We can predict these results analytically as we explain next.

\subsection{AUC}

\begin{figure*}
\includegraphics[width=3in]{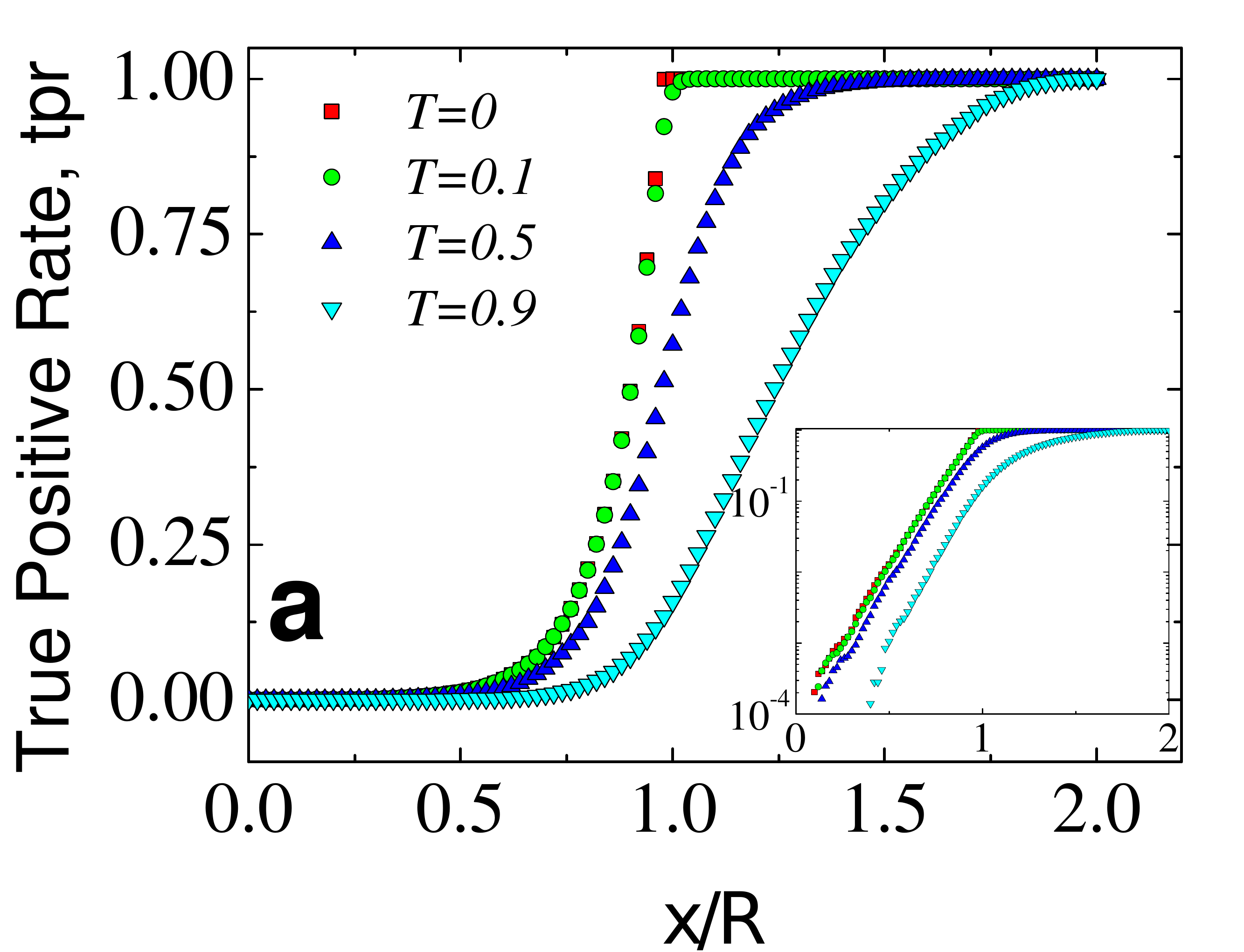}
\includegraphics[width=3in]{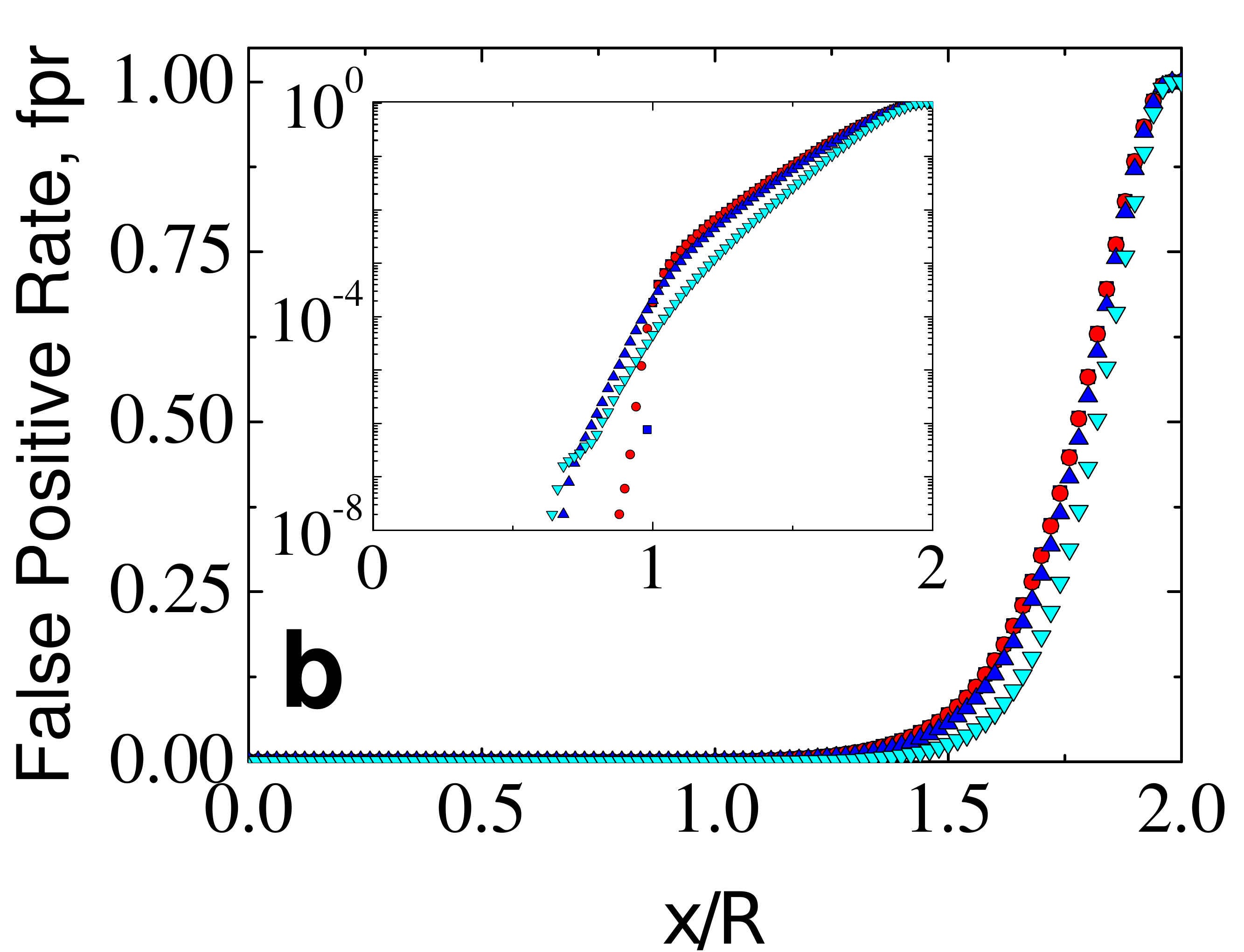}
\includegraphics[width=3in]{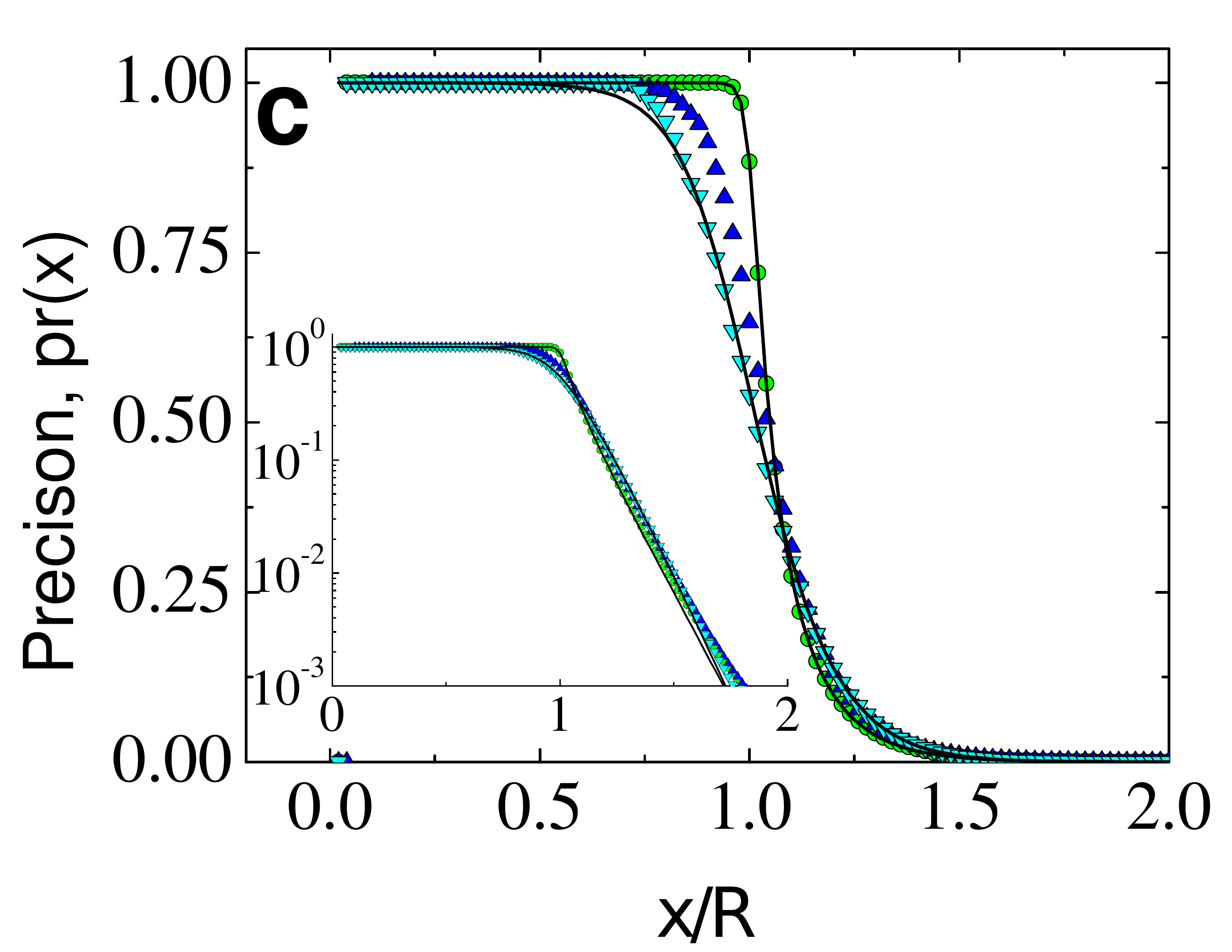}
\includegraphics[width=3in]{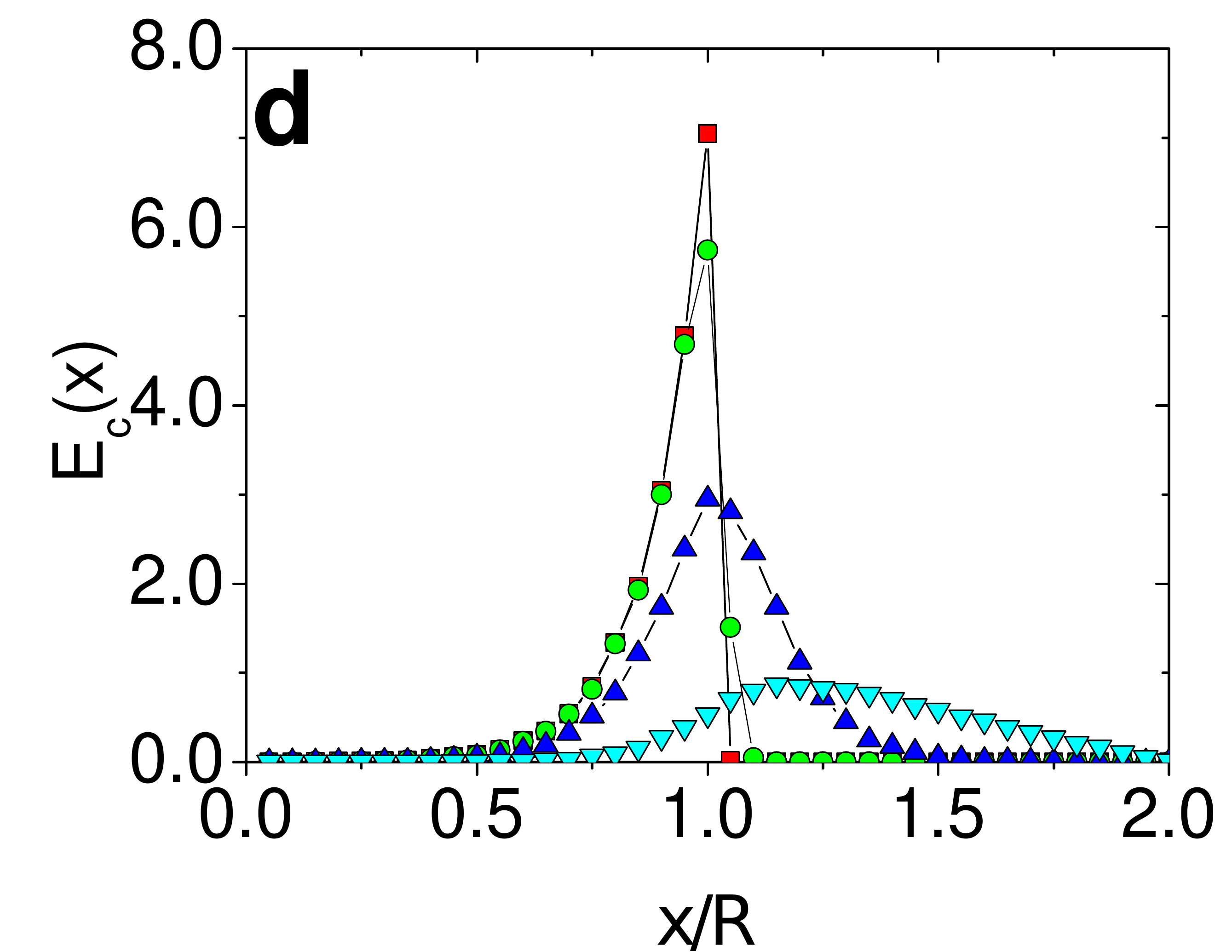}
\caption{\footnotesize {\bf Link prediction with known coordinates.} {\bf a}, true positive rate ${\rm tpr}(x)$, {\bf b}, false positive rate ${\rm fpr}(x)$,  {\bf c}, Precision ${\rm pr}(x)$ and, {\bf d}, link density $n(x)p(x)$ as a function of hyperbolic distance $x$. In all experiments we remove links uniformly at random with probability $1-q=0.5$. Then missing links are predicted using hyperbolic distances between unconnected node pairs.  All results correspond to RHGs with $N=10^{4}$ nodes, $\gamma=2.5$, and $\overline{k}=10$. The insets display the same plots as the main panels but in log-linear format. Solid lines correspond to analytical estimates.}
\label{fig:12}
\end{figure*}

The AUC score in RHGs is
\begin{equation}
{\rm AUC} = \int_{0}^{2R}  {\rm tpr}(x) {\rm fpr}'(x) {\rm d} x,
\end{equation}
where ${\rm tpr}(x)$ and ${\rm fpr}(x)$ are, respectively, distance-dependent true positive and false-positive rates among node pairs separated by distances up to $x$.
As seen from Fig.~\ref{fig:12}{\bf a}, the true positive rate grows exponentially for $x < R$ and saturates to $ {\rm tpr}(x) =1 $ as $x$ approaches $2R$. This observation is easy to predict analytically. Let $n(x)$ be the distribution of hyperbolic distances $x$ between node pairs in the RHG. It follows from the results in~\cite{Alanis-Lobato2016distance} that $n(x)$ can be approximated as
\begin{equation}
n(x) = \frac{4 \alpha^{2}}{\pi \left(2\alpha -1 \right)^{2}}\,e^{x/2-R}
\label{eq:nx}
\end{equation}
for $\alpha > \frac{1}{2}$. To be more specific, we note that $R$ in the RHG is a function of network size $N$, given by Eq.~(\ref{eq:r_sparse}) and the approximation in Eq.~(\ref{eq:nx}) holds in the large $N$ limit for any $x = c R$, where constant $c \in (0,2)$, $\lim_{N \to \infty} \frac{n(x)}{n^{\rm true}(x)} = 1$. Henceforth, we say $f(x) \approx g(x)$  if $\lim_{N \to \infty} \frac{f(x)}{g(x)} = 1$, and, more generally,  $f(x) \sim g(x)$  if $\lim_{N \to \infty} \frac{f(x)}{g(x)} = K \neq 1$.

The connection probability $p(x)$ is close to unity for $x<R$, so that, the number of true positives for $x<R$ grows proportional to the number of node pairs $N(x) \equiv \int_{0}^{x}n(y) {\rm d} y$ in the hyperbolic disk, ${\rm tpr}(x) \sim N(x) \sim e^{\frac{x}{2}}$ for $x <R$.  In the $x > R$ regime connection probability $p(x)$ decays exponentially as $p(x) \sim e^{-\frac{x}{2T}}$ faster than the exponential growth of $n(x)$, leading to the saturation of ${\rm tpr}(x) = 1$, Sec.~\ref{sec:hl_accuracy}.

The false positive rate  remains negligible for $x < R$ and grows exponentially for $x \in (R, 2R)$, Fig.~\ref{fig:12}{\bf b}. We explain this observation using similar arguments. Since $p(x)$ is close to unity for $x<R$, and all unconnected node pairs with $x<R$ are almost guaranteed to be true positives, the false positive rate is negligible for $x < R$. In the $x>R$ regime,  $p(x) \sim e^{-\frac{x}{2T}}$, and the number of unconnected node pairs is proportional to $N(x)$, resulting in ${\rm fpr}(x) \sim e^{\frac{x}{2}}$ for $x>R$, Sec.~\ref{sec:hl_accuracy}.

Taken together, ${\rm tpr}(x)$ and ${\rm fpr}(x)$ rates provide a qualitative explanation for nearly perfect ${\rm AUC}$ scores observed in Fig.~\ref{fig:11}{\bf a}.
The false positive rate ${\rm fpr}(x)$ takes large values only when $x$ approaches $2R$. At the same time, as $x$ approaches $2R$ the ${\rm tpr}(x)$ approaches 1.

Supporting this rough estimation, our more detailed analytical calculations in Sec.~\ref{sec:hl_accuracy} show that the AUC scores for RHGs with known coordinates converge to $1$ in the large $N$ limit as
\begin{equation}
1-AUC
\begin{cases}
\sim N^{-1} &\text{if $ T \in \left.\left[0, \frac{1}{2}\right. \right) $ },\\
=\mathcal{O} \left(\frac{\ln N}{N}\right) &\text{if $ T =\frac{1}{2} $ },\\
=\mathcal{O} \left(N^{1-\frac{1}{T}}\right)  &\text{if $ T \in \left(\frac{1}{2},1\right). $ }
\end{cases}
\label{eq:auc_convergence}
\end{equation}

\subsection{AUPR}

To calculate the AUPR score we need to calculate the distance-dependent precision and recall rates ${\rm pr}(x)$ and ${\rm rc}(x)$ because
\begin{equation}
{\rm AUPR} = \int_{0}^{2R}  {\rm pr}(x)  {\rm rc}'(x) {\rm d} x,
\label{eq:aupr}
\end{equation}

Since $p(x)$ is close to~$1$ for $x<R$, all unconnected node pairs at $x<R$ are true positives, resulting in ${\rm pr}(x)=1$ [see Fig.~\ref{fig:12}{\bf c} and Sec.~\ref{sec:hl_accuracy}]. The precision rate decays exponentially for $x>R$ since the true positive rate ${\rm tpr}(x)$ approaches~$1$ for $x > R$, while the number of unconnected node pairs $N_{d}(x)$ grows exponentially, $\binom{N}{2}  \int_{0}^{x}  n(y) \left[1 - q p(y)\right]{\rm d} y \sim e^{\frac{x}{2}}$ [see Fig.~\ref{fig:12}{\bf c} and Sec.~\ref{sec:hl_accuracy}].

The dependence of ${\rm AUPR}$ on $T$ arises from the recall function or its derivative, $rc'(x)$, quantifying the expected distance-dependent link density and, consequently, the density of missing links. As $T$ increases, the missing links are more likely to be located at larger distances, Fig.~\ref{fig:12}{\bf d}, where precision ${\rm pr}(x)$ is smaller, resulting in lower AUPR scores, consistent with our observations in Fig.~\ref{fig:11}.

We also note that the AUPR score depends weakly on the node density parameter $\alpha$ and consequently on the degree distribution exponent $\gamma=2\alpha +1$. Indeed, the precision and recall rates depend on $\alpha$ only via the node pair distribution $n(x)$, Sec.~\ref{sec:hl_accuracy}, which depends on $\alpha$ only in subleading terms, as shown in Ref.~\cite{Alanis-Lobato2016distance}.

\subsection{Coordinate uncertainty and link prediction accuracy}

\begin{figure}
\includegraphics[width=2.5in]{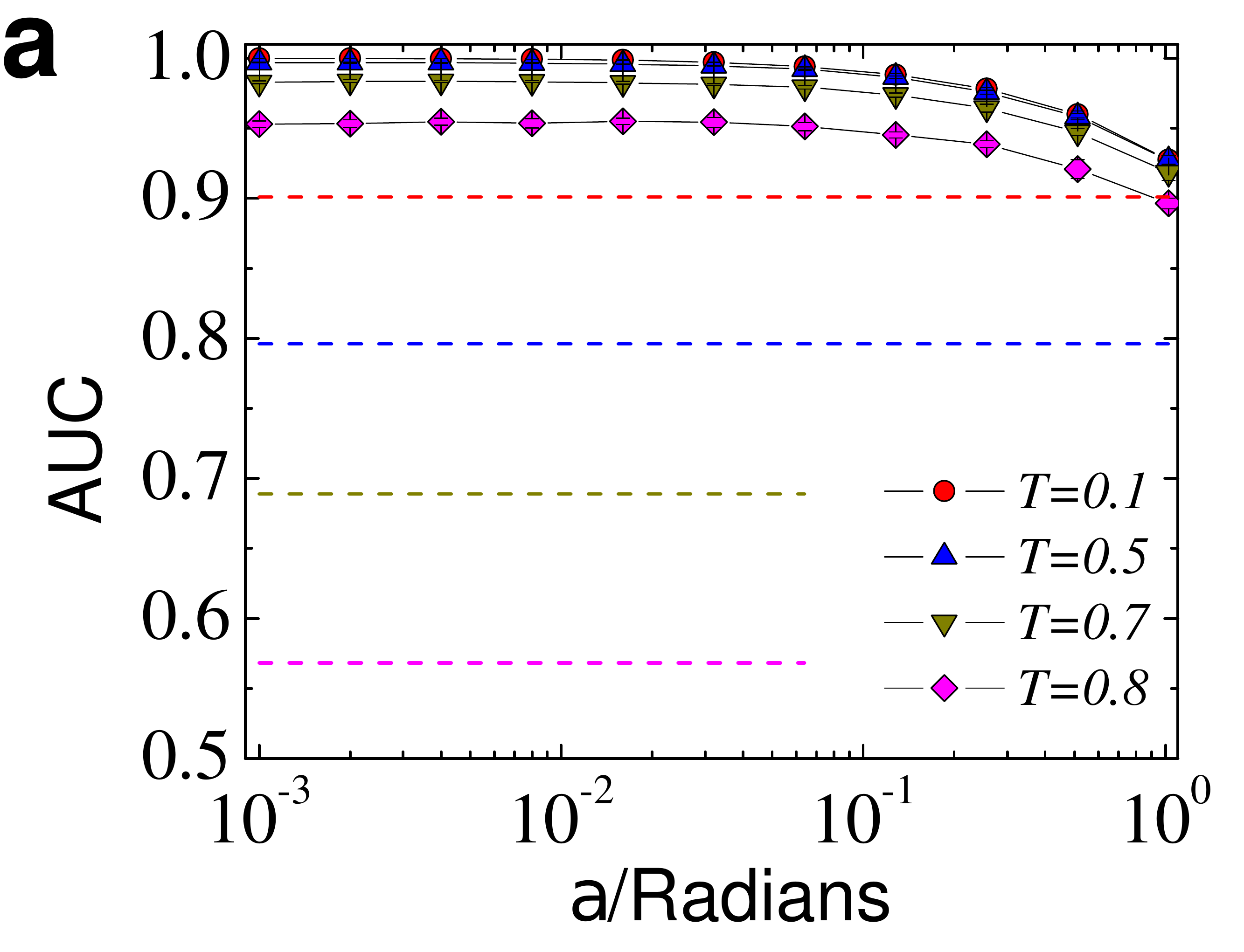}
\includegraphics[width=2.5in]{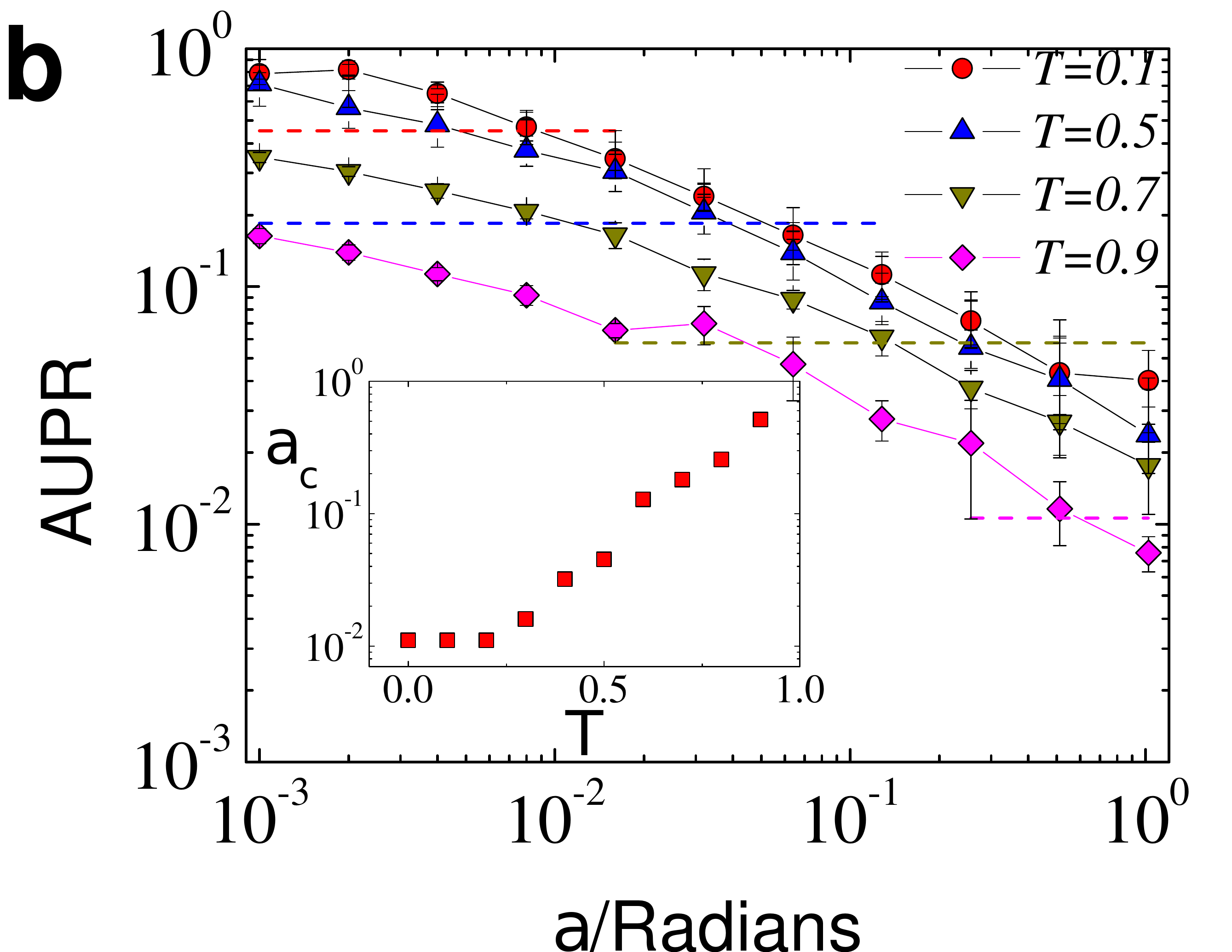}
\includegraphics[width=2.5in]{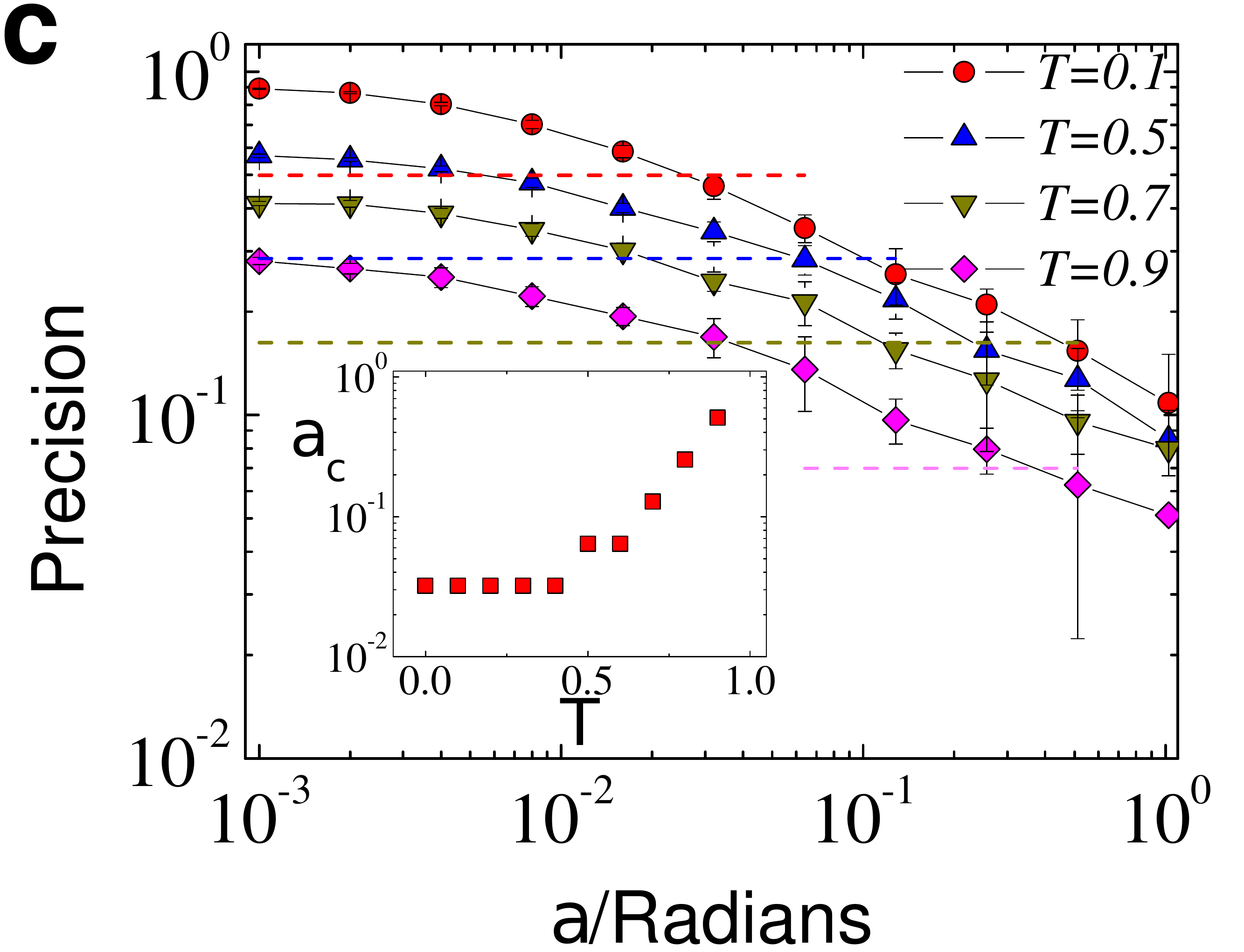}
\caption{\footnotesize {\bf Effects of synthetic noise on link prediction accuracy.} \textsc{hyperlink} accuracy quantified using, {\bf a}, AUC, {\bf b}, AUPR, and {\bf c}, Precision scores as a function of noise amplitude $a$ for RHGs  with different $T$ values. All results correspond to RHGs  with $N=10^{4}$ nodes, $\gamma=2.5$, and $\overline{k}=10$. The \textsc{hyperlink} accuracy is compared to that of RA, i.e., its top competitor according to Fig.~\ref{fig:11}. Corresponding scores of the RA index are shown with dashed lines of matching color. The insets of panels {\bf b} and {\bf c} display the maximum tolerable coordinate noise amplitude as a function of $T$, i.e., the values of $a$ corresponding to equal \textsc{hyperlink} and RA accuracy.
}
\label{fig:13}
\end{figure}

While the \textsc{hyperlink} provides the upper bound for link prediction on RHGs, it is important to note that its accuracy is comparable to that of other link prediction methods, in particular, resource allocation (RA), Adamic-Adar (AA), and stochastic block models SBM(d,n), Fig.~\ref{fig:11}. This observation motivates the question: \emph{How accurately does one need to infer node coordinates to ensure the superior performance of the \textsc{hyperlink}?}

To answer this question we analyze the impact of node coordinate uncertainty on the \textsc{hyperlink} accuracy.  To this end, we add synthetic noise to original angular node coordinates, while keeping radial node coordinates unchanged:
\begin{eqnarray}
\hat{\theta}_{i} &\leftarrow& \theta_{i} + a X_{i},\label{eq:noise2}\\
X_i &\leftarrow& U \left(-\frac{1}{2}, \frac{1}{2}\right),
\end{eqnarray}
where $a>0$ is the noise amplitude. The effects of synthetic noise on the \textsc{hyperlink} accuracy are depicted in Fig.~\ref{fig:13}. Our results indicate that AUPR and Precision scores, Fig.~\ref{fig:13}{\bf b,c}, decrease rapidly as a function of noise amplitude, while AUC scores remain largely unchanged even at $a > 1~{\rm radians}$ values.

To better understand the effects of noise on link prediction accuracy we juxtapose \textsc{hyperlink} prediction results to those of the
RA method, which is one of its leading competitors according to Fig.~\ref{fig:11}. We show RA accuracy with dashed lines of matching color in Fig.~\ref{fig:13}. Consistent with our earlier observations we find that \textsc{hyperlink} AUC scores are robust to noise, preserving its leading ranking among other link prediction methods, Fig.~\ref{fig:13}{\bf a}.

In contrast, as quantified by AUPR and Precision scores, the \textsc{hyperlink} is superior to the RA method only if coordinate uncertainty is sufficiently small. The maximum tolerable noise amplitude value $a_{c}$  increases as $T$ increases [see the inset of Fig.~\ref{fig:13}{\bf b,c}]. While noise amplitude $a$ does not exceed $10^{-2}$ radians in the case of $T=0.1$, the noise tolerance in the case of $T=0.9$ is significantly higher, $a_{c} \approx 0.5$  radians, suggesting, somewhat surprisingly, that the \textsc{hyperlink} is better off on networks characterized by larger $T$  values or, equivalently, smaller clustering coefficient.

Qualitatively, the observed fast degradation of the AUPR  and Precision scores is due to the sensitivity of the hyperbolic distance to the angular distance between the nodes $\Delta \theta$. It follows from Eq.~(\ref{eq:hypercos}) that even a small change in $\Delta \theta$ may significantly change the corresponding hyperbolic distance, adversely affecting the ranking of missing link candidates at small distances $x$, Appendix~\ref{sec:uncertainty}. Since AUPR  and Precision emphasize link prediction accuracy of most likely candidates, proper ranking of unconnected node pairs at small $x$ values is crucial. AUC scores, on the other hand, place more emphasis on less obvious link candidates and are less affected by coordinate uncertainty.
We find that the uniform synthetic noise adversely affects distance dependent true positive rate ${\rm tp}(x|a)$, which scales as
\begin{equation}
{\rm tp}(x|a) \sim
\begin{cases}
a^{1-2\gamma} & \text{if $x\leq R$},\\
a^{1-2\gamma}\left(R + 2 \ln \frac{a}{2}\right) & \text{if $x > R$},
\end{cases}
\end{equation}
see Appendix~\ref{sec:uncertainty}, leading to
\begin{equation}
{\rm AUPR}(a) \sim a^{2-4\gamma}\left(R+2\ln\frac{a}{2}\right)^{2}.
\end{equation}

The robustness of the AUC scores to synthetic noise in RHGs can be qualitatively explained by the fact that AUC scores emphasize the prediction of  missing links at large $x$ distances. Large hyperbolic distances are affected by synthetic noise to a lesser extent than small hyperbolic distances. This effect follows directly from Eq.~(\ref{eq:hypercos}) and can be observed in Fig.~\ref{fig:a1}{\bf a}, displaying the saturation of ${\rm tp}(x|a) \to 1$ as $x$ approaches $2R$, regardless of noise amplitude $a$.

Our conclusions in this section are different for AUC and AUPR/Precision metrics.

The AUPR and Precision metrics emphasize prediction of the most likely missing link candidates and are
highly sensitive to the accuracy of node coordinate inference. Synthetic noise added to original node coordinates smears hyperbolic distances among missing link candidates, adversely affecting the \textsc{hyperlink} accuracy. Our results suggest that one needs to maximize the accuracy of the network mapping in order to efficiently predict missing links.
We also find that as temperature $T$ increases, the performance of other link prediction methods, as measured by AUPR and Precision, decreases faster than that of the \textsc{hyperlink}, suggesting that the latter has a competitive advantage on networks characterized by large $T$ values.

AUC scores, on the other hand, emphasize less obvious link candidates that correspond to node pairs at larger hyperbolic distances. Since larger hyperbolic distances are affected by coordinate uncertainty to a lesser extent, the AUC scores of the \textsc{hyperlink} are robust to synthetic noise, suggesting that \textsc{hyperlink} is capable of predicting less obvious missing links even under less accurate mapping conditions.

\begin{figure*}
\includegraphics[width=7in]{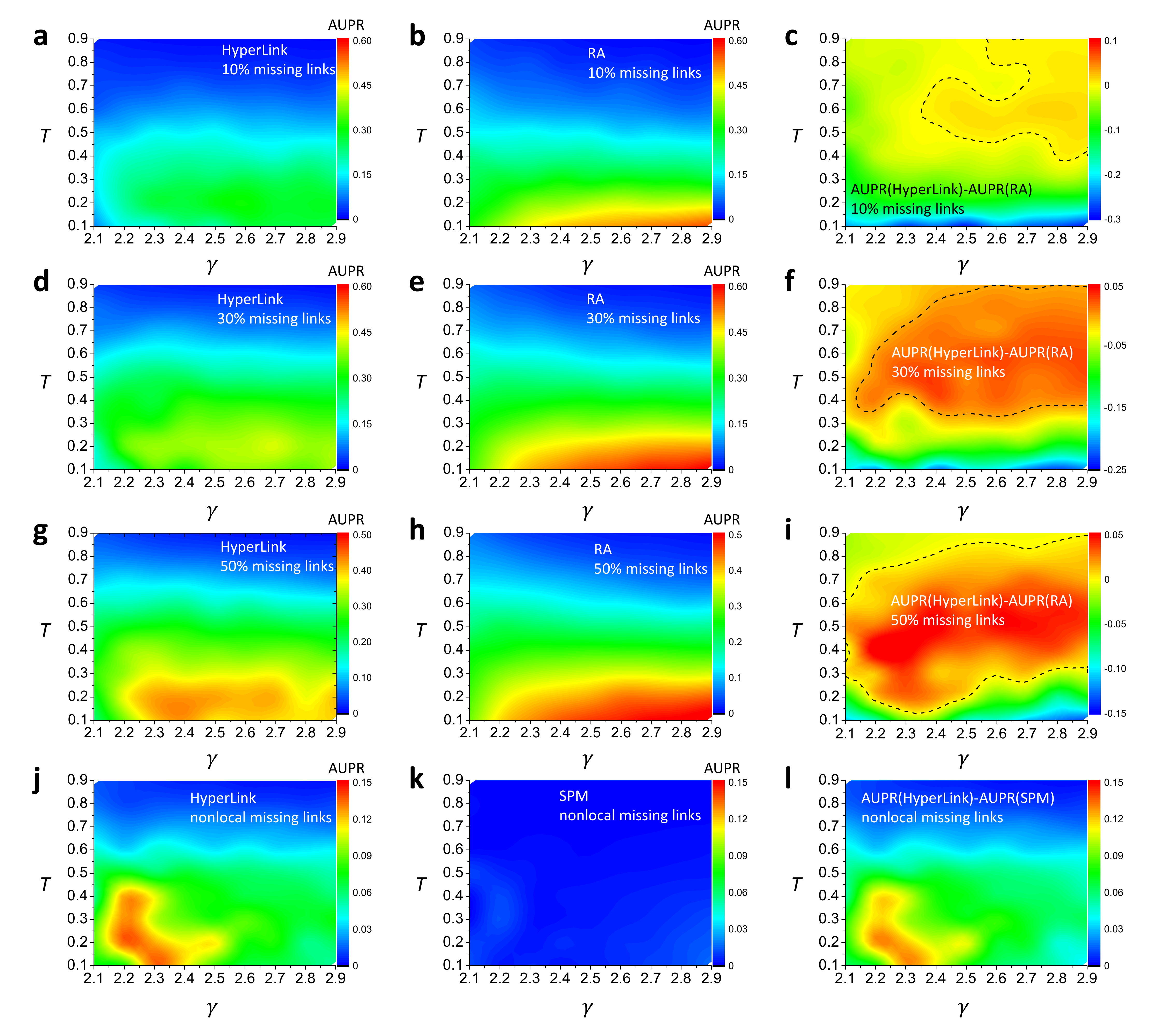}
\caption{\footnotesize  {\bf Link prediction accuracy for RHGs with inferred coordinates: AUPR.} Panels {\bf a-i} correspond to random missing links, and {\bf j-l} to nonlocal missing links. Each panel is a heatmap displaying AUPR values as functions of $T$ and $\gamma=2\alpha + 1$  parameters of the RHG. We compare link prediction accuracy of the \textsc{hyperlink} to that of the RA and SPM methods, which are its leading competitors in cases of randomly missing links and nonlocal missing links, respectively. In each random missing link experiment links are removed uniformly at random with prescribed probabilities: {\bf a-c}, $1-q = 0.1$, {\bf d-f}, $1-q = 0.3$ and {\bf g-i}, $1-q = 0.5$. Panels {\bf a, d, g} and {\bf b, e, h} show the AUPR values for \textsc{hyperlink} and RA respectively. Panels {\bf j-l} show the AUPR values of \textsc{hyperlink} and SPM, as well as their difference, for nonlocal links. i.e., links connecting nodes with no common neighbors, which comprise a subset of randomly removed links with $1-q=0.5$. The dashed curves in panels {\bf c, f, i, l} denote the regions in the $\gamma$-$T$ parameter space where the \textsc{hyperlink} accuracy is higher than that of the competitive method.}
\label{fig:2}
\end{figure*}
\begin{figure*}
\includegraphics[width=7in]{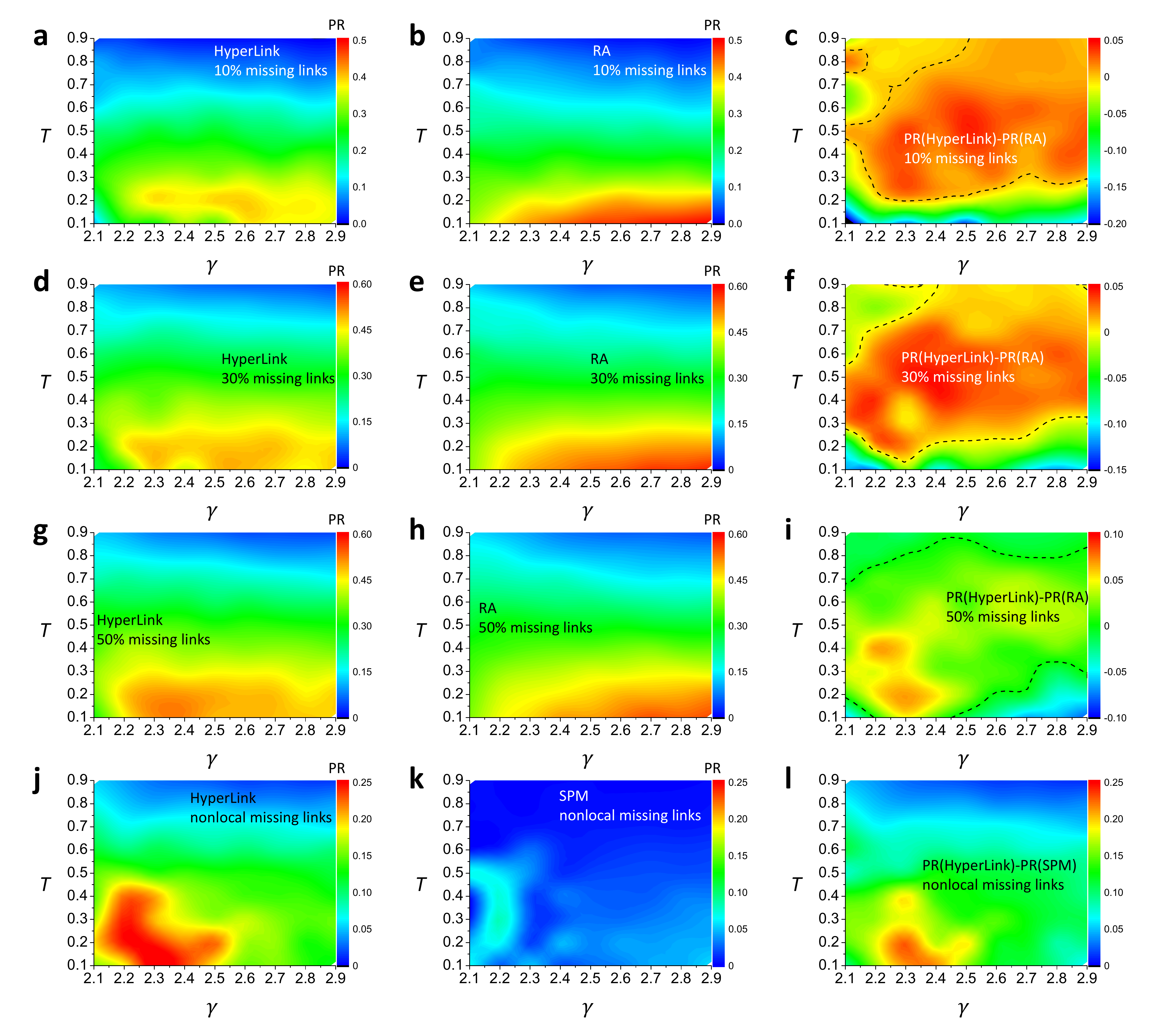}
\caption{\footnotesize  {\bf Link prediction accuracy for RHGs with inferred coordinates: Precision.} The legend is identical to that of Fig.~\ref{fig:2}.}
\label{fig:pr}
\end{figure*}
\begin{figure*}
\includegraphics[width=7in]{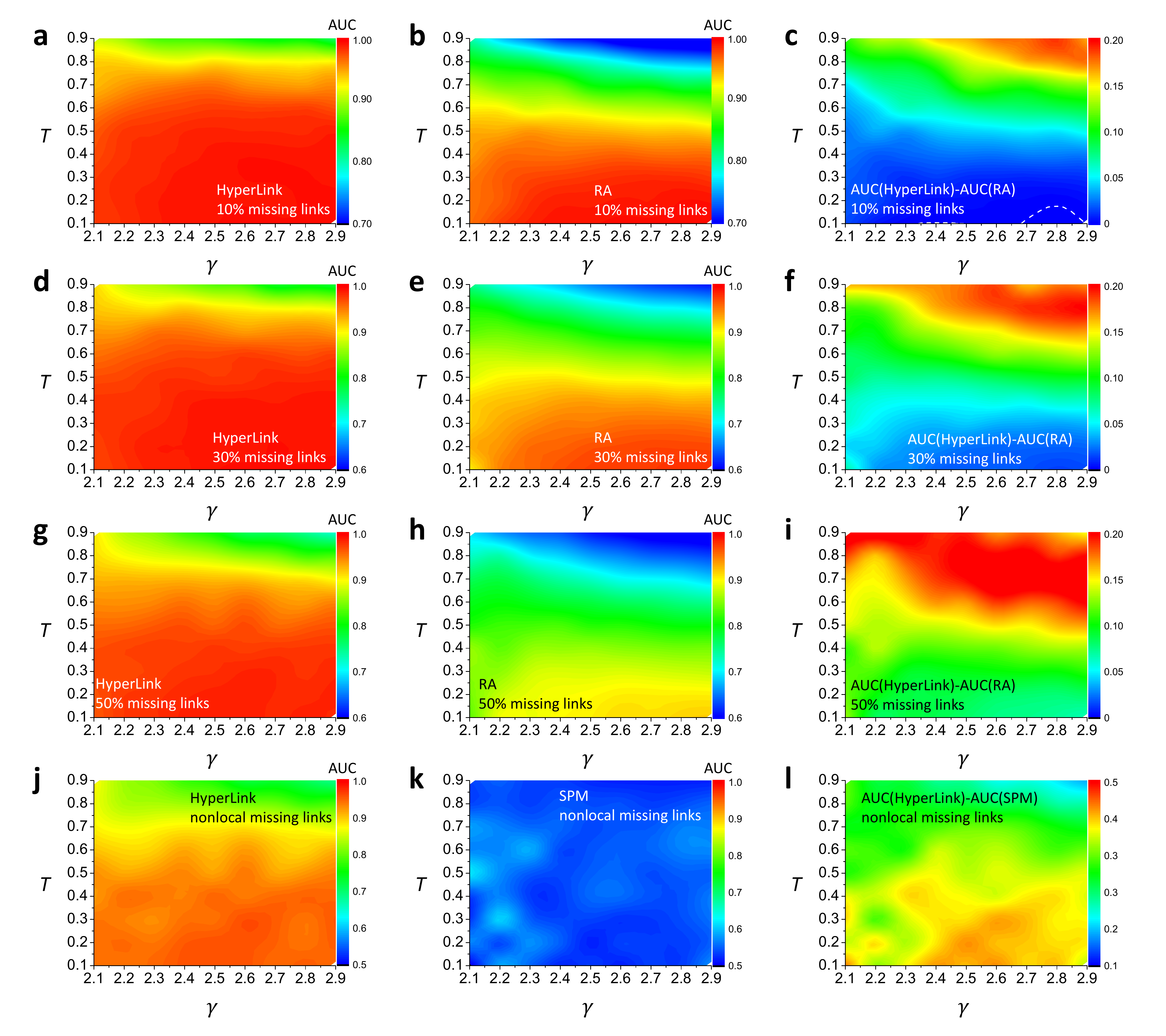}
\caption{\footnotesize  {\bf Link prediction accuracy for RHGs with inferred coordinates: AUC.} The legend is identical to that of Fig.~\ref{fig:2}. }
\label{fig:auc}
\end{figure*}

\section{Link Prediction with Inferred Coordinates}
\label{sec:inferred_coords}

In this section we build upon our results obtained in the previous section to analyze the \textsc{hyperlink} accuracy on networks with unknown node coordinates. We first conduct systematic analysis of \textsc{hyperlink} accuracy on RHGs with unknown node coordinates and then apply \textsc{hyperlink} to several real networks. In both cases network coordinates are unknown and in order to predict missing links we first infer node coordinates by mapping networks of interest to the two-dimensional hyperbolic disk. To this end, we developed a mapping algorithm, which is tailored to the link prediction problem. This algorithm is referred to as the {\it \textsc{hyperlink} embedder} and is fully described in Appendix~\ref{sec:hyper_inference}.

\subsection{Tests on RHGs with inferred coordinates}

To evaluate the \textsc{hyperlink} accuracy on RHGs  with unknown node coordinates we perform the following experiments. After generating an RHG we remove a fraction of existing missing links. As before, each existing link  is removed with probability $1-q$. Occasionally, after  links are removed, the remaining network splits into several components. If this is the case, we limit our consideration to the largest connected component of the pruned network. We refer to the resulting connected component of the pruned network as the training network. To predict missing links we erase our knowledge of the true node coordinates and then infer node coordinates by mapping the training network to the hyperbolic disk using the \textsc{hyperlink} embedder (see Appendix~\ref{sec:hyper_inference} for details on the mapping procedure). After the mapping is complete, we use the inferred node coordinates to calculate distances  between all unconnected node pairs in the training network and rank these pairs in the increasing order of distance.

Figures~\ref{fig:2}, \ref{fig:pr}, and \ref{fig:auc} show the results for the AUPR, Precision, and AUC scores, respectively. Each panel in these figures is a heatmap, aggregating the link prediction accuracy scores for RHGs with different $\gamma \in [2.1,2.9]$ and $T \in [0.1,0.9]$ values, which we change with an increment of $0.1$ each. We compare the \textsc{hyperlink} to the RA method, which is its leading competitor in these experiments [cf.~Figs.~\ref{fig:11}(a)~and~~\ref{fig:11}(b)].

The results for the AUPR and Precision scores are similar.
Quantified by these scores, the \textsc{hyperlink} accuracy is nearly independent of the degree distribution exponent~$\gamma$, and at the same time decreases rapidly as temperature $T$ increases, see Figs.~(\ref{fig:2},\ref{fig:pr}){\bf a,d,g}. This observation is consistent with our theoretical analysis in Sec.~\ref{sec:model_true_coords}, where we establish that AUPR scores decrease as $T$ increases and do not strongly depend on $\gamma$.

Even though RA performs similar to \textsc{hyperlink}, Figs.~(\ref{fig:2},\ref{fig:pr}){\bf b,e,h}, we note that RA is more accurate at lower $T$ values and less accurate than \textsc{hyperlink} for higher $T$ values. To obtain the direct comparison of the two methods we plot the difference between their AUPR (Precision) scores in Figs.~\ref{fig:2}{\bf c,f,i}(\ref{fig:pr}{\bf c,f,i}). In agreement with our theoretical considerations in Fig.~\ref{fig:13}, we find that the \textsc{hyperlink} is superior to RA in the region of $\gamma$-$T$ phase space corresponding to higher $T$ values; these regions are denoted with  dashed lines in Figs.~(\ref{fig:2},\ref{fig:pr}){\bf c,f,i}.

Compared to RA, the \textsc{hyperlink} yields better link prediction accuracy for larger fractions of missing links. In the case $1-q = 0.1$, for instance, \textsc{hyperlink} is better than RA in a small upper right corner region of the $\gamma$-$T$ phase space, Fig.~(\ref{fig:2},\ref{fig:pr}){\bf c}. On the other hand, in the case $50\%$ of links are missing, $1-q=0.5$, the \textsc{hyperlink} outperforms RA for the majority of $\gamma$-$T$ values with the exception of smallest, $T = 0.1$, and largest, $T = 0.9$, temperature values, Fig.~(\ref{fig:2},\ref{fig:pr}){\bf i}.

The better, compared to RA, performance of the \textsc{hyperlink} in Fig.~(\ref{fig:2},\ref{fig:pr}){\bf i} is  the result of two effects. On one hand, the \textsc{hyperlink} accuracy appears to increase as $1-q$ increases. This effect is consistent with a recent observation in Ref.~\cite{garcia-perez2020precision} that the upper bound of link predictability in edge-independent graphs increases with $1-q$. On the other hand, as $1-q$ increases, the accuracy of RA decreases. RA, as well as  other similarity-based methods, e.g., RA, Cannistraci resource allocation (CRA), AA, common neighbors (CN), and Jaccard's index (JC), predict missing links based on the similarity of node neighborhoods, e.g, the number of common neighbors; the higher the similarity the higher the probability of a missing link, Appendix~\ref{sec:lp}. Neighborhood similarities are local measures, reflecting network structure in the network-based vicinity of the node pair of interest, and ignoring the structure of the remaining network. The larger the fraction of missing links, the smaller the fraction of links in the training network and, as a result, the poorer the link prediction results. While this is true for all link prediction methods, the similarity-based methods are the ones that suffer  most. Since links are established independently in RHGs, and each link is removed with probability $ p = 1-q$, the number of common neighbors between any node pairs on average decreases proportionally to  $p^{2}$. All extensive RHG properties, on the other hand, depend on $p$ linearly. \textsc{hyperlink} as a global method uses the structure of the entire network to map it, so that it is less sensitive to network incompleteness.

An attractive feature of a global method is that it is capable of predicting \emph{nonlocal missing links}, i.e., links between node pairs with no common neighbors. To quantify \textsc{hyperlink} accuracy for nonlocal links we consider the subset of nonlocal links within the set of links removed with probability $1-q=0.5$,  Fig.~(\ref{fig:2},\ref{fig:pr}){\bf j}, which comprise from $20\%$ (for $\gamma = 2.1$, $T=0.9$) to $86\%$ (for $\gamma=2.9$, $T=0.1$) of all removed links.

Similarity-based methods, RA, AA, CN, and JC, cannot predict nonlocal missing links since corresponding node pairs have no common neighbors at all, and, consequently, have zero similarity. Therefore, in nonlocal link prediction experiments we compare \textsc{hyperlink} to the structural perturbation method (SPM) index, which is a global method and the leading competitor to \textsc{hyperlink} for nonlocal links. As seen in Figs.~(\ref{fig:2},\ref{fig:pr}){\bf k,l}, the SPM index yields substantially lower link prediction accuracy than \textsc{hyperlink} for all the considered values of $\gamma$ and $T$.

Overall, we observe that according to the AUPR and Precision scores \textsc{hyperlink}'s competitive advantage is higher the more incomplete the network is, and the \textsc{hyperlink} is particularly strong in prediction of nonlocal links.

According to AUC scores, the \textsc{hyperlink} offers superior link prediction accuracy across the entire $\gamma$-$T$ parameter space, surpassing its leading competitors---RA for all links, and SPM for nonlocal links, Fig.~\ref{fig:auc}. This result is again consistent with our calculations in Sec.~\ref{sec:model_true_coords} showing that RHG-based AUC scores are robust with respect to coordinate uncertainty.

\subsection{Tests on real networks}
\label{sec:real_infer_coords}

Finally, we apply the \textsc{hyperlink} to real networks:  the network of human metabolism~\cite{ma2003reconstruction}, the Internet at the autonomous system level~\cite{routeviews}, and the Pretty-Good-Privacy (PGP) web of trust~\cite{openpgp}. Basic properties of these networks as well as the data curation steps are documented in Appendix~\ref{sec:real_nets}.

\begin{figure*}
\includegraphics[width=7in]{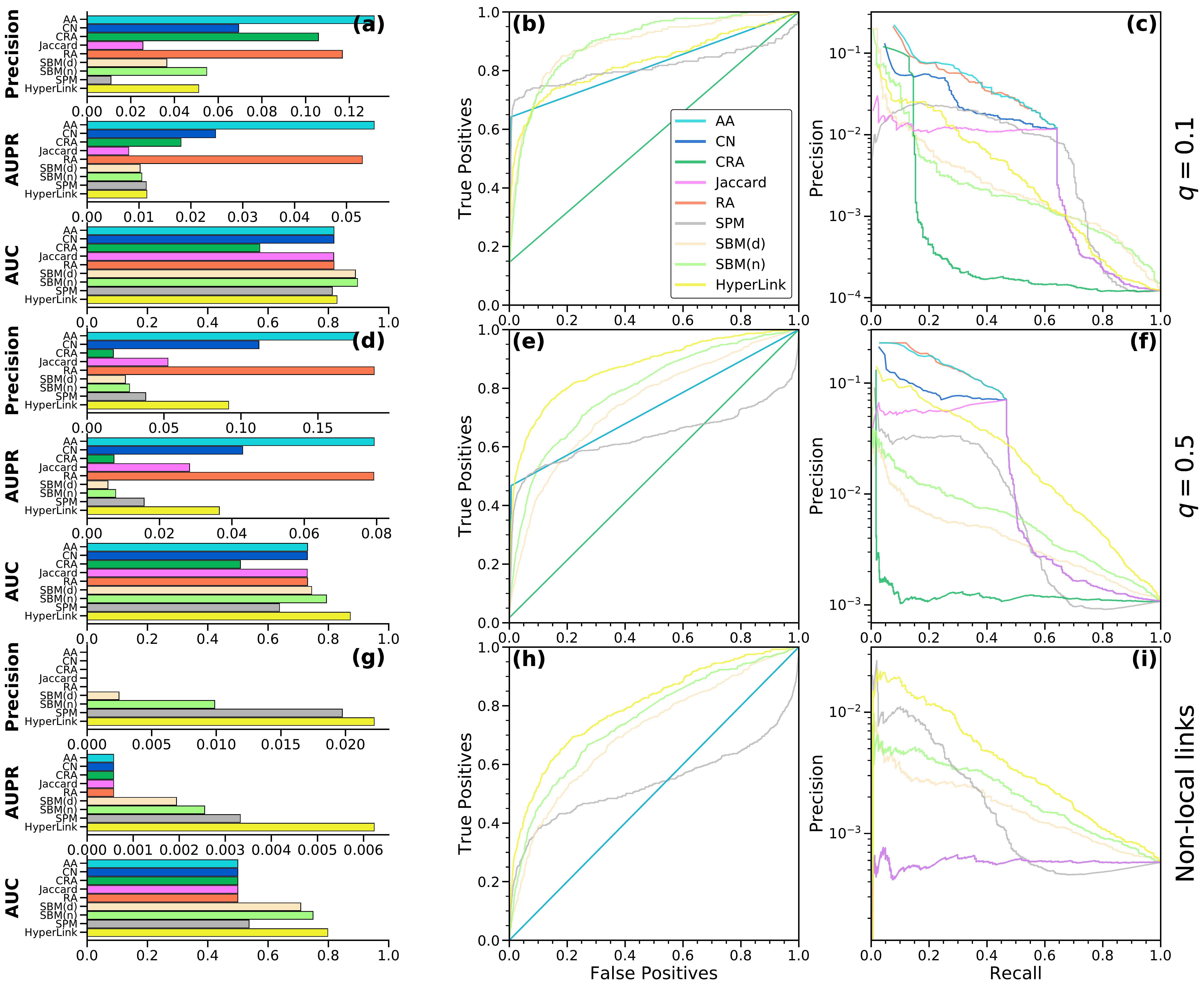}
\caption{\footnotesize  Link prediction accuracy for the Metabolic network with {\bf a-c} $10\%$ ($q=0.9$) randomly missing links, {\bf d-f} $50\%$ ($q=0.5$) randomly missing links, and {\bf g-i} nonlocal missing links, i.e., links connecting node pairs that have no common neighbors. Nonlocal links constitute $20\%$ of the $q=0.5$ missing links set. Panels {\bf a, d, g, j} depict Precision, AUPR, and AUC  link prediction scores. Panels {\bf b, e, h, k} and {\bf c, f, i, l} show, respectively, the ROC and PR curves.
}
\label{fig:3}
\end{figure*}

\begin{figure*}
\includegraphics[width=7in]{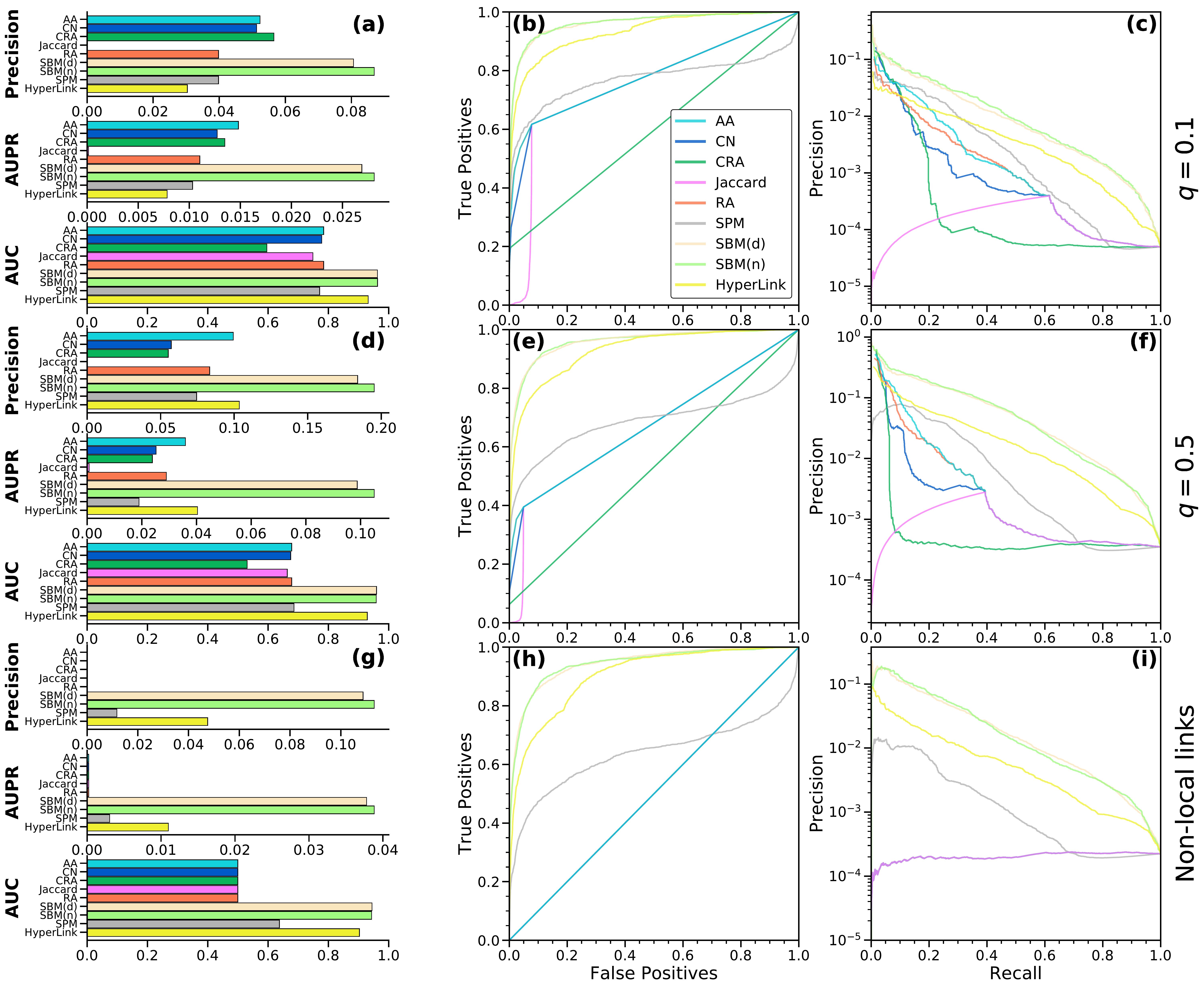}
\caption{\footnotesize  Link prediction accuracy for the Internet. Nonlocal links constitute $32\%$ of the $q=0.5$ missing links set.}
\label{fig:4}
\end{figure*}

\begin{figure*}
\includegraphics[width=7in]{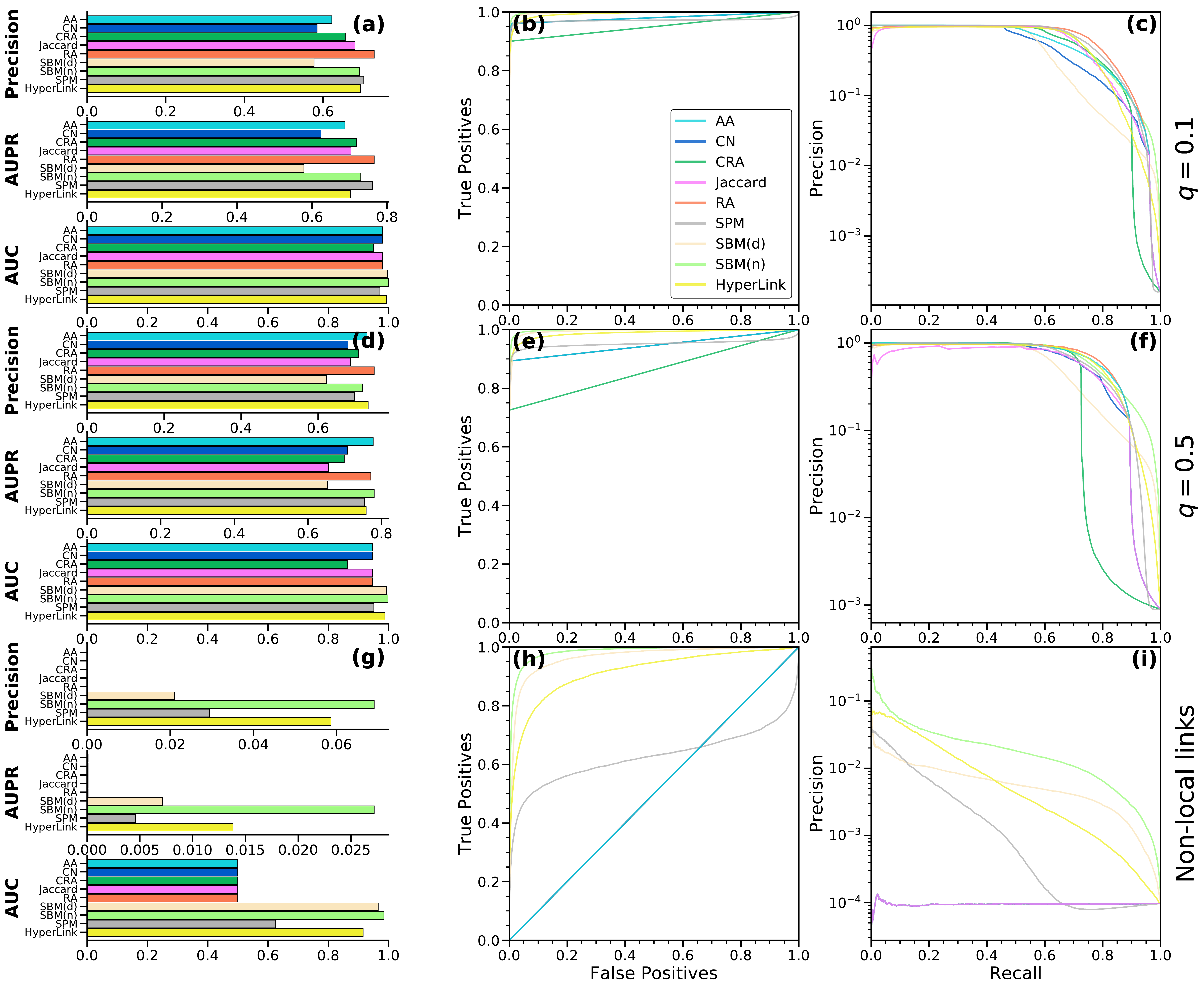}
\caption{\footnotesize  Link prediction accuracy for the PGP network. Panels are identical to those of Fig.~\ref{fig:3}. Nonlocal links constitute $10\%$ of the $q=0.5$ missing links set.}
\label{fig:5}
\end{figure*}

Our link prediction experiments on real networks are performed identically to those on RHGs with inferred coordinates, and the results are shown in Figs.~\ref{fig:3}-\ref{fig:5}.

According to AUPR and Precision metrics, the \textsc{hyperlink} offers competitive performance in random link removal experiments, Figs.~\ref{fig:3}-\ref{fig:5}{\bf(a-f)}, but, at the same time, is not the most accurate. We do note that the relative performance of the \textsc{hyperlink} is better in cases of higher missing link rate, $1-q=0.5$, which is consistent with our results in Sec.~\ref{sec:model_true_coords}.

We also note that the \textsc{hyperlink} offers superior performance in prediction of nonlocal links where it is  either the winner or runner-up, with the SBM methods being its leading competitors, Figs.~\ref{fig:3}-\ref{fig:5}{\bf(g-i)}. This observation comes in sharp contrast with nearly random performance of similarity based methods, RA, AA, CN, JC, and CRA, in nonlocal link prediction.

In contrast to AUPR-based rankings where the \textsc{hyperlink} is rarely the most accurate method, it is either the winner or runner-up in all the experiments according to the AUC metric, in agreement with all the AUC-related results above. In particular, it is the winner in predicting nonlocal links in the most challenging human metabolic network. This network is the most challenging because it is the sparsest and has the lowest clustering, Appendix~\ref{sec:real_nets}, thus providing the least amount of local information for link prediction.

\section{Summary, Discussion, and Conclusion}
\label{sec:discussion}

\begin{table}[!ht]
\begin{center}
\begin{tabular}
 {|c|c|c|}
 \hline Parameter & AUC &  AUPR, Precision \\
 \hline Exponent $\gamma \in (2,3)$  & $\approx {\rm const}$ & $\approx {\rm const}$   \\
 \hline Temperature $T \in (0,1)$ & $\approx {\rm const}$ & decreasing  \\
 \hline Fraction of missing links $1-q$ & increasing & increasing   \\
 \hline Noise amplitude $a$ & $\approx {\rm const}$ & decreasing  \\
 \hline
\end{tabular}
\caption{\footnotesize The summary of the results in Sec.~\ref{sec:model_true_coords}: \textsc{hyperlink}'s measures of accuracy of link prediction in RHGs with known node coordinates as functions of the parameters in Sec.~\ref{sec:model_true_coords}.
\label{table:result1}
}
\end{center}
\end{table}

\begin{table}[!ht]
\begin{center}
\begin{tabular}
 {|c|c|c|}
 \hline Scenario & AUC & AUPR, Precision \\
 \hline \makecell[{t}{c}]{RHGs with \\ inferred coordinates} & \makecell[{t}{c}]{Winner} & \makecell[{t}{c}]{Winner if $T$, $\gamma$, \\ or $1-q$ is large}\\
 \hline \makecell[{t}{c}]{Real networks} & \makecell[{t}{c}]{Winner/ \\ Runner-up} & \makecell[{t}{c}]{The more competitive, \\ the larger the $1-q$}\\
 \hline \makecell[{t}{c}]{Nonlocal links in \\ RHGs and real networks} & \makecell[{t}{c}]{Winner/ \\ Runner-up} & \makecell[{t}{c}]{Winner/ \\ Runner-up}\\
 \hline
\end{tabular}
\caption{\footnotesize The summary of the results in Sec.~\ref{sec:inferred_coords}: \textsc{hyperlink}'s measures of accuracy of link prediction in RHGs with inferred coordinates and in real networks, as well as those for nonlocal links, compared to other methods. The parameters are the same as in Table~\ref{table:result1}.
\label{table:result2}
}
\end{center}
\end{table}

Tables~\ref{table:result1}~and~\ref{table:result2} summarize the results in Secs.~\ref{sec:model_true_coords} and~\ref{sec:inferred_coords}, respectively. We see that when it comes to predicting obvious missing links that are easy to predict employing hyperbolic geometry may be an overkill. In fact, one should consider using much simpler local methods instead of any global ones, according to the AUPR or Precision results presented here.  This is because according to these results the local methods appear to be nearly as good as the global ones at predicting easy links. In particular, the \textsc{hyperlink} method cannot be the best at predicting the most obvious missing links because such links are the links between closest nodes in the latent hyperbolic space, and to rank them exactly at the top of the disconnected node pair list one has to infer the coordinates nearly exactly, Sec.~\ref{sec:model_true_coords}.

However, if the task is to identify missing links that are really hard to predict, then this is  the situation where one should consider using global methods in general and hyperbolic geometry in particular. The most striking example is the prediction of missing links between the nodes that do not share any common neighbors. Here the \textsc{hyperlink} is either the winner or runner-up to the SBM methods, according to all the AUC, AUPR, and Precision measures, in all the considered real and synthetic networks. It is not surprising that local methods do a poor job in predicting such links---they are simply not designed to do so. In contrast,  the \textsc{hyperlink}, SBM, and SPM are global methods that base their decisions on the global structure of the whole network, which helps enormously to predict nonlocal and other hard-to-predict links. The SBM and SPM methods were reported to outperform a vast collection of other methods~\cite{guimera2009missing,peixoto2018reconstructing,Lu2015toward}. Here we see that the \textsc{hyperlink} outperforms even these powerful methods in many cases. In particular, the \textsc{hyperlink} is the winner according to all the scores in the most challenging considered case, which is nonlocal links in the sparsest lowest-clustering network of human metabolic reactions.

We also see that according to the AUC measure, the \textsc{hyperlink} is  either the winner or runner-up in all the considered situations. This is because the AUC does not care that much about false positives, and \textsc{hyperlink} achieves (nearly) the best balance between the true and false positive rates by finding missing links between highly dissimilar nodes located at large distances in the latent hyperbolic space.

We have also shown that the \textsc{hyperlink} is better off the weaker the clustering (the higher the $T$) is, and the larger the fraction of missing links $1-q$ in RHGs with inferred coordinates. This does not mean that \textsc{hyperlink}'s link prediction accuracy scores are getting better in these more difficult conditions; its scores do degrade. But the speeds of the degradation of these that the other methods experience are higher than \textsc{hyperlink}'s.

Our results also resolve the controversy among earlier reports on link prediction using hyperbolic geometry~\cite{Serrano2012uncovering,Papadopoulos2015network1,Papadopoulos2015network,Alessandro2018leveraging,Muscoloni2018minimum,garcia-perez2020precision}. These reports approached link prediction using different measures of link prediction accuracy. To reiterate, if applied to sparse networks, the AUPR emphasizes the prediction of a small fraction of the most likely missing links and, as a result, is extremely sensitive to inaccuracies in the node coordinate inference. On the other hand, the AUC is more robust to coordinate uncertainties as it emphasizes the prediction of less likely missing links between dissimilar nodes at large latent distances.

To maximize \textsc{hyperlink}'s link prediction accuracy, we have developed a hyperbolic network mapping method, the \textsc{hyperlink} embedder, that maximizes the accuracy of coordinate inference. Its accuracy comes at the computational complexity cost of $O\left(n^{2}\right)$. While faster methods for hyperbolic mapping have been developed recently~\cite{Blasius2016efficient,Wang2016link,Alanis-Lobato2016efficient,Alanis-Lobato2016manifold,Muscoloni2017machine}, an optimal balance between the accuracy and speed of hyperbolic mapping is still to be found. Ideally, it would be highly desirable to have a method that would be as accurate as at least the \textsc{hyperlink} embedder, and that would run in $O\left(n\right)$ time.

\begin{figure*}
\includegraphics[width=7in]{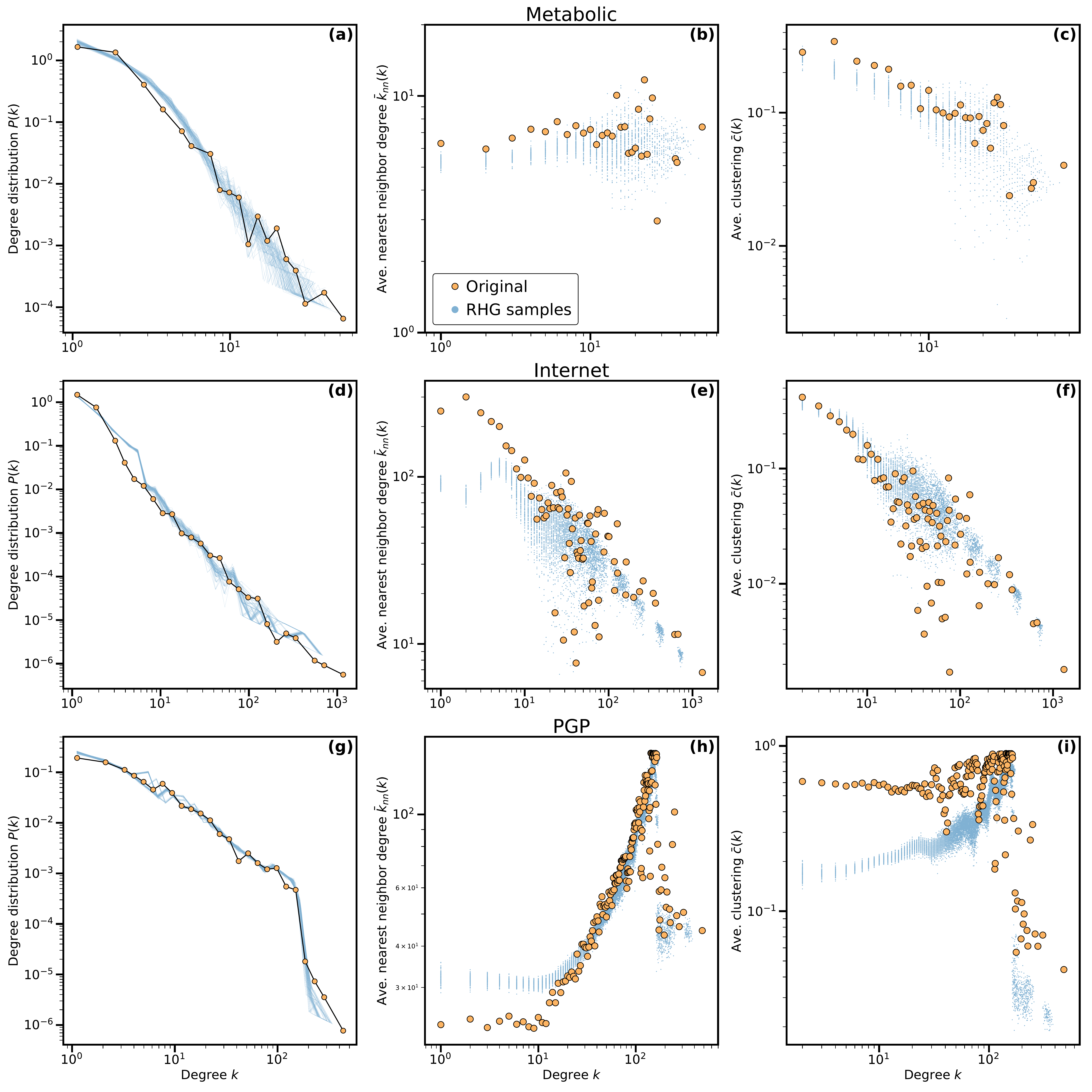}
\caption{\footnotesize The comparison of main structural properties of the three pruned real networks and corresponding 100 pruned RHGs generated using the hyperbolic coordinates learned by the \textsc{hyperlink} embedder. In both real and synthetic networks the pruning is the random link removal with rate $1-q=0.9$. To generate an RHG for a real network of interest we use its parameters $R$ and $T$ and node coordinates that are learned by the HL embedder. For each real network we generate $100$ i.i.d. instances of RHGs by connecting node pairs with probabilities given by Eq.~(\ref{eq:pijq}), where $1 = q = 0.9$ and $p(x)$ is given by Eq.~(\ref{eq:conn}). Panels {\bf a, d, g} show the degree distribution $P(k)$, {\bf b, e, h} the degree-dependent average nearest-neighbor degree $\bar{k}_{nn}(k)$, and {\bf c, f, i} the degree-dependent average local clustering coefficient $\bar{c}(k)$.  Since RHG parameters $R$ and $T$ are inferred using the assumption of uniform angular coordinate distribution, $\rho(\theta) = 1/2\pi$, which is not the case in real networks, we had to adjust the hyperbolic disk radius $R$ in RHGs to match the average degrees in the pruned real and synthetic networks.
}
\label{fig:si-frozen-coords}
\end{figure*}

We emphasize that link prediction using latent hyperbolic geometry is expected to yield good results only if this geometry is there in a given network. That is, the network structure must be consistent with the existence of this geometry. It is well known that RHGs are characterized by sparsity, self-similarity, scale-free degree distributions, and strong clustering, meaning that these properties are necessary conditions for hyperbolic geometry presence. It is also well known that many real networks do possess these properties as well. The results in~\cite{Krioukov2016Clustering} suggest that clustering is also a sufficient condition for network geometricity, but these results apply only to homogeneous large-world networks, and ignore coordinate entropy. That is, in theory, the detailed sufficient conditions for the presence of latent hyperbolic geometry are currently unknown, remaining a subject of ongoing research.  Experimentally it is known however that random hyperbolic graphs are good descriptors of the structure of many real networks. In particular, we are not aware of any other model capable of reproducing self-similarity of real networks, a highly nontrivial property~\cite{Serrano2008}. As far as the more standard structural properties of real networks are concerned, the adequacy of hyperbolic geometry to model them has been documented many times, as early as in~\cite{Papadopoulos2012popularity}. Here we report similar results in Fig.~\ref{fig:si-frozen-coords}.

\begin{figure}[H]
\centering
\includegraphics[width=3.2in]{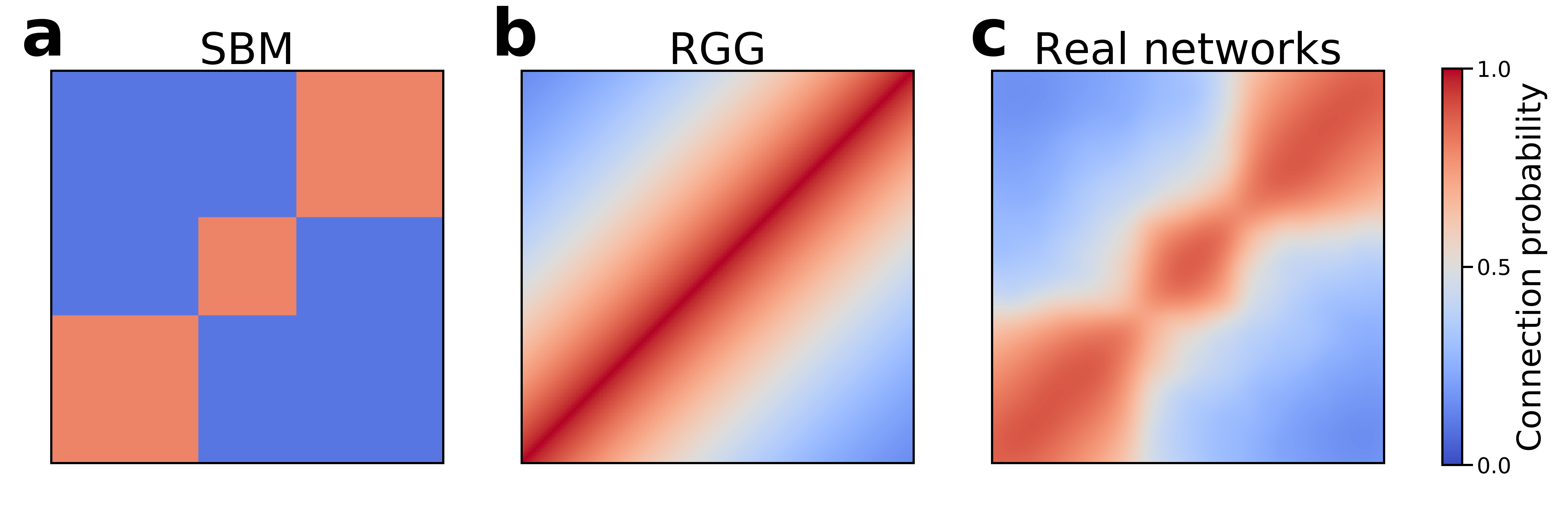}
\caption{\footnotesize Schematic illustration of the connection probabilities as functions of latent variables/coordinates of pairs of nodes in \textbf{(a)}~the stochastic block model (SBM), \textbf{(b)}~random geometric graphs (RGG), and \textbf{(c)}~real networks.}
\label{fig:sbm-rgg-real}
\end{figure}

Overall, it appears that the harder a specific link prediction task, the better the \textsc{hyperlink} is at this task.  Yet the \textsc{hyperlink} is not always the winner even at such hard tasks. In particular, in application to real networks it is often a close runner-up to the stochastic block model methods. These results are consistent with the findings in Ref.~\cite{faqeeh2018scharacterizing}, where the RHG and SBM were compared across a variety of properties. In the SBM the connection probability has a block structure, while in RHGs it is a function of the latent distance, Fig.~\ref{fig:sbm-rgg-real}{\bf a,b}. Clearly neither model can pretend to describe the connection probability in real networks exactly---at least because the RHGs have no communities, while the SBM has no clustering in the large-network limit. In view of the results in Refs.~\cite{colomer2013clustering,faqeeh2018scharacterizing}, the connection probability in real networks is likely to be some nontrivial mixture of the two pictures, Fig.~\ref{fig:sbm-rgg-real}{\bf c}, with geometry appearing as a mesoscopic structure gluing community blocks together. In short, the RHGs and SBM are complementary models capturing different aspects of the structure of real networks, and the link prediction accuracy of a model-based method depends on how prominent and prevalent the model's features are in a given real network.

\section{Acknowledgements}
We thank R.~Aldecoa, F.~Papadopoulos, C.~V.~Cannistraci, M.~\'{A}.~Serrano, T.~Peixoto, and A.~Clauset for useful discussions and suggestions. This work was supported by Army Research Office (ARO) Grant No.~W911NF-17-1-0491 and National Science Foundation (NSF) Grant No.~IIS-1741355. M.~Kitsak was additionally supported by the NExTWORKx project.

\onecolumngrid
\appendix

\section{Real Networks}
\label{sec:real_nets}

\subsection{Metabolic network}
The metabolic network is based on the dataset of metabolic interactions of $107$ organisms constructed by Ma and Zeng~\cite{ma2003reconstruction}. The original network is bipartite and consists of metabolites (top domain) connected to chemical reactions (bottom domain). We consider the unipartite projection of the network on the top domain. Basic properties of the metabolic network are summarized in Table~\ref{table:1}.
\subsection{Internet}
The Internet network is a snapshot of the autonomous system level Internet taken from the University of Oregon Route Views Project~\cite{routeviews}. The full dataset contains $733$ daily instances which span an interval of $785$ days from November 8 1997 to January 2 2000. Here we use a network instance as of January 2, 2000~\cite{metabolicdetails}.
\subsection{PGP web of trust}
PGP is a data encryption and decryption computer program that provides cryptographic privacy and authentication for data communication~\cite{openpgp}. The data is collected and maintained by Cederl\"{o}f~\cite{cederlof}. In the paper we use the PGP snapshot taken in April of 2003. The PGP web of trust is a directed network where nodes are certificates consisting of public PGP keys and owner information. A directed link in the web of trust pointing from certificate A to certificate B represents a digital signature by the owner of A endorsing the owner/public key association of B. We construct the undirected PGP graph by taking into account only bi-directional trust links between the certificates. Further, we only consider the giant connected component of the resulting undirected PGP web of trust network. Basic properties of the PGP network are summarized in Table~\ref{table:1}.

\begin{table}[!ht]
\begin{center}
\begin{tabular}
 {|c|c|c|c|c|c|c|}
 \hline Network name& $N$ &  $E$ & $\overline{k}$ & $\gamma$ & $\overline{c}$ & $T$  \\
 \hline Internet  & $6,474$ & $13,234$ & $4.09$ &     $2.1$ &  $0.51$ & $0.7$  \\
 \hline Metabolic network & $2,732$ & $4,040$ & $2.96$ & $2.9$ & $0.29$ & $0.6$  \\
 \hline PGP web of trust & $14,138$ & $160,080$  & $22.65$ & $2.1$ & $0.66$ & $0.8$   \\
 \hline
\end{tabular}
\caption{\footnotesize Basic properties of the considered real networks. $N$ is the number of top nodes; $E$ is the number of edges; $\overline{k}$ is the average degree; $\gamma$ is the degree distribution exponent, which we estimated using methods from Ref.~\cite{voitalov2019scale}; $\overline{c}$ is the average degree-dependent clustering coefficient; and $T$ is the corresponding RHG temperature.
\label{table:1}}
\end{center}
\end{table}

\section{Link Prediction: Alternative Methods and Scoring Techniques}
\label{sec:lp}

We compare the accuracy of the \textsc{hyperlink} link prediction method against the following set of link prediction methods: CN~\cite{Liben2003link},  AA~\cite{Adamic2003friends},  RA~\cite{Zhou2009predicting}), CRA~\cite{Cannistraci2013b}, JC~\cite{Jaccard1901}, SPM~\cite{Lu2015toward}, and SBM~\cite{guimera2009missing,peixoto2018reconstructing} methods.

All these methods, as well as the \textsc{hyperlink}, assign scores to (a subset of) all not directly connected pairs of nodes (nonlinks), and all such pairs are then ranked according to these scores from the most to least likely interaction prediction. To briefly describe these methods, it is thus sufficient to tell how these scores are calculated, for which we use the following notations: $k_{i}$ is the degree of node $i$; $\Gamma(i)$ is the set of $i$'s neighbors (directly connected nodes); $\gamma_{ij}(s)$ is the subset of all $\Gamma(s)$ that are neighbors of both $i$ and $j$; $e_{i}^{j}$ is $i$'s \emph{$j$-external degree}, the number of $i$'s neighbors that are \emph{not} $j$'s neighbors; $\pmb{A}$ is the network adjacency matrix.\\

\subsection{Common Neighbors (CN)}

The score for a pair of nodes $i$ and $j$ is defined as the cardinality of the intersection of their sets of neighbors,
\begin{equation}\label{eq:cn}
    s_{ij}^{CN} = |\Gamma(i) \cap \Gamma(j)|.
\end{equation}

\subsection{Jaccard's index (JC)}

The score is a normalized measure of the overlap of $i$'s and $j$'s sets of neighbors,
\begin{equation}\label{eq:jc}
    s_{ij}^{JC} = \frac{|\Gamma(i) \cap \Gamma(j)|}{|\Gamma(i) \cup \Gamma(j)|}.
\end{equation}

\subsection{Adamic-Adar index (AA)}

The score assigns more weight to the less-connected neighbors,
\begin{equation}\label{eq:aa}
    s_{ij}^{AA} = \sum_{s \in \Gamma(i) \cap \Gamma(j)} \frac{1}{\log{k_{s}}}.
\end{equation}

\subsection{Resource Allocation index (RA)}

The score is similar to the AA score, but punishes high-degree nodes more strongly,
\begin{equation}\label{eq:ra}
    s_{ij}^{RA} = \sum_{s \in \Gamma(i) \cap \Gamma(j)} \frac{1}{k_{s}}.
\end{equation}

\subsection{Cannistraci Resource Allocation index (CRA)}

The score is similar to the RA score, but takes into account the subset of nodes shared between nodes $i$, $j$ and their common neighbors $s$:
\begin{equation}\label{eq:cra}
    s_{ij}^{RA} = \sum_{s \in \Gamma(i) \cap \Gamma(j)} \frac{\gamma_{ij}(s)}{k_{s}}.
\end{equation}

\subsection{Structural Perturbation Method (SPM)}

This method is based on repetitive perturbations of the adjacency matrix $\pmb{A}$ by removals of small fractions of links that we denote by $\Delta E$. The original adjacency matrix can then be written as $\pmb{A} = \pmb{A'} + \pmb{\Delta A}$, where $\pmb{A'}$ is the adjacency matrix of the network after removal of links $\Delta E$, and $\pmb{\Delta A}$ is the adjacency matrix constructed on the set of removed links $\Delta E$. Denoting eigenvectors and eigenvalues of $\pmb{A'}$  by $x_k$ and $\lambda_k$, the perturbations of the original eigenvalues $\lambda_k$ using the perturbation matrix $\pmb{\Delta A}$, are
\begin{equation}\label{eq:spm:perturbation}
    \Delta \lambda_k \approx \frac{x_{k}^{T} \pmb{\Delta A} x_{k}}{x_{k}^{T} x_{k}},
\end{equation}
so that the perturbed adjacency matrix is
\begin{equation}\label{eq:spm:pertmatrix}
    \pmb{\widetilde{A}} = \sum_{k = 1}^{N} (\lambda_k + \Delta \lambda_k) x_k x_k^T.
\end{equation}
All nonlinks $i,j$ are then ranked by $\widetilde{A}_{ij}$. In our experiments we repeat this perturbation procedure ten times, and then average perturbed matrices over these trials, thus obtaining an averaged perturbed matrix $\langle \pmb{\widetilde{A}} \rangle$, so that the SPM score is
\begin{equation}
    s_{ij}^{SPM}=\langle \widetilde{A}_{ij} \rangle.
\end{equation}

\subsection{Stochastic Block Model (SBM)}

The stochastic block model is a generative network model designed to model community structure. Nodes are partitioned into groups (blocks) forming a node partition $\boldsymbol{b}$. The number of links between blocks is given by a matrix $\boldsymbol{e}$ the elements $e_{rs}$ of which are the numbers of links between blocks $r$ and $s$. If the observed degree sequence of a network $\boldsymbol{k}$ is used as an additional model parameter, the model is called \textit{degree-corrected} SBM~\cite{peixoto2017nonparametric}. Moreover, if node blocks are themselves clustered into groups, and these groups are organized into higher-level groups, and so on recursively up to some nestedness level $l$, the model is called \textit{nested} SBM. It can capture hierarchical and fine-grained structural properties of a given network~\cite{Peixoto2014}. In our experiments, we use both degree-corrected and nested SBMs denoted as \textit{SBM(d)} and \textit{SBM(n)} in the figures. We rely on the \texttt{graph-tool} library~\cite{peixoto_graph-tool_2014} in the procedures below. Given the observed data (network adjacency matrix) $\boldsymbol{\mathcal{D}}$, and the prior probability density $P(\boldsymbol{A}, \boldsymbol{b})$ given by the network block structure $\boldsymbol{b}$ that produces a network with adjacency matrix $\boldsymbol{A}$ in the model, we reconstruct the full network using the posterior distribution:
\begin{equation}\label{eq:sbm:posterior}
    P(\boldsymbol{A}, \boldsymbol{b} | \boldsymbol{\mathcal{D}}) = \frac{P(\boldsymbol{\mathcal{D}} | \boldsymbol{A}) P(\boldsymbol{A}, \boldsymbol{b})}{P(\boldsymbol{\mathcal{D}})},
\end{equation}
where $P(\boldsymbol{\mathcal{D}}|\boldsymbol{A})$ describes the measurement process of a network. We avoid computing the normalization factor $P(\boldsymbol{\mathcal{D}})$ by Markov Chain Monte Carlo (MCMC) sampling from the joint posterior distribution $P(\boldsymbol{A}, \boldsymbol{b} | \boldsymbol{\mathcal{D}})$ as described in~\cite{peixoto2018reconstructing}. In our experiments, we assume that each network link is observed and measured once. A possible link $(i,j)$ then has marginal probability
\begin{equation}\label{eq:sbm:marginal}
    \pi_{ij} = \sum\limits_{\boldsymbol{A},\boldsymbol{b}} a_{ij} P(\boldsymbol{A}, \boldsymbol{b} | \boldsymbol{\mathcal{D}}).
\end{equation}
To sample over $(\boldsymbol{A},\boldsymbol{b})$ configurations, the MCMC algorithm is initialized with the block structure obtained by the procedure from~\cite{peixoto2014efficient}. The MCMC is equilibrated using $10|E|$ equilibration steps, where $E$ is the set of links in a given network. For the PGP network, due to its large size, we use only $2|E|$ MCMC equilibration steps. Then $T = 10$ epochs of MCMC iterations, $1,000$ swaps each, are performed to sample different block-network configurations. After each epoch, marginal link probabilities from Eq.~\eqref{eq:sbm:marginal} are collected. These probabilities are then averaged over the epochs to obtain a single score used for link prediction:
\begin{equation}\label{eq:sbm:score}
    s_{ij}^{SBM} = \frac{1}{T} \sum_{t = 1}^{T} \pi_{ij}^{(t)}.
\end{equation}

\section{Basic properties of the RHG}
\label{sec:hrg_prop}
The hyperbolic geometry inference algorithm relies on several properties of the RHG, which we review in this section.

\emph{Degree distribution.} RHGs  are characterized by scale-free degree distributions, $P(k) \sim k^{-\gamma}$, where $\gamma = 2\alpha + 1$. Indeed, the expected degree of a node located at $(r,\theta)$ is independent of its angular coordinate $\theta$, $\overline{k}(r, \theta)= \overline{k}(r, 0) = \overline{k}(r)$, and   is given by
\begin{eqnarray}
\overline{k}(r) &=& (N-1)\int {\rm d} r' \rho(r') \int {\rm d} \theta' \rho(\theta')   p\left[ x (r, 0, r', \theta')
 \right]  \nonumber \\
 &\approx& \frac{4N\alpha}{2\alpha - 1} \frac{T}{\sin \pi T}  e^{-r/2},
\label{eq:avgk_r}
\end{eqnarray}
see~\cite{Krioukov2010hyperbolic}. The average degree of the model is given by
\begin{equation}
\overline{k} = \int{\rm d} r \rho(r) \overline{k}(r) = \frac{8N\alpha^{2}}{ \left(2\alpha - 1\right)^{2}} \frac{ T}{\sin \pi T}  e^{-R/2}.
\label{eq:avgk}
\end{equation}
As seen from Eq.~(\ref{eq:avgk}), $\overline{k}$ in the most general case depends on the network size $N$.

To achieve sparse models with $\overline{k}$ independent of $N$ one sets the radius of the hyperbolic disk to
\begin{equation}
R(N) = 2 \ln \left(N/\nu \right),
\label{eq:r_sparse}
\end{equation}
where $\nu > 0$ is the tuning parameter, directly related to $\overline{k}$. Indeed, with $R(N)$ given by (\ref{eq:r_sparse})
\begin{equation}
\overline{k} =  \frac{8\nu\alpha^{2}}{\left(2\alpha - 1\right)^{2}} \frac{T}{\sin \pi T},
\label{eq:avgk2}
\end{equation}
prescribing the value of $\nu$ for the target values of $\overline{k}$, $\alpha$, and $T$.

It has been shown in \cite{boguna2003class} that in the sparse limit the probability of a node located at $(r,\theta)$ to have $k$ connections can be approximated with the Poisson distribution with the mean of $\overline{k}(r)$:
\begin{equation}
P(k|r) = e^{-\overline{k}(r)} \frac{\left[\overline{k}(r)\right]^{k}}{k!}.
\label{eq:cond_pkr}
\end{equation}
Then the degree distribution of the RHG is
\begin{eqnarray}
P(k) &=& \int {\rm d} r \rho(r) P(k|r) \sim k^{-\gamma},\\
\gamma &=& 2\alpha +1.
\label{eq:gamma}
\end{eqnarray}
It follows from  Eqs.~(\ref{eq:avgk_r}) and (\ref{eq:avgk2}) that model parameters $\alpha$ and $R$ can be used to control degree distribution exponent $\gamma$ and the average degree of the model, respectively.

\emph {Clustering coefficient.} As seen from Eq.~(\ref{eq:conn}), connection probability $p(x)$ decreases exponentially for distances $x > R$ with the rate of $\frac{1}{2}T$. Thus, the temperature parameter $T$ tunes the role of large distances in the formation of links: the higher the $T$ the more likely are long-distance connections. As a result, $T$ controls the clustering coefficient of the RHG. In the $T \to 0$ limit connections are only possible at hyperbolic distances $x < R$ and the clustering coefficient is maximized. Conversely, the clustering coefficient decreases  as $T$ increases and vanishes asymptotically in the $T \geq 1$ case~\cite{Krioukov2010hyperbolic}.

\section{HYPERLINK Accuracy}
\label{sec:hl_accuracy}

In this section we calculate analytically the \textsc{hyperlink} accuracy, in terms of AUC and AUPR, on RHGs with known coordinates. Our results in this section are confirmed by the numerical experiments in Sec.~\ref{sec:model_true_coords} and build our intuition for Sec.~\ref{sec:inferred_coords},
where we analyze \textsc{hyperlink} on RHGs and real networks with inferred coordinates.

\subsection{AUC}
To understand the behavior of AUC  scores as a function of RHG parameters we define distance-dependent true positive ${\rm tpr}(x)$ and false positive ${\rm fpr}(x)$ rates as the fractions of true and false positives, respectively, contained among unconnected node pairs separated by distances up to $x$:
\begin{eqnarray}
{\rm tpr}(x) &=& \frac{{\rm tp}(x)}{(1-q) E} = \frac{1}{E} \binom{N}{2} \int_{0}^{x} n(y) p(y){\rm d} y,
\label{eq:tprx}\\
{\rm fpr}(x) &=& \frac{ \binom{N}{2} \int_{0}^{x} n(y)\left(1- p(y)\right){\rm d} y}{\binom{N}{2}-E},
\label{eq:fprx}
\end{eqnarray}
where $E$ is the true number of links in the network, $E = |\Omega_{E}\cup \Omega_{R}|$, $p(y)$ is the connection probability in the RHG given by Eq.~(\ref{eq:conn}) and $n(y)$ is the distance distribution for node pairs in the RHG, given by Eq.~(\ref{eq:nx}).

It is seen from Eqs.~(\ref{eq:tprx}) and (\ref{eq:fprx}) that in the $T\to 0$ limit $p(y) = \Theta(R-y)$, resulting in ${\rm fpr}(x) = 0$ for $x \leq R$ and  ${\rm tpr}(x) = 1$ for $x \geq R$, resulting in the ideal ROC curve, Fig.~\ref{fig:typical_roc_pr}{\bf a}, and  ${\rm AUC}=1$.

Using the expression for $n(y)$ from  Eq.~(\ref{eq:nx}), we can evaluate true and false positive rates, up to the proportionality coefficient, as:
\begin{eqnarray}
{\rm tpr}(x) &\approx& \frac{4 \alpha^{2}}{\pi \left(2 \alpha - 1\right)^{2} } \frac{N^{2}}{E} e^{-\frac{R}{2}} I\left(e^{\frac{x-R}{2}};T\right),\\
\label{eq:tprx1}
{\rm fpr}(x) & \approx& \frac{8 \alpha^{2}}{\pi \left(2 \alpha - 1\right)^{2} } e^{-R} \left[  e^{\frac{x}{2}} - e^{\frac{R}{2}}I\left(e^{\frac{x-R}{2}};T\right)\right],
\label{eq:fprx1}
\end{eqnarray}
where
\begin{equation}
I\left(z;T\right) \equiv \int_{0}^{z} \frac{{\rm d} x}{1 + x^{1/T}} = z~_{2}F_{1}\left(1, T, 1+T, -z^{1/T} \right),
\label{eq:int}
\end{equation}
and$~_{2}F_{1}$ is the Gaussian hypergeometric function. In the $z \ll 1$ regime $I\left(z;T\right) \approx z$ and, thus, ${\rm tpr}(x) \sim e^{\frac{x-R}{2}}$  and  ${\rm fpr}(x) \approx 0$  for $x < R$, Fig.~\ref{fig:12}{\bf a,b}.

In the $z \gg 1$ regime $I\left(z;T\right) \approx I(T) $, where $I(T) = \frac{\pi}{T \sin \left(\pi/T\right)}$, explaining the saturation of the true positive rate, ${\rm tpr}(x) \to 1$ as $x$ approaches $2R$, and the exponential growth of the false positive rate, ${\rm fpr}(x) \sim e^{\frac{x}{2}}$ for $x>R$, Fig.~\ref{fig:12}{\bf a,b}.

To obtain the analytical estimate of the ${\rm AUC}$ as a function of RHG parameters we represent it as
\begin{equation}
{\rm AUC} = \int_{0}^{2R}  {\rm tpr}(x) {\rm fpr}'(x) {\rm d} x
\end{equation}
By making use of Eqs.~(\ref{eq:tprx}) and (\ref{eq:fprx}) we arrive at
\begin{eqnarray}
{\rm AUC} & =& 1 - \Delta_{1} - \Delta_{2},\\
\Delta_{1} &=& \frac{E}{\binom{N}{2}},\\
\Delta_{2} &=& - \frac{1}{E}\binom{N}{2} \int_{0}^{2R} \left[n^{c}(x)\right]^{2} p'(x) {\rm d}x,
\label{eq:auc_correction}
\end{eqnarray}
where $n^{c}(x) \equiv \int_{0}^{x} n(y) {\rm d} y$.

In the case of sparse networks the first correction term $\Delta_{1} \sim N^{-1}$ and can be ignored in the large $N$ limit. The second correction term requires further analysis. It is straightforward to verify that in the $T\to 0$ limit $\Delta_{2} \sim N^{-1}$ and can also be ignored. Indeed, in this case $p'(x)=-\delta(x-R)$, and
\begin{equation}
\Delta_{2}(T=0) = \frac{1}{E}\binom{N}{2}\left[ n^{c}(R)\right]^{2}.
\end{equation}
Since $\binom{N}{2}\left[ n^{c}(R)\right]$ equals the number of node pairs in the hyperbolic disk with distances up to $R$ and all these node pairs are connected in the $T\to 0$ case, $\binom{N}{2}\left[ n^{c}(R)\right] = E$, resulting in $\Delta_{2}(T=0) = \frac{E}{\binom{N}{2}}  \sim N^{-1}$.

To estimate the behavior of $\Delta_{2}$ in the case of $T > 0$ we need to understand the behavior of its integrand in Eq.~(\ref{eq:auc_correction}). Since $n^{c}(x) \sim e^{\frac{x}{2}}$  and $-p'(x) = \frac{1}{2T} \exp\left(\frac{x-R}{2T}\right) \left[p(x)\right]^{2}$, the integrand is sharply peaked at $x=R+ 2T \ln \left( \frac{1+T}{1-T}\right)$ in the case of $T \in \left(0, \frac{1}{2}\right)$, resulting in $\Delta_{2} \sim N^{-1}$, similar to the $T\to 0$ case.

Conversely, the integrand in Eq.~(\ref{eq:auc_correction}) grows monotonously as a function of $x$ in the case of $T \in \left(\frac{1}{2}, 1\right)$. The evaluation of $\Delta_{2}$ in this regime is quite involved and is not informative. Instead, we elect to compute the upper bound for $\Delta_{2}$, which also provides the lower bound for AUC scores. In doing so we note that the leading term behavior of $n(x)$ given by Eq.~(\ref{eq:nx}) is also its upper bound, see Ref.~\cite{Alanis-Lobato2016distance}. Then
\begin{equation}
\Delta_{2} \leq \frac{2 \alpha^{2} e^{-R} \binom{N}{2}}{\pi T \left(2\alpha -1 \right)^{2} E}  \int_{0}^{2R} \frac{e^{\left(x-R\right)\left(1 + \frac{1}{2 T}\right)}{\rm d} x}{\left[1 + e^{\frac{x-R}{2 T}}\right]^{2}} \sim N^{1-\frac{1}{T}},
\label{eq:delta2_scaling}
\end{equation}
since $e^{\frac{R}{2}} \sim N$ in the case of sparse RHGs, see Eq.~(\ref{eq:r_sparse}). In the case of $T=\frac{1}{2}$ Eq.~(\ref{eq:delta2_scaling}) simplifies to
\begin{equation}
\Delta_{2} \leq \frac{4 \alpha^{2} e^{-R} \binom{N}{2}}{\pi \left(2\alpha -1 \right)^{2} E}  \int_{0}^{2R} \frac{  e^{2\left(x-R\right)}  {\rm d} x }{\left[1 + e^{x-R}\right]^{2}} \sim \frac{\ln N}{N}.
\end{equation}

Taken together, the results above show that  the AUC scores for RHGs with known coordinates converge to $1$ in the large $N$ limit as
\begin{equation}
1-AUC
\begin{cases}
\sim N^{-1} &\text{if $ T \in \left.\left[0, \frac{1}{2}\right. \right) $ },\\
=\mathcal{O} \left(\frac{\ln N}{N}\right) &\text{if $ T =\frac{1}{2} $ },\\
=\mathcal{O} \left(N^{1-\frac{1}{T}}\right)  &\text{if $ T \in \left(\frac{1}{2},1\right) $ }
\end{cases}
\end{equation}

\subsection{AUPR}
AUPR scores  can be evaluated in a similar fashion:
\begin{equation}
{\rm AUPR} = \int_{0}^{2R}  {\rm pr}(x)  {\rm rc}'(x) {\rm d} x,
\end{equation}
where ${\rm pr}(x)$ and $ {\rm rc}(x)$ are, respectively, distance-dependent precision and recall functions for hyperbolic distances up to $x$:
\begin{eqnarray}
{\rm pr}(x) &\equiv& \frac{{\rm tp}(x)}{N_{d}(x)},
\label{eq:pr}\\
{\rm rc}(x)  &\equiv& {\rm tpr}(x) =  \frac{{\rm tp}(x)}{(1-q)E},
\label{eq:rc}
\end{eqnarray}
where $N_{d}(x)$ is the number of disconnected node pairs with distances up to $x$:
\begin{equation}
N_{d}(x) = \binom{N}{2}  \int_{0}^{x}  n(y) \left[1 - q p(y)\right]{\rm d} y.
\label{eq:nd}
\end{equation}

Using Eqs.~(\ref{eq:tprx}) and (\ref{eq:nd}) we obtain
\begin{eqnarray}
{\rm pr}(x)  & =& (1-q)\frac{\int_{0}^{x} n(y) p(y){\rm d} y}{\int_{0}^{x}  n(y) \left[1 - q p(y)\right]{\rm d} y} \label{eq:pr2}, \\
{\rm rc}(x)  &=& \frac{1}{E} \binom{N}{2} \int_{0}^{x} n(y) p(y){\rm d} y.
\label{eq:rc2}
\end{eqnarray}

In the $T\to 0$ limit ${\rm pr}(x) = 1$ for all $x<R$, while ${\rm rc}(x) = E(x)/E$, resulting, as expected, in ${\rm AUPR} = 1$. Here $E(x)$ is the cumulative number of links between the node pairs with distances up to $x$.

In the $T> 0$ case we rely on Eqs.~ (\ref{eq:tprx}), (\ref{eq:nx}), and (\ref{eq:nd}) to obtain
\begin{eqnarray}
{\rm tp}(x) & \approx& \frac{4 \alpha^{2} (1-q)}{\pi \left(2 \alpha - 1\right)^{2} } \binom{N}{2} e^{-\frac{R}{2}} I\left(e^{\frac{x-R}{2}};T\right),\\
{\rm pr}(x) & \approx& \frac{(1-q) I\left(e^{\frac{x-R}{2}};T\right)} {e^{\frac{x-R}{2}} - q I\left(e^{\frac{x-R}{2}};T\right)},\label{eq:pr3}
\end{eqnarray}
where $I\left(z;T\right)$ is given by Eq.~(\ref{eq:int}).

In the $x \ll R$ regime $I\left(e^{\frac{x-R}{2}};T\right) \sim e^{\frac{x-R}{2}}$ and ${\rm pr}(x) \to 1$. In the $x \gg R$ case $I\left(e^{\frac{x-R}{2}};T\right)  \sim \frac{\pi}{T \sin \left(\pi/T\right)}$, and as a result, precision decays exponentially, ${\rm pr}(x) \sim e^{-x/2}$, independent of $T$, Fig.~\ref{fig:12}{\bf c}.

The dependence of ${\rm AUPR}$ on $T$ arises from the recall function or its derivative, $rc'(x)$, quantifying the
expected distance-dependent link density and, consequently, the density of missing links.
\begin{equation}
{\rm rc}'(x)  = \frac{1}{E} \binom{N}{2}  n(x) p(x).
\label{eq:ecx}
\end{equation}
$E_{c}(x)$ grows exponentially as $e^{x/2}$ for $x \ll R$ values and decays as $e^{x\left(1- \frac{1}{T}\right)}$ for $x \gg R$, reaching the maximum at $x^{*} = R - 2T \ln \left(\frac{1}{T}-1\right)$, Fig.~\ref{fig:12}{\bf d}.  Thus, as $T$ increases, the missing links are more likely to be located at larger distances where precision ${\rm pr}(x)$ is smaller, resulting in lower AUPR scores, consistent with the observations in Fig.~\ref{fig:11}.

\section{Effects of Coordinate Uncertainty on HYPERLINK Accuracy}
\label{sec:uncertainty}
To understand the effects of coordinate uncertainties on \textsc{hyperlink} accuracy we model coordinate inference uncertainty as synthetic noise that we add to true angular coordinates of the RHG. In the following we first generate RHG as described in Sec.~\ref{sec:hrg} and then simulate uncertainties of angular coordinates by adding synthetic noise to original angular coordinates:
\begin{eqnarray}
\hat{\theta}_{i} &=& \theta_{i} + a X_{i},\\
X_{i}  &\leftarrow& U\left(-\frac{\pi} {2}, \frac{\pi}{2}\right),
\end{eqnarray}
where $a > 0$ is the noise amplitude. Further, we conduct link prediction experiments by calculating latent distances with uncertain coordinates:
\begin{equation}
\hat{x}_{ij} = x(r_i, \hat{\theta}_{i}, r_j, \hat{\theta}_{j}),
\end{equation}
where $x$ is calculated according to the hyperbolic law of cosines, Eq.~(\ref{eq:hypercos}).

\subsection{Link prediction with noise}
\label{sec:auprnoise}
In the case of synthetic noise, the AUPR scores are still given by Eq.~(\ref{eq:aupr}) with effective precision and recall rates ${\rm pr}(y|a)$ and ${\rm rc}(y|a)$ evaluated in the presence of noise. To calculate these rates we start with the effective true positive rate ${\rm tp}(y|a)$.

To this end, we first define the subgraph $G_{y}$ obtained from the RHG $G$ by keeping only links between node pairs separated by distances at most $y$. Then, it is easy to realize that the true positive rate ${\rm tp}(y)$  is proportional to the expected degree $\overline{k}_{y}$ of the $G_y$:
\begin{equation}
{\rm tp}(y) = (1-q)\frac{N}{2} \overline{k}_{y}.
\end{equation}

$\overline{k}_{y}$ can be calculated using the hidden variable formalism
\begin{equation}
\overline{k}_{y} = (N-1) \idotsint_{x \left( r_1, \theta_1, r_2,\theta_2 \right) \leq y} {\rm d} r_1 {\rm d} r_2 {\rm d} \theta_1 {\rm d} \theta_2 \rho(r_1) \rho(r_2) \rho(\theta_1) \rho(\theta_2) p\left[x \left( r_1, \theta_1, r_2,\theta_2 \right) \right]
\end{equation}

To account for noise we next define noisy subgraph $G_{y}(a)$ as follows. First, noise is added to node coordinates of the original RHG as prescribed by Eq.~(\ref{eq:noise2}) and hyperbolic distances between nodes are recalculated using the updated coordinates. Second, $G_{y}(a)$ is formed from RHG by keeping connections at \emph{recalculated} distances up to $y$. It is then easy to see that the thought true positive rate is given by
\begin{equation}
{\rm tp}(y|a) = (1-q)\frac{N}{2} \overline{k}_{y}(a),
\end{equation}
where $\overline{k}_{y}(a)$ is the average degree of noisy subgraph $G_{y}(a)$.

\begin{figure}
\includegraphics[width=3in]{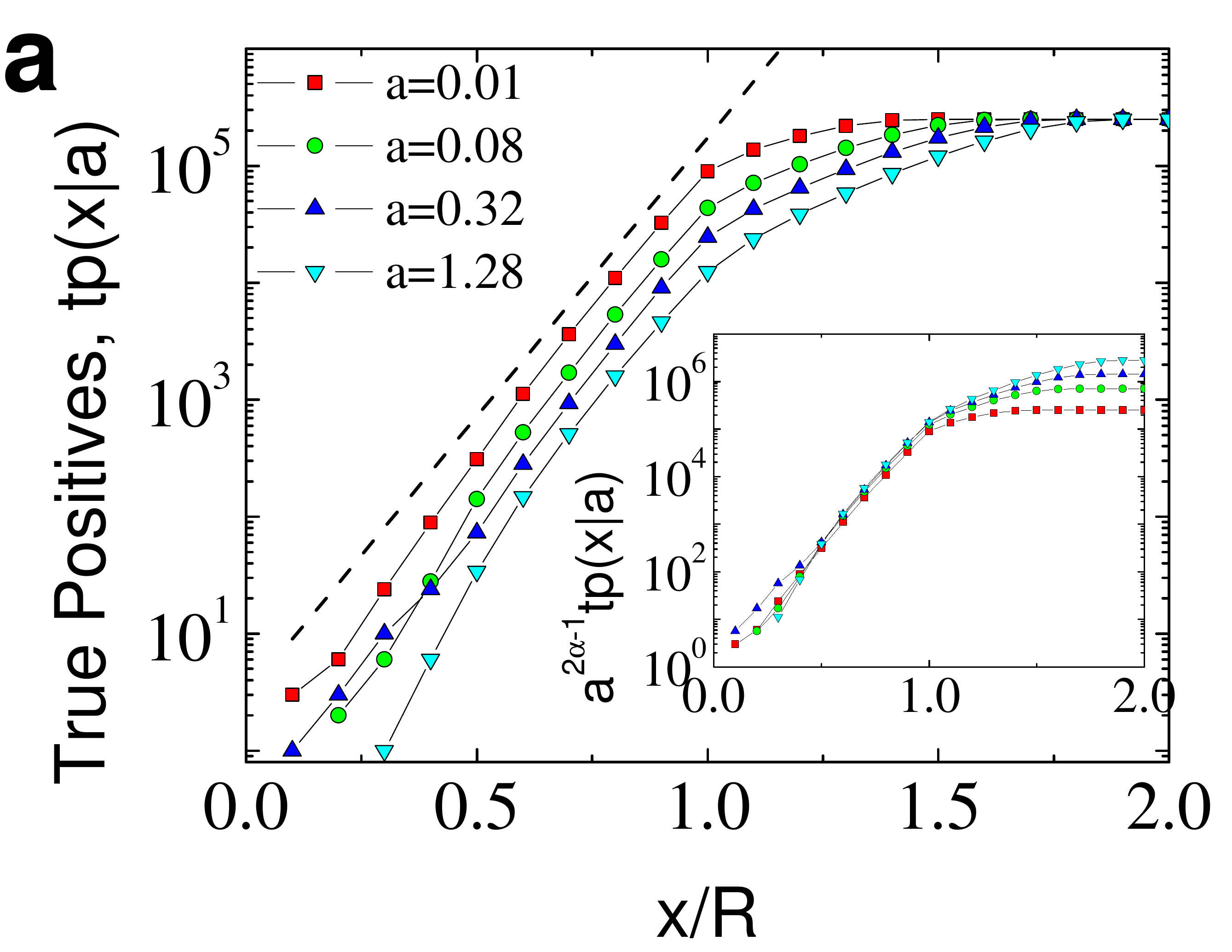}
\includegraphics[width=3in]{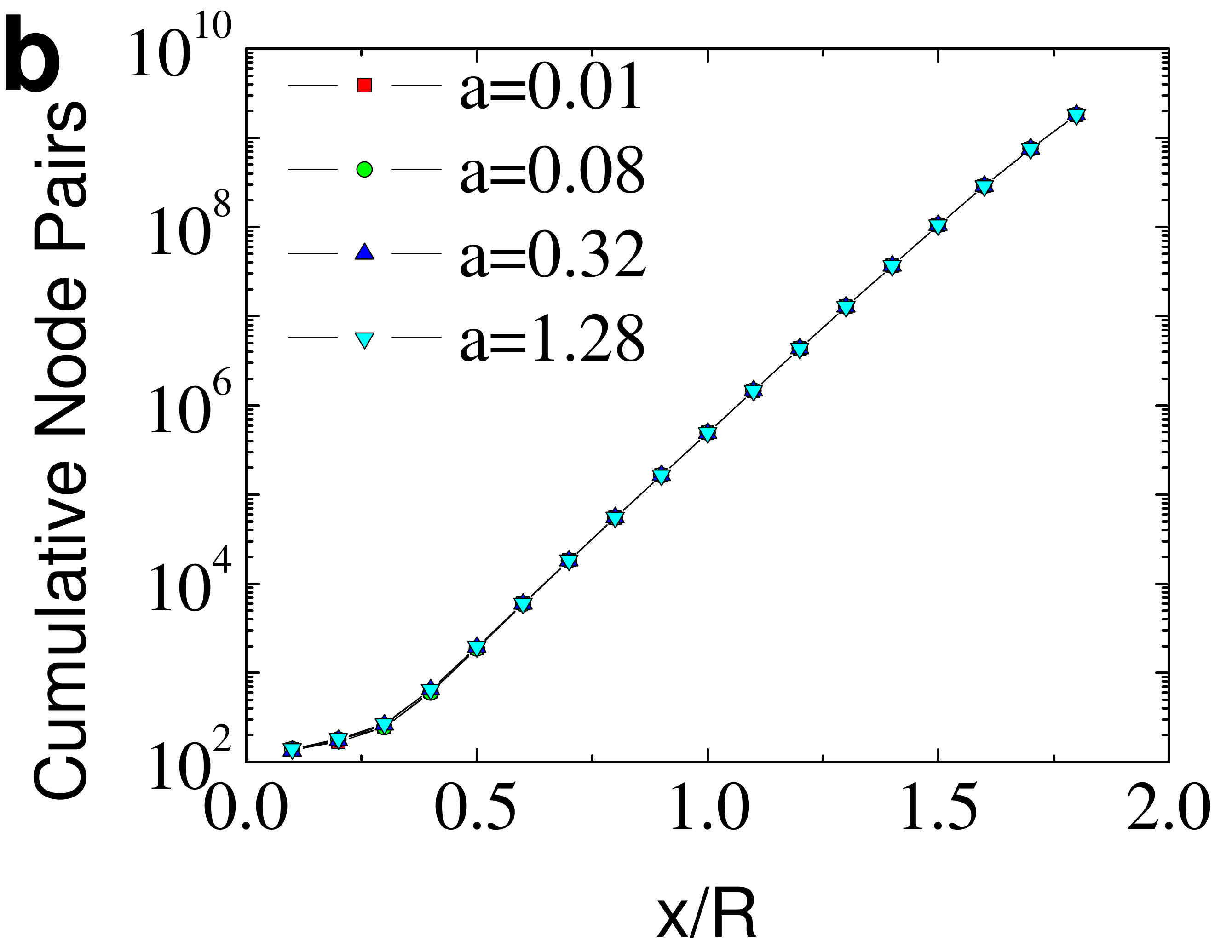}
\includegraphics[width=3in]{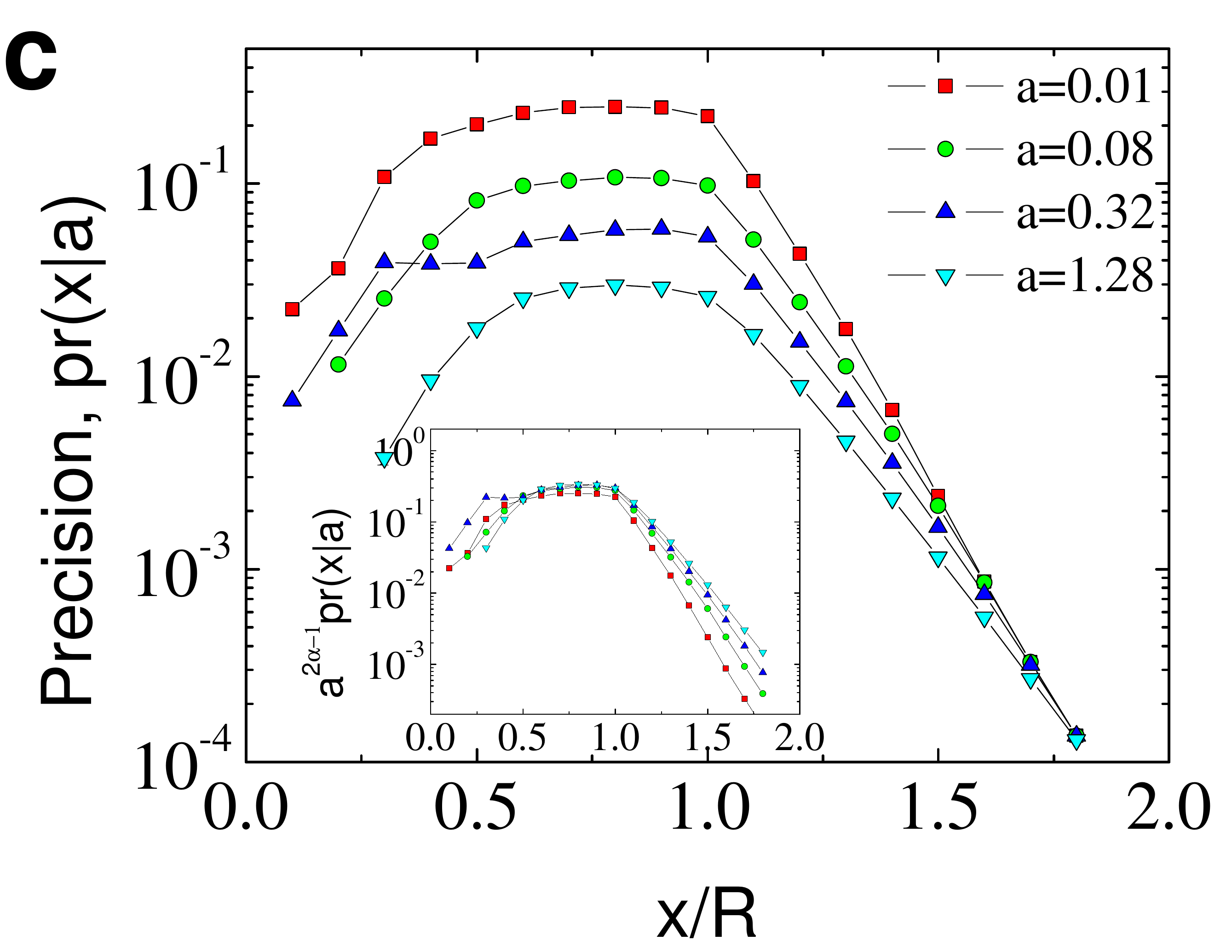}
\includegraphics[width=3in]{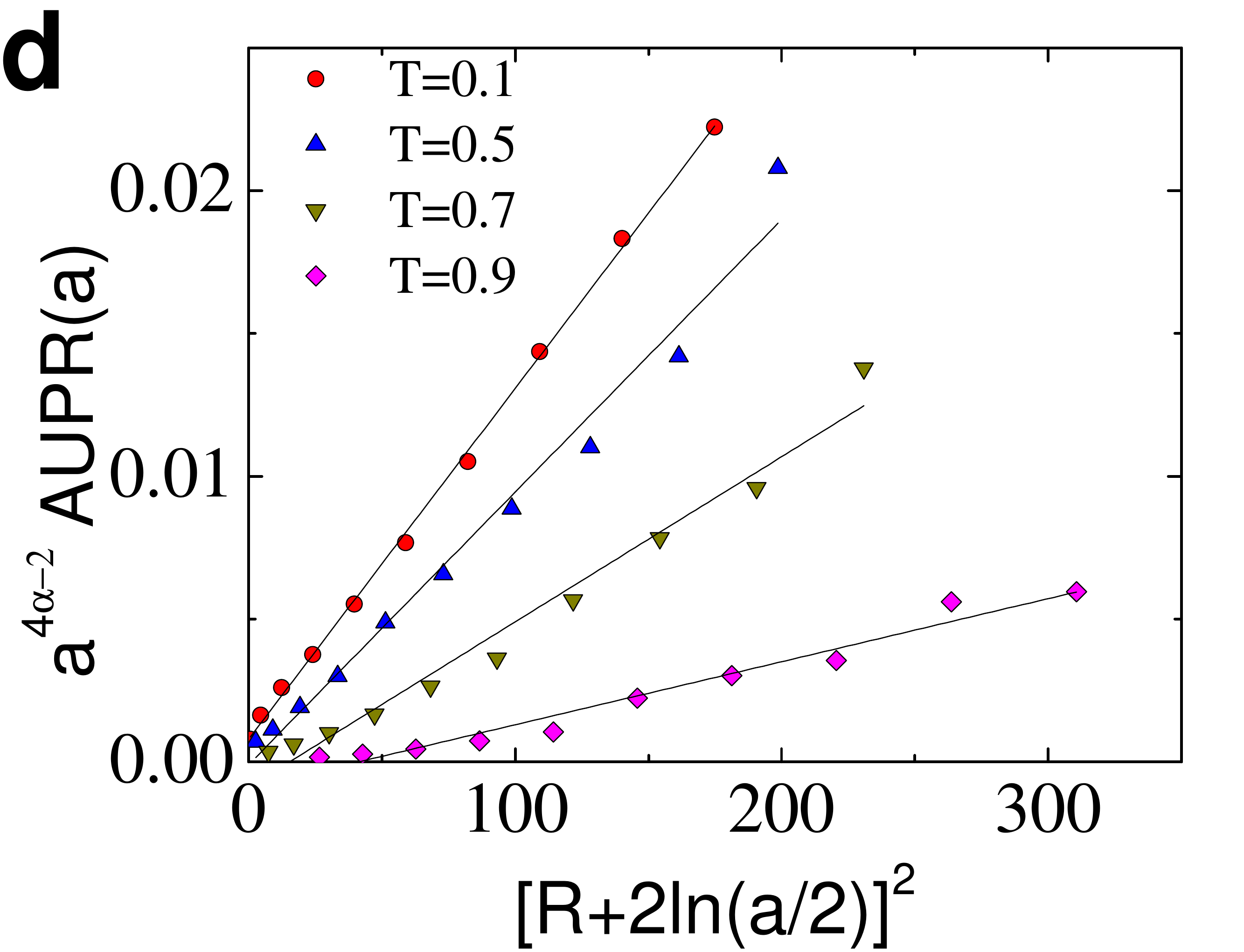}
\caption{\footnotesize {\bf HYPERLINK accuracy in case of  coordinate uncertainty.} All plots correspond to RHGs of $N=10^{5}$ nodes, $\gamma = 2.5$ ($\alpha = 0.75$), $T=0.1$, and $\overline{k}=10$. {\bf a}, Distance-dependent true positive rate ${\rm tp}(x|a)$ evaluated for different noise amplitude values. For $x<R$, ${\rm tp}(x|a)$ grows as $e^{x/2}$ (see the dashed line for the reference). The inset tests the scaling of ${\rm tp}(x|a) \sim a^{1-2\alpha}$ for $x<R$. {\bf b}, The cumulative number of node pairs in the hyperbolic disk as a function of hyperbolic distance between the nodes. Note that the cumulative number of node pairs is independent of noise amplitude. {\bf c}, Distance-dependent precision rate ${\rm pr}(x|a)$ for different $a$ values. ${\rm pr}(x|a)$ is nearly constant for $x<R$ since both ${\rm tp}(x|a)$ and $n(x|a)$ grow as $e^{x/2}$. ${\rm pr}(x|a)$ decays as $e^{-x/2}$ for $x>R$. The inset tests the scaling of  ${\rm pr}(x|a) \sim a^{1-2\alpha}$ for $x<R$. {\bf d}. The scaling test for ${\rm AUPR}(a)$ of the RHG with $N=5000$, $\gamma=2.5$, and $\overline{k}=10$. Note that $a^{4\alpha - 2} {\rm AUPR}(a)$ grows linearly as a function of
$\left(R+2\ln\frac{a}{2}\right)^{2}$, confirming Eq.~(\ref{eq:aupr_scaling}). }
\label{fig:a1}
\end{figure}

After a series of tedious calculations, which we detail in the Subsection~\ref{sec:kyra}, we obtain the leading order behavior of $\overline{k}_{y}(a)$:
\begin{equation}
\overline{k}_{y}(a) \sim
\begin{cases}
N g(y)a^{1-2\alpha} &\text{if $\frac{R}{2} \leq y \leq R$},\\
N g(y)a^{1-2\alpha}\left[R + 2 \ln \frac{a}{2}\right] &\text{if $y > R$},
\end{cases}
\label{eq:kx_a}
\end{equation}
where $\alpha \in \left(\frac{1}{2},1 \right)$ is the radial node density parameter in Eq.~(\ref{eq:rho_r}) corresponding to degree distribution exponent $\gamma = 2\alpha+1$.
Similar to the noiseless case, $g(y)$ grows as $\exp\left(\frac{y}{2}\right)$ for $y \leq R$ and saturates to a constant value, corresponding to $\overline{k}_{y}(a) = \overline{k}$ as $ y\to 2R$, Fig.~\ref{fig:a1}{\bf a}.

Using Eq.(\ref{eq:pr2}) one can rewrite the distance-dependent precision function as
\begin{equation}
{\rm pr}(y|a)   = \frac{{\rm tp}(y|a)}{\binom{N}{2} \int_{0}^{y} n(y'|a) {\rm d} y'  - \frac{q}{1-q} {\rm tp}(y|a)},
\end{equation}
where $n(y|a)$ is the node pair distribution in the hyperbolic disk with coordinate noise.

Due to the uniform initial angular distribution $\rho(\theta)$, the node pair distribution is independent of noise, $n(y|a) = n(y)$, Fig.~\ref{fig:a1}{\bf b}. Further, in the case of  sufficiently large noise amplitude $a$, ${\rm tp}(y|a) \ll \binom{N}{2} \int_{0}^{y} n(y'|a) {\rm d} y' $ and
\begin{equation}
{\rm pr}(y|a)   \approx \frac{{\rm tp}(y|a)}{\binom{N}{2} \int_{0}^{y} n(y') {\rm d} y' }.
\end{equation}

As a result, in the case $y \leq R$, ${\rm pr}(y|a) \sim a^{1-2\alpha}$, see Fig.~\ref{fig:a1}{\bf c}.

Since the distance-dependent recall function is proportional to the true positive rate,
\begin{equation}
{\rm rc}(y|a) = \frac{{\rm tp}(y|a)}{(1-q)E}.
\end{equation}

The resulting AUPR score scales as
\begin{equation}
{\rm AUPR}(a) \sim a^{2-4\alpha} \left[A + B \left(R + 2\ln \frac{a}{2}\right)^{2}\right],
\label{eq:aupr_scaling}
\end{equation}
where

\begin{eqnarray}
A &=& \frac{1-q}{E}\binom{N}{2} \int_{\frac{R}{2}}^{R} \frac{{\rm d} y g(y)g'(y)}{n^{c}(y)}, \\
B &=& \frac{1-q}{E}\binom{N}{2} \int_{R}^{2R} \frac{{\rm d} y g(y)g'(y)}{n^{c}(y)}
\end{eqnarray}
see Fig.~\ref{fig:a1}{\bf d}.

This result suggests that the impact of coordinate uncertainty on link prediction is higher in RHG with larger $\gamma = 2\alpha + 1$ values. Intuitively, this is the case since networks with larger $\gamma$ values have larger fractions of small degree nodes. Small degree nodes in the RHG are characterized by large radial coordinates, and the hyperbolic distance between the points with large radial coordinates is most affected by angular coordinate uncertainties.

\subsection{The average degree of the noisy subgraph}
\label{sec:kyra}
Here we derive the leading term behavior of the average degree of the noisy  subgraph $G_{y}(a)$ as a function of noise amplitude $a$
\begin{figure*}
\includegraphics[width=7in]{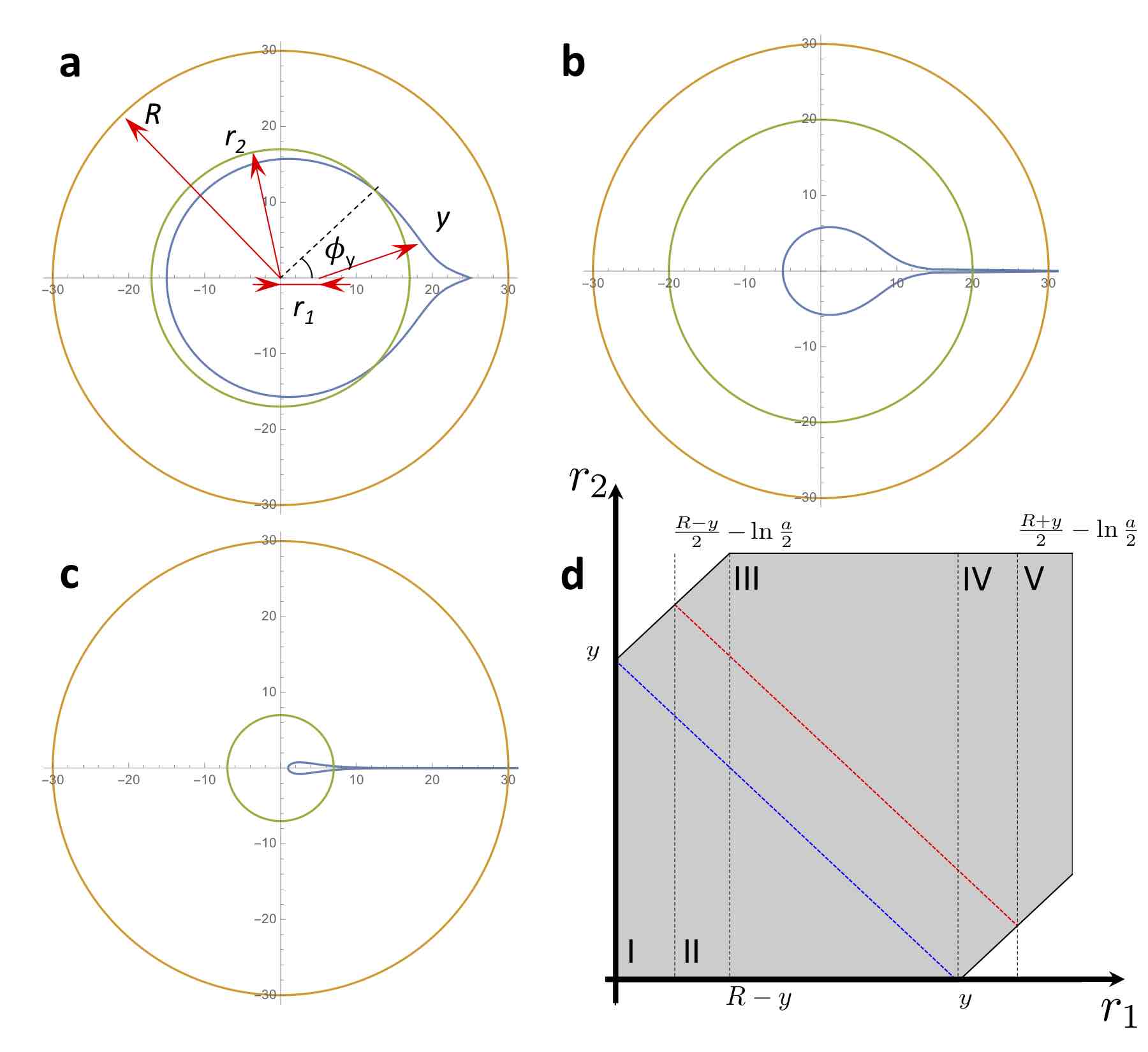}
\caption{\footnotesize {\bf Integration domain for $\overline{k}_{y}(a)$ in the case $y < R$.} The integration is performed at the intersection of two hyperbolic disks. The first disk (yellow) corresponds to the latent space of the RHG, has radius $R$ and is centered at the origin. The second disk (blue) has radius $y$ and is centered at $(r_1,0)$. The third disk depicts the integration radius $r_{2}$ that sweeps the integration domain. Angle $\phi_{y} \approx 2 e^{y-r_1-r_2}$  corresponds to the intersection of disks $y$ and $r_2$. Based  on $R$, $y$, and $r_1$ values we distinguish three configurations. {\bf a}, Disk $y$ contains the origin and is fully contained within $R$, regions I and II. {\bf b}, Disk $y$ contains the origin and is partially contained within $R$, region III. {\bf c}, Disk $y$ does not contain the origin and is partially contained within $R$, regions IV and V. {\bf d}, The shaded region corresponds to the integration domain for $\overline{k}_{y}(a)$. Vertical dashed lines separate the five integration regions. Phase space below the blue dashed line corresponds to the case of the disk $r_2$ fully contained within the disk $y$. Phase space above the blue line corresponds to the case of disk $r_{2}$ intersecting disk $y$. The red dashed line is given by $r_{2}+r_{1}=R - 2\ln \left(\frac{a}{2}\right)$ and corresponds to the loci of the integrand maxima in regions II, III, and IV.}
\label{fig:spheres}
\end{figure*}

As shown in the subsection above, the number of true positives ${\rm tp}(y|a)$ is related to the average degree of noisy  subgraph  $G_{y}(a)$. To define $G_{y}(a)$ we add uniform noise of amplitude $a$ to original angular coordinates of the RHG and calculate noisy hyperbolic distances $\hat{x}_{ij}$  between all node pairs using noisy coordinates.  $G_{y}(a)$ is the RHG subgraph formed by node pairs with noisy hyperbolic distances $\hat{x}_{ij} < y$. The average degree of $G_{y}(a)$ is given by
\begin{equation}
\overline{k}_{y}(a) = (N-1) \idotsint_{x \left( r_1, \hat{\theta}_1, r_2, \hat{\theta}_2 \right) \leq y} {\rm d} r_1 {\rm d} r_2 {\rm d} \hat{\theta}_1 {\rm d} \theta_1 {\rm d}  \hat{\theta}_2 {\rm d} \theta_2 \rho(r_1) \rho(r_2) \rho(\hat{\theta}_1) \rho(\theta_1|\hat{\theta}_{1}) \rho(\hat{\theta}_2) \rho(\theta_1|\hat{\theta}_{2}) p\left[x \left( r_1, \theta_1, r_2,\theta_2 \right) \right].
\label{eq:noisy_degree}
\end{equation}
Here $\rho(r)$ is given by Eq.~(\ref{eq:rho_r}), and $\rho(\theta|\hat{\theta})$ is the conditional probability of the true angle $\theta$, given inferred angle $\hat{\theta}$. In case of the uniform noise, $\rho(\theta|\hat{\theta})$ is also a uniform distribution centered at $\hat{\theta}$:
\begin{equation}
\rho(\theta|\hat{\theta}) = U\left(\hat{\theta} - a/2, \hat{\theta} + a/2\right),
\end{equation}
while
\begin{equation}
\rho(\hat{\theta}) = \rho(\theta) = \frac{1}{2 \pi}.
\end{equation}

Throughout the calculation of $k_{y}(a)$ we will rely on the number of assumptions. We are primarily interested in RHGs with $2 < \gamma < 3$, which correspond to $\frac{1}{2} < \alpha < 1 $.  To identify leading terms we will also recall on the scaling of $R$ with the system size, $N \sim e^{\frac{R}{2}}$.

Since hyperbolic distance $x$ in Eq.~(\ref{eq:hypercos}) depends on $\theta_1$ and $\theta_2$ only through their difference,
\begin{eqnarray}
x \left( r_1, \theta_1, r_2,\theta_2 \right) &=& x \left( r_1,r_2, \Delta \theta_{12}\right),\\
\Delta \theta_{12} &\equiv& \pi - | \pi - |\theta_1 - \theta_2||,
\end{eqnarray}
and angles distributed uniformly on $[-\pi,\pi]$, $\rho(\hat{\theta}_{1,2}) = \frac{1}{2\pi}$,  we can simplify Eq.~(\ref{eq:noisy_degree}) as
\begin{equation}
\overline{k}_{y}(a) = \frac{N}{\left(2\pi\right)^{2}} \idotsint_{x \left( r_1, r_2, \Delta \hat{\theta}_{12} \right) \leq y} {\rm d} r_1 {\rm d} r_2 \rho(r_1) \rho(r_2)  {\rm d} \hat{\theta}_{1}  {\rm d}  \hat{\theta}_2 {\rm d} \Delta \theta_{12} \tilde{\rho}(\Delta \theta_{12} | \Delta
\hat{\theta}_{12})   p\left[x \left( r_1, r_2, \Delta \theta_{12}\right) \right],
\label{eq:noisy_degree2}
\end{equation}
where
\begin{equation}
\tilde{\rho}(\Delta \theta_{12} | \Delta \hat{\theta}_{12}) = \frac{1}{a^{2}}\Theta \left(a - |\Delta \theta_{12} - \Delta\hat{\theta}_{12}|\right),
\end{equation}
and $\Theta[x]$ is the Heaviside theta function. Similar to the calculation of $\overline{k}$  in the RHGs, Ref.~\cite{Krioukov2010hyperbolic}, we can rewrite Eq.~(\ref{eq:noisy_degree2}) as
\begin{equation}
\overline{k}_{y}(a) =  \int_{0}^{R}{\rm d} r_1 \rho (r_1) \overline{k}_{y}(r_1|a),\\
\label{eq:noisy_degree3}
\end{equation}
where $\overline{k}_{y}(r|a)$ is the average degree of node with radial coordinate $r$ in noisy  subgraph  $G_{y}(a)$:
\begin{equation}
\overline{k}_{y}(r_1|a) =  \frac{N}{\left(2\pi\right)} \idotsint_{x \left( r_1, r_2, \hat{\phi}\right) \leq y} {\rm d} r_2 \rho(r_2)  {\rm d} \hat{\phi}  {\rm d}  \phi \tilde{\rho}(\phi | \hat{\phi})   p\left[x \left( r_1, r_2, \phi\right) \right],
\label{eq:noisy_degree4}
\end{equation}
and angles $\phi \equiv \Delta\theta_{12}$ and $\hat{\phi}\equiv \Delta\hat{\theta}_{12}$ are introduced to ease the notation.

To evaluate $\overline{k}_{y}(r_1|a)$ we note that the integration region in Eq.~(\ref{eq:noisy_degree4}) is given by intersection of two hyperbolic disks. The first one is of radius $R$ and is centered at the coordinate system origin, $(0,0)$. The second disk is of radius $y$ and is centered at $\left(r_{1},0\right)$.

We perform the integration for the two regimes of $y \in \left[\frac{R}{2},R \right]$ and  $[R,2R]$ separately. We do not perform the integration for the $y \in \left[0,\frac{R}{2}\right]$ regime since the number of true positives here is much smaller than that in the other two regimes. This is the case since $n(y)$ grows exponentially with $y$, $n(y) \sim e^{\frac{y}{2}}$. Consequently, the number of possible true positives in the $ y \in \left[0,\frac{R}{2}\right]$ regime is much smaller than that in the $y \in \left[\frac{R}{2},R \right]$ regime.

\subsubsection{$y\in\left[\frac{R}{2},R\right]$}

To evaluate $\overline{k}_{y}(r_1|a)$  we perform the integration over $r_{1}$  and $r_{2}$ values over the domain shown in Fig.~\ref{fig:spheres}{\bf d}.  Based on this domain, it is convenient to split the integration over $r_1$ into three regions, $0 \leq r_1 \leq R-y$, $R-y \leq r_1 \leq y$, and $y \leq r_1 \leq R$. However, due to specifics of the approximation techniques, it is more convenient to split the integration not into three but into five regions --- (i) $0 \leq r_1 \leq \frac{R-y}{2} -  \ln \frac{a}{2} $, (ii) $\frac{R-y}{2} -  \ln \frac{a}{2}  \leq r_1 \leq R-y$, (iii) $R-y \leq r_1 \leq y$, (iv) $y \leq r_1 \leq \frac{R+y}{2} - \ln \frac{a}{2}$, and (v) $\frac{R+y}{2} - \ln \frac{a}{2} \leq R  $ --- which we depict for convenience in Fig.~\ref{fig:spheres}{\bf d} with vertical dashed lines. We evaluate the contributions to $\overline{k}_{y}(r_1|a)$ from each of these five regions below.

{\it Region I:} $0 \leq r_1 \leq \frac{R-y}{2} -  \ln \frac{a}{2}$.
In this region the disk $y$ is fully contained within the disk $R$. Further, since $y>R/2$,  disk $y$ is guaranteed to include the coordinate system origin for all $r_1 \in [0, R-y]$ values,  Fig.~\ref{fig:spheres}{\bf a}. In this case the integral in $\overline{k}_{y}(r_1|a)$ can be evaluated as
\begin{eqnarray}
\overline{k}_{y}(r_1|a) &=& \mathfrak{I}_{1} + \mathfrak{I}_{2}, \\
\mathfrak{I}_{1} & = & \frac{N}{2 \pi} \int_{0}^{y-r_1}  {\rm d} r_2 \rho(r_2) \int_{0}^{2\pi}  {\rm d} \hat{\phi} \int_{\hat{\phi} - a}^{\hat{\phi} + a}  {\rm d} \phi \tilde{\rho}(\phi | \hat{\phi})  p\left[x \left( r_1, r_2, \phi\right) \right],\\
\mathfrak{I}_{2} & = & \frac{N}{\pi} \int_{y-r_1}^{y+r_1}  {\rm d} r_2 \rho(r_2) \int_{0}^{\phi_{y}}  {\rm d} \hat{\phi} \int_{\hat{\phi} - a}^{\hat{\phi} + a}  {\rm d} \phi \tilde{\rho}(\phi | \hat{\phi})  p\left[x \left( r_1, r_2, \phi\right) \right],
\end{eqnarray}
where $\phi_{y}$ is the angle given by the intersection of the disk with radius $r_2$ centered at $r=0$ and that of radius $y$, centered at $r=r_1$.  To estimate $\phi_{y}$ we consider the triangle formed by the origin $(0,0)$, disk $y$ centered at $(r_1,0)$, and the intersection of $r_{2}$ with $y$. The triangle has sides equal to $r_1$, $r_2$, and $y$ with $\phi_{y}$ being the angle between $r_1$ and $r_2$. Thus, $\phi_y$ is given by the hyperbolic law of cosines:
\begin{equation}
\cosh y = \cosh r_1 \cosh r_2 -  \sinh r_1 \sinh r_2 \cos \phi_y,
\label{eq:phi_y1}
\end{equation}
In the case of sufficiently large $r_1$, $r_2$, and $y$ values we can approximate $\cos \phi_y$ as
\begin{equation}
\cos \phi_y \approx 1 - 2 e^{y - r_1 - r_2}
\label{eq:phi_y2}
\end{equation}

Since $\hat{\phi}$ in the first integral sweeps the entire $2\pi$ angle, $\mathfrak{I}_{1}$ is given by
\begin{equation}
\mathfrak{I}_{1}  =  \frac{N}{2 \pi} \int_{0}^{y-r_1}  {\rm d} r_2 \rho(r_2) \int_{0}^{2\pi}  {\rm d} \hat{\phi}  p\left[x \left( r_1, r_2, \hat{\phi} \right) \right]
\end{equation}
Then, since $x \left( r_1, r_2, \hat{\phi} \right) \leq r_1+r_2 \leq R$, $p\left[x \left( r_1, r_2, \hat{\phi} \right) \right] \approx 1$,
leading to
\begin{equation}
\mathfrak{I}_{1}  =  N e^{\alpha\left(y- r_1 -R\right)},
\end{equation}
The evaluation of $\mathfrak{I}_{2}$ is more involved and requires further approximations. We notice that  $\phi_{y} \ll 1$ since $r_{2} \in [ y - r_1, y+r_1]$, which can be further approximated as
\begin{equation}
\phi_y \approx 2 e^{\frac{y-r_1-r_2}{2}}.
\label{eq:phi_y}
\end{equation}
Then, for sufficiently large noise amplitudes $a \gg \phi_{y}$, we can approximate the integral  $\int_{\hat{\phi} - a}^{\hat{\phi} + a}  {\rm d} \phi$ as $2 \int_{0}^{a}  {\rm d} \phi$, resulting in
\begin{equation}
\mathfrak{I}_{2}  = \frac{2 N}{\pi a^{2}} \int_{y-r_1}^{y+r_1}  {\rm d} r_2 \rho(r_2) \phi_{y} \int_{0}^{a}  \frac{{\rm d} \phi \left(a - \phi \right)}{1 + {\rm exp}\left(\frac{x \left(r_1, r_2, \phi\right) - R}{2T}\right)}
\end{equation}
Since $r_1 < \frac{R-y}{2} - \ln {\frac{a}{2}}$, $r_2 < y+ r_1$, and $y < R$, it follows that $x\left(r_1, r_2, \phi\right) < r_1+r_2  + 2 \ln \frac{a}{2}< R$, and, as a result, ${\rm exp}\left(\frac{x \left(r_1, r_2, \phi\right) - R}{2T}\right)\ll 1$, resulting in
\begin{equation}
\mathfrak{I}_{2}  = \frac{4 \alpha  N}{\pi \left(2 \alpha - 1\right)} e^{ - \alpha R } e^{\alpha y}  e^{\left(\alpha -1 \right)r_1}.
\end{equation}
Since  $\gamma > 2$ case ($\alpha > \frac{1}{2}$), $\mathfrak{I}_{2} \gg \mathfrak{I}_{1}$, and
\begin{equation}
\overline{k}_{y}(r_1|a) \approx \mathfrak{I}_{2} = \frac{4 \alpha  N}{\pi \left(2 \alpha - 1\right)} e^{ - \alpha R } e^{\alpha y}  e^{\left(\alpha -1 \right)r_1}.
\label{eq:region1}
\end{equation}

{\it Region II:} $\frac{R-y}{2} -  \ln \frac{a}{2} \leq r_1 \leq R-y$.

Similar to region I, the hyperbolic disk $y$ fully lies within disk $R$, Fig.~\ref{fig:spheres}{\bf a}. Thus, $\overline{k}_{y}(r_1|a)$ is given by the same expression:
\begin{eqnarray}
\overline{k}_{y}(r_1|a) &=& \mathfrak{I}_{3} + \mathfrak{I}_{4}, \\
\mathfrak{I}_{3} & = & \frac{N}{2 \pi} \int_{0}^{y-r_1}  {\rm d} r_2 \rho(r_2) \int_{0}^{2\pi}  {\rm d} \hat{\phi} \int_{\hat{\phi} - a}^{\hat{\phi} + a}  {\rm d} \phi \tilde{\rho}(\phi | \hat{\phi})  p\left[x \left( r_1, r_2, \phi\right) \right],\\
\mathfrak{I}_{4} & = & \frac{N}{\pi} \int_{y-r_1}^{y+r_1}  {\rm d} r_2 \rho(r_2) \int_{0}^{\phi_{y}}  {\rm d} \hat{\phi} \int_{\hat{\phi} - a}^{\hat{\phi} + a}  {\rm d} \phi \tilde{\rho}(\phi | \hat{\phi})  p\left[x \left( r_1, r_2, \phi\right) \right].
\end{eqnarray}
The calculation of $\mathfrak{I}_{3}$ is identical to that of $\mathfrak{I}_{1}$, resulting in
\begin{equation}
\mathfrak{I}_{3}  = \mathfrak{I}_{1} = N e^{\alpha\left(y- r_1 -R\right)}.
\end{equation}

Different from region I is the calculation of $\mathfrak{I}_{4}$. Indeed, in the case $r_1 \geq \frac{R-y}{2} +\ln \frac{a}{2}$, and $r_2 \in [y-r_1, y+r_1]$ hyperbolic distance $x \left( r_1, r_2, \phi\right)$ is no longer guaranteed to be smaller than $R$, and $p\left[x \left( r_1, r_2, \phi\right)\right]$ can no longer be approximated by unity. We first split $\mathfrak{I}_{4}$ into two parts and calculate them separately:
\begin{equation}
\mathfrak{I}_{4} = \mathfrak{I}_{4,1} - \mathfrak{I}_{4,2},
\end{equation}
where
\begin{eqnarray}
\mathfrak{I}_{4,1}  &=& \frac{N}{\pi a} \int_{y-r_1}^{y+r_1}  {\rm d} r_2 \rho(r_2) \phi_{y} \int_{0}^{a}  \frac{{\rm d} \phi }{1 + {\rm exp}\left(\frac{x \left(r_1, r_2, \phi\right) - R}{2T}\right)},\\
\mathfrak{I}_{4,2}  &=& \frac{N}{\pi a^{2}} \int_{y-r_1}^{y+r_1}  {\rm d} r_2 \rho(r_2) \phi_{y} \int_{0}^{a}  \frac{  \phi {\rm d} \phi}{1 + {\rm exp}\left(\frac{x \left(r_1, r_2, \phi\right) - R}{2T}\right)}.
\end{eqnarray}
By approximating the hyperbolic law of cosines in Eq.~(\ref{eq:hypercos}) as, $x \left(r_1, r_2, \phi\right) \approx r_1 + r_2 + 2 \ln \frac{\phi}{2}$  and making use of  Eq.~(\ref{eq:phi_y}) we obtain for $\mathfrak{I}_{41}$
\begin{equation}
\mathfrak{I}_{4,1}  = \frac{4\alpha  N}{\pi a} e^{\left( \frac{1}{2} - \alpha\right) R } e^{\frac{y}{2}}  e^{-r_1} \int_{y-r_1}^{y+r_1}  {\rm d} r_2 e^{\left(\alpha -1 \right)r_2} I\left(\frac{a}{2} e^{\frac{r_1+r_2-R}{2}};T\right),
\end{equation}
where $I(z;T) \equiv \int_{0}^{z}\frac{{\rm d}x}{1 + x^{\frac{1}{T}}}$ is the same function as in Eq.~(\ref{eq:int}).

Recall that for small $z\ll1$ function $I(z;T)\approx z$,  while, for $z\gg 1$, $I(z;T)\approx I(T)=\frac{\pi}{T \sin\left(\frac{\pi}{T}\right)}$.   With these approximations in mind we split the integration in $\mathfrak{I}_{41}$ into two subregions :
\begin{equation}
\int_{y-r_1}^{y+r_1}{\rm d} r_2 = \int_{y-r_1}^{R-r_1-2\ln \frac{a}{2}} {\rm d} r_2 + \int_{R-r_1-2\ln \frac{a}{2}}^{y+r_1} {\rm d} r_2
\end{equation}
In the first subregion, $r_2 \in \left[y-r_1, R-r_1-2\ln \frac{a}{2} \right]$, and  $\frac{a}{2} e^{\frac{r_1+r_2-R}{2}} \leq 1$, which allows us to approximate  $I\left(\frac{a}{2} e^{\frac{r_1+r_2-R}{2}};T\right) \approx \frac{a}{2} e^{\frac{r_1+r_2-R}{2}}$.  In the second subregion, $r_2 \in [R-r_1-2\ln \frac{a}{2}, y+r_1]$, $\frac{a}{2} e^{\frac{r_1+r_2-R}{2}} \geq 1$, and $I\left( \frac{a}{2} e^{\frac{r_1+r_2-R}{2}};T\right) \approx  I\left(T\right)$.  Using these approximations we obtain, to the leading order,
\begin{equation}
\mathfrak{I}_{4,1} = \frac{2N \alpha}{\pi} \left[ \frac{2}{2\alpha - 1}  + \frac{I(T)}{1 - \alpha }  \right] e^{\frac{y-R}{2}}
 e^{-\alpha r_1} \left(\frac{a}{2}\right)^{1-2\alpha}.
 \end{equation}
Following the same approximation steps,
\begin{equation}
\mathfrak{I}_{4,2} = \frac{2N \alpha}{\pi} \left[ \frac{1}{ 2\alpha - 1}  + \frac{2 \tilde{I}(T)}{3 - 2\alpha }  \right] e^{\frac{y-R}{2}}
 e^{-\alpha r_1} \left(\frac{a}{2}\right)^{1-2\alpha}
 \end{equation}
where
\begin{equation}
\tilde{I}(T) \equiv \int_{0}^{\infty} \frac{x {\rm d} x}{1 + x^{\frac{1}{T}}} = \frac{\pi T}{{\rm sin} (2\pi T)}
\end{equation}
in the case $T<1/2$.

Taken together, $\mathfrak{I}_{4,1}$ and $\mathfrak{I}_{4,2}$ result in
\begin{equation}
\mathfrak{I}_{4} = \frac{2N \alpha}{\pi} \left[ \frac{1}{2\alpha - 1}  + \frac{2I(T)}{1 - \alpha } -\frac{8 \tilde{I}(T)}{3 - 2\alpha } \right] e^{\frac{y-R}{2}}
 e^{-\alpha r_1} \left(\frac{a}{2}\right)^{1-2\alpha}.
 \end{equation}
Finally, since $y <R$, we conclude that $\mathfrak{I}_{3} \ll \mathfrak{I}_{4}$, resulting in
\begin{equation}
\overline{k}_{y}(r_1|a) \approx \frac{2N \alpha}{\pi} \left[ \frac{1}{2\alpha - 1}  + \frac{I(T)}{1 - \alpha } -\frac{2 \tilde{I}(T)}{3 - 2\alpha } \right] e^{\frac{y-R}{2}}
 e^{-\alpha r_1} \left(\frac{a}{2}\right)^{1-2\alpha}
 \label{eq:region2}
\end{equation}
for $\frac{R-y}{2} - \ln\frac{a}{2} \leq r_1 \leq R-y$.

{\it Region III:} $ R-y \leq r_1 \leq y$.

In this region  disk $y$ is partially contained within the disk $R$. Since $r_1 \leq y$, disk $y$ still contains the coordinate system origin, Fig.~\ref{fig:spheres}{\bf b}. Similar to regions I and II, we split the calculation of $\overline{k}_{y}(r_1|a)$ into two parts:
\begin{eqnarray}
\overline{k}_{y}(r_1|a) &=& \mathfrak{I}_{5} + \mathfrak{I}_{6}, \\
\mathfrak{I}_{5} & = & \frac{N}{2 \pi} \int_{0}^{y-r_1}  {\rm d} r_2 \rho(r_2) \int_{0}^{2\pi}  {\rm d} \hat{\phi} \int_{\hat{\phi} - a}^{\hat{\phi} + a}  {\rm d} \phi \tilde{\rho}(\phi | \hat{\phi})  p\left[x \left( r_1, r_2, \phi\right) \right],\\
\mathfrak{I}_{6} & = & \frac{N}{\pi} \int_{y-r_1}^{R}  {\rm d} r_2 \rho(r_2) \int_{0}^{\phi_{y}}  {\rm d} \hat{\phi} \int_{\hat{\phi} - a}^{\hat{\phi} + a}  {\rm d} \phi \tilde{\rho}(\phi | \hat{\phi})  p\left[x \left( r_1, r_2, \phi\right) \right],
\end{eqnarray}
where $\phi_y \ll 1$ is the intersection angle of disk $r_2$ with that of $y$, see Fig.~\ref{fig:spheres}{\bf b}, and is given by Eq.~(\ref{eq:phi_y}).

We first note that the integral in $\mathfrak{I}_{5}$ is identical to those in $\mathfrak{I}_{3}$ and $\mathfrak{I}_{1}$:
\begin{equation}
\mathfrak{I}_{5}  = \mathfrak{I}_{1} = N e^{\alpha\left(y- r_1 -R\right)}.
\end{equation}
The integration in $\mathfrak{I}_{6}$ is very similar to that in $\mathfrak{I}_{4}$ with the only difference in the upper integration bound of $r_{2} \leq R$. The evaluation of $\mathfrak{I}_{6}$ is, therefore, straightforward and requires the same approximation steps as in $\mathfrak{I}_{4}$. A quicker estimate can be obtained by noting that the
upper bound for $r_{2}$ in $\mathfrak{I}_{4}$ does not contribute to the leading term. The reason is that $\mathfrak{I}_{42}$ is dominated by $r_{2}$ in the vicinity of
the $r_{2} = R-r_1-2\ln \frac{a}{2}$ point.

Since $R > R-r_1-2\ln \frac{a}{2} > y-r_1$
\begin{equation}
\mathfrak{I}_{6} = \int_{y-r_1}^{R-r_1-2\ln \frac{a}{2}}  {\rm d} {r_2} + \int_{R-r_1-2\ln \frac{a}{2}}^{R} {\rm d} {r_2}
\end{equation}
with integrands identical to those of $\mathfrak{I}_{41}$ and $\mathfrak{I}_{42}$. Since the integrand in $\mathfrak{I}_{42}$ is dominated by smaller $r_{2}$ values
we conclude that
\begin{equation}
\mathfrak{I}_{6} = \mathfrak{I}_{4}
\end{equation}
Finally, $\mathfrak{I}_{6}$ dominates $\mathfrak{I}_{5}$ for $\alpha > \frac{1}{2}$, resulting in
\begin{equation}
\overline{k}_{y}(r_1|a) \approx \frac{2N \alpha}{\pi} \left[ \frac{1}{2\alpha - 1}  + \frac{I(T)}{1 - \alpha } -\frac{2 \tilde{I}(T)}{3 - 2\alpha } \right] e^{\frac{y-R}{2}}
 e^{-\alpha r_1} \left(\frac{a}{2}\right)^{1-2\alpha}
 \label{eq:region3}
\end{equation}
for $R-y \leq r_1 \leq y$.

{\it Region IV:} $ y \leq r_1 \leq \frac{R+y}{2} - \ln \frac{a}{2}$.

In this region, hyperbolic disk $y$ is partially contained within $R$ and does not include the  origin, Fig.~\ref{fig:spheres}{\bf c}. Therefore, in this region
\begin{equation}
\overline{k}_{y}(r_1|a)  =  \frac{N}{\pi} \int_{r_1-y}^{R}  {\rm d} r_2 \rho(r_2) \int_{0}^{\phi_{y}}  {\rm d} \hat{\phi} \int_{\hat{\phi} - a}^{\hat{\phi} + a}  {\rm d} \phi \tilde{\rho}(\phi | \hat{\phi})  p\left[x \left( r_1, r_2, \phi\right) \right],
\end{equation}
Using the arguments similar to that of region III, we obtain
\begin{equation}
\overline{k}_{y}(r_1|a) \approx \frac{2N \alpha}{\pi} \left[ \frac{1}{2\alpha - 1}  + \frac{I(T)}{1 - \alpha } -\frac{2 \tilde{I}(T)}{3 - 2\alpha } \right] e^{\frac{y-R}{2}}
 e^{-\alpha r_1} \left(\frac{a}{2}\right)^{1-2\alpha}
 \label{eq:region4}
\end{equation}
for $y \leq r_1 \leq \frac{R+y}{2} - \ln \frac{a}{2}$.

{\it Region V:} $ \frac{R+y}{2} - \ln \frac{a}{2}\leq r_1 \leq R$.

Similar to the situation in region IV, hyperbolic disk $y$ intersects disk $R$ and does not include the coordinate system origin. Different from region IV is the
$r_{2} = R-r_1 -2\ln \frac{a}{2} $ point that lies outside the $r_{2}$ integration region and we can no longer relate $\overline{k}_{y}(r_1|a)$ to those in other regions.

To evaluate
\begin{equation}
\overline{k}_{y}(r_1|a)  =  \frac{N}{\pi} \int_{r_1-y}^{R}  {\rm d} r_2 \rho(r_2) \int_{0}^{\phi_{y}}  {\rm d} \hat{\phi} \int_{\hat{\phi} - a}^{\hat{\phi} + a}  {\rm d} \phi \tilde{\rho}(\phi | \hat{\phi})  p\left[x \left( r_1, r_2, \phi\right) \right]
\end{equation}
we recall that $\phi_{y} \ll 1$, and for sufficiently large $a \gg \phi_y$ we obtain
\begin{eqnarray}
\overline{k}_{y}(r_1|a)  &=&  \mathfrak{I}_{7} - \mathfrak{I}_{8},\\
 \mathfrak{I}_{7}  &=& \frac{N}{\pi a} \int_{r_1-y}^{R}  {\rm d} r_2 \rho(r_2) \phi_{y} \int_{0}^{a}  \frac{{\rm d} \phi }{1 + {\rm exp}\left(\frac{x \left(r_1, r_2, \phi\right) - R}{2T}\right)},\\
\mathfrak{I}_{8}  &=& \frac{N}{\pi a^{2}} \int_{r_1-y}^{R}  {\rm d} r_2 \rho(r_2) \phi_{y} \int_{0}^{a}  \frac{  \phi {\rm d} \phi}{1 + {\rm exp}\left(\frac{x \left(r_1, r_2, \phi\right) - R}{2T}\right)},
\end{eqnarray}
After straightforward approximations we obtain
\begin{equation}
\overline{k}_{y}(r_1|a) = \frac{4 \alpha N}{\pi a}   e^{\left(\frac{1}{2} - \alpha\right)R}e^{\left(\frac{3}{2} - \alpha\right)y}e^{\left(\alpha-2\right)r_1}\left[ \frac{I(T)}{1-\alpha} - \frac{4 \tilde{I}(T)}{a\left(3-2\alpha\right)} e^{\frac{R+y}{2}-r_1} \right]
\label{eq:region5}
\end{equation}
for $ \frac{R+y}{2} - \ln \frac{a}{2}\leq r_1 \leq R$

Merged together, Eqs.~(\ref{eq:region1}),(\ref{eq:region2}), (\ref{eq:region3}), (\ref{eq:region4}), and (\ref{eq:region5}) provide the solution for $\overline{k}_{y}(r_1|a)$:
\begin{equation}
\overline{k}_{y}(r_1|a) \approx
\begin{cases}
\frac{4 \alpha  N}{\pi \left(2 \alpha - 1\right)} e^{ - \alpha R } e^{\alpha y}  e^{\left(\alpha -1 \right)r_1}&\text{if $ 0\leq r_1 \leq\frac{R-y}{2} - \ln \frac{a}{2}$},\\
\frac{2N \alpha}{\pi} \left[ \frac{1}{2\alpha - 1}  + \frac{I(T)}{1 - \alpha } -\frac{2 \tilde{I(T)}}{3 - 2\alpha } \right] e^{\frac{y-R}{2}}
 e^{-\alpha r_1} \left(\frac{a}{2}\right)^{1-2\alpha}&\text{if $ \frac{R-y}{2} - \ln \frac{a}{2}\leq r_1 \leq\frac{R+y}{2} - \ln \frac{a}{2}$},\\
\frac{4 \alpha N}{\pi a}   e^{\left(\frac{1}{2} - \alpha\right)R}e^{\left(\frac{3}{2} - \alpha\right)y}e^{\left(\alpha-2\right)r_1}\left[ \frac{I(T)}{1-\alpha} - \frac{4 \tilde{I}(T)}{a\left(3-2\alpha\right)} e^{\frac{R+y}{2}-r_1} \right] & \text{if $ \frac{R+y}{2} - \ln \frac{a}{2}\leq r_1 \leq R$}.
\end{cases}
\label{eq:k1ya}
\end{equation}

Using Eq.~(\ref{eq:k1ya}) together with Eq.~(\ref{eq:noisy_degree3}) we finally obtain
\begin{equation}
k_{y}(a) \sim N e^{-\left( \alpha + \frac{1}{2}\right)R} e^{\frac{y}{2}} a ^{1-2\alpha},
\label{eq:kya}
\end{equation}

\subsubsection{$y \in [R,2R]$}
\begin{figure*}
\includegraphics[width=3in]{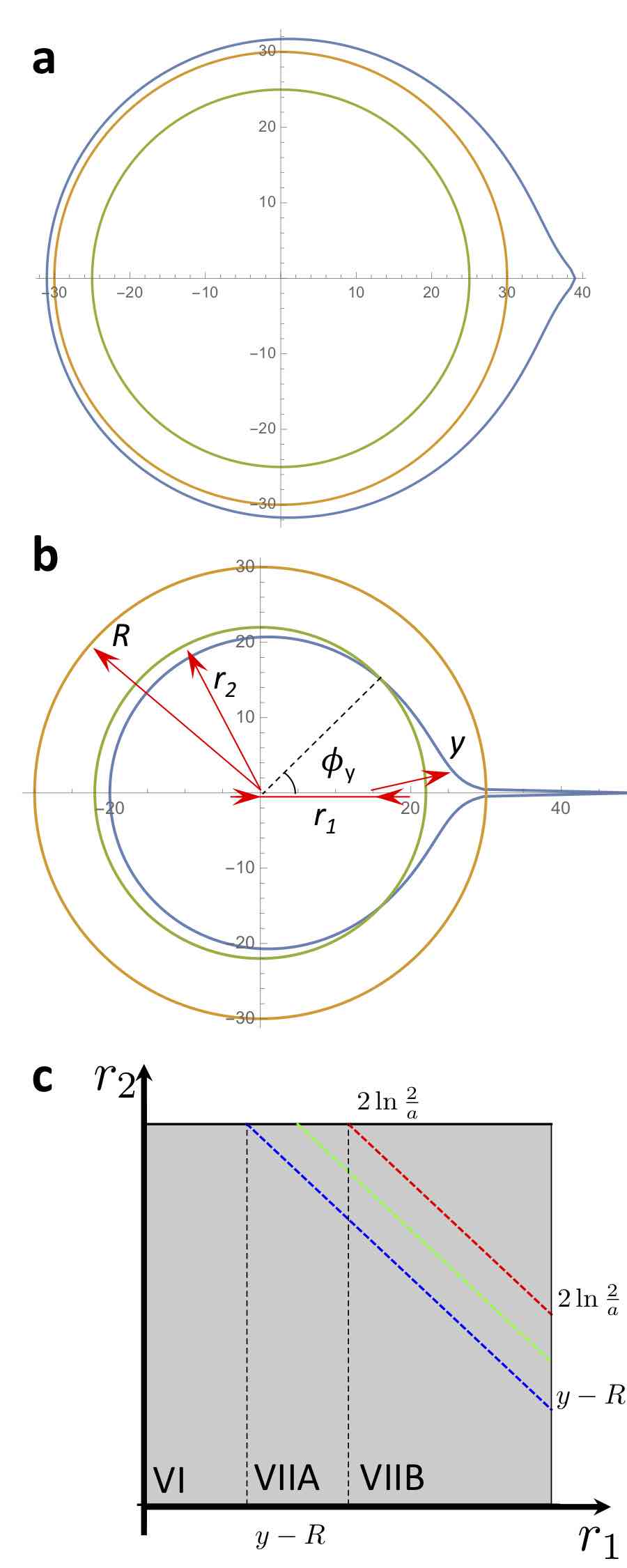}
\caption{\footnotesize {\bf Integration domain for $\overline{k}_{y}(a)$ at $y > R$.} The integration is performed at the intersection of two hyperbolic disks. The first disk (yellow) corresponds to the latent space of the RHG, has radius $R$ and is centered at the origin. The second disk (blue) has radius $y$ and is centered at $(r_1,0)$. The third disk (green) depicts the integration radius $r_{2}$ that sweeps the integration domain. Angle $\phi_{y}$ corresponds to the intersection of disks $y$ and $r_2$. Based on $R$, $y$, and $r_1$ values, we distinguish two configurations. {\bf a}, Disk $y$ fully contains disk $R$, regions VI. {\bf b}, Disk $y$ overlaps within $R$, region VII. {\bf c}, The integration domain $\overline{k}_{y}(a)$ is shown by the shaded region. Vertical dashed lines separate the domain into two integration regions, VI and VII. Region VII further splits into subregions VIIA and VIIB. Phase space below the blue dashed line corresponds to the case of disk $r_2$ fully contained within disks $y$ and $R$. Phase space above the blue line corresponds to the case of disk $r_{2}$ intersecting disk $y$. The red dashed line is given by $r_{2}+r_{1}=R - 2\ln \frac{a}{2}$ and corresponds to the loci of the integrand maxima in region VII. The green dashed line corresponds to the $\tilde{R}(r_1)$ line. By construction,  $\phi_{y} \ll 1$ for $r_2 \geq \tilde{R}(r_1)$.}
\label{fig:spheres2}
\end{figure*}
In the regime $y \geq R$ hyperbolic disk $y$ always contains the origin, Fig.~\ref{fig:spheres2}. To evaluate $\overline{k}_{y}(r_1|a)$ in this regime we need to distinguish two cases, (VI) $0 \leq r_1 \leq y-R$, and (VII) $ y-R \leq r_1 \leq R$.

{\it Region VI:} $ 0 \leq r_1 \leq y-R$.

In this regime hyperbolic disk $R$ is fully contained within hyperbolic disk $y$, Fig.~\ref{fig:spheres2}{\bf a}, and $\overline{k}_{y}(r_1|a) = \overline{k}(r_1)$, where $\overline{k}(r_1)$ is the average degree of a node at $r_1$ in the RHG.  Indeed, radial coordinates of all points are within disk $R$, and all distances from point $(r_1,0)$ to any point within disk $R$ are guaranteed to be smaller than $y$, $x(r_1,0,r_2,\theta) < y$ for any $\theta \in [0,2 \pi]$. Therefore in this regime
\begin{equation}
\overline{k}_{y}(r_1|a)  =  \frac{N}{2\pi} \int_{0}^{R}  {\rm d} r_2 \rho(r_2) \int_{0}^{2\pi}  {\rm d} \hat{\phi} \int_{\hat{\phi} - a}^{\hat{\phi} + a}  {\rm d} \phi \tilde{\rho}(\phi | \hat{\phi})  p\left[x \left( r_1, r_2, \phi\right) \right],
\end{equation}
Since the integral over $\hat{\phi}$ sweeps the entire circle, $\hat{\theta} \in [0,2 \pi]$, synthetic noise does not affect the integration:
\begin{equation}
\overline{k}_{y}(r_1|a)  =  \frac{N}{2\pi} \int_{0}^{R}  {\rm d} r_2 \rho(r_2) \int_{0}^{2\pi}  {\rm d} \phi  p\left[x \left( r_1, r_2, \phi\right) \right] = \overline{k}(r_1),
\end{equation}
resulting in
\begin{equation}
\overline{k}_{y}(r_1|a)  =  \frac{4 \alpha N I(T)}{\left(2 \alpha - 1\right) \pi}e^{-\frac{r_1}{2}}
\end{equation}
in the case $ 0 \leq r_1 \leq y-R$.

{\it Region VII:} $ R-y \leq r_1 \leq R$. In this regime hyperbolic disk $R$ is partially contained within $y$ and the calculation of  $\overline{k}_{y}(r_1|a)$ splits into two integrals:
\begin{eqnarray}
\overline{k}_{y}(r_1|a) &=&  \mathfrak{I}_{9} + \mathfrak{I}_{10},\\
\mathfrak{I}_{9}  & =&  \frac{N}{2\pi} \int_{0}^{y-r_1}  {\rm d} r_2 \rho(r_2) \int_{0}^{2\pi}  {\rm d} \phi  p\left[x \left( r_1, r_2, \phi\right) \right] = \overline{k}(r_1),\\
 \mathfrak{I}_{10}  &=&  \frac{N}{\pi} \int_{y-r_1}^{R}  {\rm d} r_2 \rho(r_2) \int_{0}^{\phi_y}  {\rm d} \hat{\phi} \int_{\hat{\phi} - a}^{\hat{\phi} + a}  {\rm d} \phi \tilde{\rho}(\phi | \hat{\phi})  p\left[x \left( r_1, r_2, \phi\right) \right],
\end{eqnarray}
where $\phi_y$ is the angle of intersection of disks $R$ and $y$, Fig.~\ref{fig:spheres2}{\bf b}.

We note that the integration region for  $\mathfrak{I}_{9}$ is identical to that of $\mathfrak{I}_{1}$. Different from the case of $\mathfrak{I}_{1}$ is the condition that $y > R$. In this case $x \left( r_1, r_2, \phi\right)$ is no longer guaranteed to be less than $R$, and $p\left[x \left( r_1, r_2, \phi\right) \right]$ cannot be approximated by $1$. We start evaluating $\mathfrak{I}_{9}$ by performing the integration over $\phi$, which leads to
\begin{equation}
\mathfrak{I}_{9}  =  \frac{2 \alpha N}{\pi} e^{-\left(\alpha - \frac{1}{2}\right)R} e^{-\frac{r_1}{2}} \int_{0}^{y-r_1}  {\rm d} r_2 e^{\left(\alpha - \frac{1}{2}\right) r_2} I\left(\frac{\pi}{2}e^{\frac{r_1+r_2-R}{2}};T\right),
\end{equation}
where $I(z;T)$ is given by Eq.~(\ref{eq:int}). Recall that $I(z;T) \approx z$  if $z\ll 1$ and  $I(z;T)\approx I(T)$  in case $x \gg 1$.
Thus, to evaluate   $\mathfrak{I}_{9}$ we split the integration over $r_{2}$ in to two integrals, $\int_{0}^{y-r_1}= \int_{0}^{R-r_1-2\ln\frac{\pi}{2}} + \int_{R-r_1-2\ln\frac{\pi}{2}}^{y-r_1}$. In the first integral $\frac{\pi}{2}e^{\frac{r_1+r_2-R}{2}} < 1$ and we approximate $I\left(\frac{\pi}{2}e^{\frac{r_1+r_2-R}{2}};T\right) \approx \frac{\pi}{2}e^{\frac{r_1+r_2-R}{2}}$, while in the second integral  $\frac{\pi}{2}e^{\frac{r_1+r_2-R}{2}} > 1$ and $I\left(\frac{\pi}{2}e^{\frac{r_1+r_2-R}{2}};T\right) \approx I(T)$.  The remaining integration steps in $\mathfrak{I}_{9}$ are straightforward, resulting in
\begin{equation}
\mathfrak{I}_{9}  \approx  \frac{2 \alpha N}{\pi} e^{-\left(\alpha - \frac{1}{2}\right)R} e^{-\alpha r_1} \left[ \frac{1}{\alpha} \left(\frac{\pi}{2}\right)^{1-2\alpha}e^{R\left(\alpha - \frac{1}{2}\right)} + \frac{2 I(T)}{2\alpha - 1}e^{y\left(\alpha - \frac{1}{2}\right)} \right].
\end{equation}

Finally, since $y>R$ and $\alpha > \frac{1}{2}$, we get
\begin{equation}
\mathfrak{I}_{9}  \approx  \frac{4 \alpha I(T) N}{\left(2 \alpha -1 \right)\pi} e^{\left(\alpha - \frac{1}{2}\right)(y-R)}e^{-\alpha r_1}.
\end{equation}

In order to calculate $\mathfrak{I}_{10}$ we first need to estimate the cutoff angle $\phi_{y}$, which is given by the intersection of disks $R$ and $y$, and is given
by Eq.~(\ref{eq:phi_y2}). $\phi_y$ takes values from $\phi_y \approx 2 e^{\frac{y-2R}{2}}$  at $r_1=r_2=R$ to   $\phi_y = \pi$ at $r_2 = y-r_1$. Thus, we can no longer use the  $\phi_y \ll a$ approximation, as in $\mathfrak{I}_{2}$.

To proceed further we note that the integration domain in $\mathfrak{I}_{10}$ is given by the area above the $r_2=y-r_1$ line, Fig.~\ref{fig:spheres2}{\bf c}. We recall that the integration in the case $y<R$ is dominated by points in the vicinity of the $r_1+r_2 = R - 2 \ln \frac{a}{2}$ line [see red dashed line in Fig.~\ref{fig:spheres}{\bf c}]. Let us assume that this is also the case in the $y \geq R$ regime [see red dashed line in Fig.~\ref{fig:spheres2}{\bf c}]. We next note that in the vicinity of the $r_1+r_2 = R - 2 \ln \frac{a}{2}$ line  $\cos \phi_y \approx 1 - 2 e^{y-R-2\ln \frac{a}{2}}$.  For sufficiently small noise amplitude, such that $y < R-2\ln \frac{a}{2}$, the cutoff angle $\phi_y \ll 1$ and can be approximated by Eq.~(\ref{eq:phi_y}), and we can employ the same approximation techniques as in $\mathfrak{I}_{2}$.

Our strategy now is to split the integration domain of $\mathfrak{I}_{10}$ into two parts by the curve $r_2=\tilde{R}(r_1)$ such that (i) this curve is below the $r_1+r_2 = R - 2 \ln \frac{a}{2}$ line, and (ii) above this curve, $r_{2} > \tilde{R}(r_1)$,  the cutoff angle $\phi_{y} \ll 1$. One possibility for such a curve is the $\tilde{R}(r_1)=A-r_{1}$ line, where
$A = \frac{y+R}{2} - \ln \frac{a}{2}$ [see green dashed curve in Fig.~\ref{fig:spheres2}{\bf c}].

Then region VII splits into two subregions, VIIA and VIIB, corresponding to $r_1 \in \left[y-R, 2\ln \frac{2}{a}\right]$ and $r_1 \in \left[2\ln \frac{2}{a}, R\right]$, respectively, see Fig.~\ref{fig:spheres2}{\bf c}. We expect that the contribution to $k_{y}(a)$ from VIIA to be much smaller than that from VIIB since the latter contains the $r_1+r_2 = R - 2 \ln \frac{a}{2}$ line and the former does not. Therefore, we will estimate the upper bound for  $k_{y}(r_1|a)$ in VIIA by replacing $\phi_y$ with $\pi$. In subregion VIIB we split the integration over $r_{2}$ into two intervals, $r_2 \in \left[0, \tilde{R}(r_1)\right]$ and $r_2 \in \left[\tilde{R}(r_1), R\right]$.

{\it Subregion VIIA:} $y-R \leq r_1 \leq 2\ln\frac{2}{a}$.

Here the integral splits into
\begin{eqnarray}
\overline{k}_{y}(r_1|a) &=& \mathfrak{I}_{11} + \mathfrak{I}_{12},\\
\mathfrak{I}_{11}  & =&  \frac{N}{2\pi} \int_{0}^{y-r_1}  {\rm d} r_2 \rho(r_2) \int_{0}^{2\pi}  {\rm d} \phi  p\left[x \left( r_1, r_2, \phi\right) \right] = \overline{k}(r_1),\\
 \mathfrak{I}_{12}  &=&  \frac{N}{\pi} \int_{y-r_1}^{R}  {\rm d} r_2 \rho(r_2) \int_{0}^{\phi_y}  {\rm d} \hat{\phi} \int_{\hat{\phi} - a}^{\hat{\phi} + a}  {\rm d} \phi \tilde{\rho}(\phi | \hat{\phi})  p\left[x \left( r_1, r_2, \phi\right) \right],
\end{eqnarray}
Following our strategy, we evaluate the upper bound for $\mathfrak{I}_{12}$ by replacing the integration limit of ${\phi_y}$ with $\pi$:
\begin{equation}
\mathfrak{I}_{12}  \leq  \frac{N}{\pi} \int_{y-r_1}^{R}  {\rm d} r_2 \rho(r_2) \int_{0}^{\pi}  {\rm d} \hat{\phi} \int_{\hat{\phi} - a}^{\hat{\phi} + a}  {\rm d} \phi \tilde{\rho}(\phi | \hat{\phi})  p\left[x \left( r_1, r_2, \phi\right) \right].
\end{equation}
Then,
\begin{equation}
\overline{k}_{y}(r_1|a) \leq \overline{k}(r_1) = \frac{4 \alpha N I( T)}{\left(2 \alpha - 1\right) \pi}e^{-\frac{r_1}{2}}
\end{equation}
for $y-R \leq r_1 \leq 2\ln\frac{2}{a}$.

{\it Subregion VIIB:} $2\ln\frac{2}{a} \leq r_1 \leq R$.

Here we distinguish three intervals:
\begin{eqnarray}
\overline{k}_{y}(r_1|a) &=& \mathfrak{I}_{13} + \mathfrak{I}_{14} + \mathfrak{I}_{15},\\
\mathfrak{I}_{13}  & =&  \frac{N}{2\pi} \int_{0}^{y-r_1}  {\rm d} r_2 \rho(r_2) \int_{0}^{2\pi}  {\rm d} \phi  p\left[x \left( r_1, r_2, \phi\right) \right],\\
 \mathfrak{I}_{14}  &=&  \frac{N}{\pi} \int_{y-r_1}^{\tilde{R}(r_1)}  {\rm d} r_2 \rho(r_2) \int_{0}^{\phi_y}  {\rm d} \hat{\phi} \int_{\hat{\phi} - a}^{\hat{\phi} + a}  {\rm d} \phi \tilde{\rho}(\phi | \hat{\phi})  p\left[x \left( r_1, r_2, \phi\right) \right],\\
 \mathfrak{I}_{15}  &=&  \frac{N}{\pi} \int_{\tilde{R}(r_1)}^{R}  {\rm d} r_2 \rho(r_2) \int_{0}^{\phi_y}  {\rm d} \hat{\phi} \int_{\hat{\phi} - a}^{\hat{\phi} + a}  {\rm d} \phi \tilde{\rho}(\phi | \hat{\phi})  p\left[x \left( r_1, r_2, \phi\right) \right].
\end{eqnarray}
where $\tilde{R}(r_1) =  \frac{y+R}{2} - \ln \frac{a}{2} - r_1$.

We evaluate the upper bound for $\mathfrak{I}_{14}$ by replacing the $\phi_y$ cutoff with $\pi$:
\begin{equation}
\mathfrak{I}_{14}  \leq  \frac{N}{2 \pi} \int_{y-r_1}^{\tilde{R}(r_1)}  {\rm d} r_2 \rho(r_2) \int_{0}^{2 \pi}  {\rm d} \hat{\phi} \int_{\hat{\phi} - a}^{\hat{\phi} + a}  {\rm d} \phi \tilde{\rho}(\phi | \hat{\phi})  p\left[x \left( r_1, r_2, \phi\right) \right],\\
\end{equation}
leading to
\begin{equation}
\mathfrak{I}_{13} + \mathfrak{I}_{14} \leq  \frac{N}{2\pi} \int_{0}^{\tilde{R}(r_1)}  {\rm d} r_2 \rho(r_2) \int_{0}^{2\pi}  {\rm d} \phi  p\left[x \left( r_1, r_2, \phi\right) \right].
\end{equation}
After the same calculation steps as in $\mathfrak{I}_{9}$ we obtain
\begin{equation}
\mathfrak{I}_{13} + \mathfrak{I}_{14} \leq  \frac{4 \alpha I(T) N}{\left(2 \alpha -1 \right)\pi} e^{\left(\alpha - \frac{1}{2}\right)(\tilde{R}(r_1)-R)}e^{-\alpha r_1}.
\label{eq:i1314}
\end{equation}

To evaluate $\mathfrak{I}_{15}$ we use the $\phi_y \ll 1$ assumption, which enables us to use Eq.~(\ref{eq:phi_y}). This approximation holds since $r_{2} > \tilde{R}(r_1)$. Then, by following the same simplification steps as in $\mathfrak{I}_{4}$ we obtain
\begin{eqnarray}
\mathfrak{I}_{15} &=& \mathfrak{I}_{151} - \mathfrak{I}_{152},\\
\mathfrak{I}_{151}  &=& \frac{N}{\pi a} \int_{\tilde{R}(r_1)}^{R}  {\rm d} r_2 \rho(r_2) \phi_{y} \int_{0}^{a}  \frac{{\rm d} \phi }{1 + {\rm exp}\left(\frac{x \left(r_1, r_2, \phi\right) - R}{2T}\right)},\\
\mathfrak{I}_{152}  &=& \frac{N}{\pi a^{2}} \int_{\tilde{R}(r_1)}^{R}  {\rm d} r_2 \rho(r_2) \phi_{y} \int_{0}^{a}  \frac{  \phi {\rm d} \phi}{1 + {\rm exp}\left(\frac{x \left(r_1, r_2, \phi\right) - R}{2T}\right)}.
\end{eqnarray}
Following the same evaluation steps as in $ \mathfrak{I}_{4}$ we confirm that both $\mathfrak{I}_{151}$ and $\mathfrak{I}_{152}$ are dominated by points in the vicinity of $r_{1}+r_{2} = R-2 \ln \frac{a}{2}$, resulting in
\begin{equation}
\mathfrak{I}_{15} = \mathfrak{I}_{4} = \frac{2N \alpha}{\pi} \left[ \frac{1}{2\alpha - 1}  + \frac{I(T)}{1 - \alpha } -\frac{2 \tilde{I}(T)}{3 - 2\alpha } \right] e^{\frac{y-R}{2}}
 e^{-\alpha r_1} \left(\frac{a}{2}\right)^{1-2\alpha}.
 \label{eq:i15}
\end{equation}
By comparing Eqs.~(\ref{eq:i15}) and Eqs.~(\ref{eq:i1314}) we establish that $\mathfrak{I}_{15} \gg \mathfrak{I}_{13} + \mathfrak{I}_{14}$ since $\tilde{R(r_{1})} < R $ and $y>R$, confirming our hypothesis  and resulting in

\begin{equation}
\overline{k}_{y}(r_1|a) \approx \mathfrak{I}_{4} = \frac{2N \alpha}{\pi} \left[ \frac{1}{2\alpha - 1}  + \frac{2I(T)}{1 - \alpha } -\frac{8 \tilde{I}(T)}{3 - 2\alpha } \right] e^{\frac{y-R}{2}}
 e^{-\alpha r_1} \left(\frac{a}{2}\right)^{1-2\alpha}
\end{equation}
in case $2\ln\frac{2}{a} \leq r_1 \leq R$.

Taken together, our results for regions VI and VII read
\begin{equation}
\overline{k}_{y}(r_1|a)
\begin{cases}
\approx \frac{4 \alpha  N I(T))}{\pi \left(2 \alpha - 1\right)} e^{-\frac{r_1}{2}}&\text{if $ 0\leq r_1 \leq y-R$},\\
\leq \frac{4 \alpha N I( T)}{\left(2 \alpha - 1\right) \pi}e^{-\frac{r_1}{2}}  &\text{if $ y-R \leq r_1 \leq 2 \ln \frac{2}{a}$},\\
\approx \frac{2N \alpha}{\pi} \left[ \frac{1}{2\alpha - 1}  + \frac{I(T)}{1 - \alpha } -\frac{2 \tilde{I}(T)}{3 - 2\alpha } \right] e^{\frac{y-R}{2}}
 e^{-\alpha r_1} \left(\frac{a}{2}\right)^{1-2\alpha} & \text{if $2 \ln \frac{2}{a}\leq r_1 \leq R$}.
 \end{cases}
\label{eq:k1ya2}
\end{equation}

Using Eq.~(\ref{eq:k1ya2}) together with Eq.~(\ref{eq:noisy_degree3}) we finally obtain
\begin{eqnarray}
\overline{k}_{y}(a) &=& k^{1}_{y}(a) + k^{2}_{y}(a),\\
k^{1}_{y}(a) &\leq& \frac{8 \alpha^{2}  N I(T))}{\pi \left(2 \alpha - 1\right)^{2}} e^{-\alpha R}  \left( \frac{a}{2}\right)^{1-2\alpha},\\
k^{2}_{y}(a) &\approx&  \frac{2N \alpha^{2}}{\pi} \left[ \frac{1}{2\alpha - 1}  + \frac{2I(T)}{1 - \alpha } -\frac{8 \tilde{I}(T)}{3 - 2\alpha } \right] e^{\frac{y-R}{2}} e^{-\alpha R} \left(\frac{a}{2}\right)^{1-2\alpha}\left[R + 2 \ln \frac{a}{2}\right]
\end{eqnarray}

Finally, we conclude that $k^{2}_{y}(a) \gg k^{1}_{y}(a)$ since $y > R$, which allows us to establish
\begin{equation}
\overline{k}_{y}(a) \sim N  e^{-\left(\alpha +\frac{1}{2}\right)R}  e^{\frac{y}{2}}  \left(\frac{a}{2}\right)^{1-2\alpha}\left[R + 2 \ln \frac{a}{2}\right]
\label{eq:kya2}
\end{equation}
for $y > R$. Equation~(\ref{eq:kya}) together with Eq.~(\ref{eq:kya2}) establish the baseline for calculation of ${\rm AUPR}(a)$ in Subsection~\ref{sec:auprnoise}.

\section{HYPERLINK Embedder}
\label{sec:hyper_inference}
The original hyperbolic geometry inference algorithm  was developed in Ref.~\cite{Boguna2010sustaining} and is based on MLE. While the algorithm is rather slow with the overall computational complexity of $\mathcal{O}\left(N^{3}\right)$, it has been shown to accurately infer node coordinates in $\mathbb{H}^{2}$ leading to a number of promising applications ranging from interdomain Internet routing~\cite{Boguna2010sustaining} to understanding the growth of large-scale networks~\cite{Papadopoulos2012popularity}.

In recent years hyperbolic geometry inference has become an active area of research and  a collection of alternative inference methods has been developed by different research teams based on the MLE~\cite{Blasius2016efficient,Wang2016link}, Laplacian eigenmaps~\cite{Alanis-Lobato2016efficient,Alanis-Lobato2016manifold,Muscoloni2017machine} and \textsc{isomap}~\cite{Muscoloni2017machine}.  Even though most of these methods are characterized by relatively small computational complexity,  $\mathcal{O}\left(N\right)$ - $\mathcal{O}\left(N^{2}\right)$, their inference accuracy has not been well explored.

At the same time, our initial experiments indicate that even small node coordinate uncertainties drastically reduce  link prediction accuracy (Fig.~\ref{fig:13}). Therefore, to optimize link prediction results one needs to maximize the accuracy of node coordinate inference. To this end, we developed an enhanced MLE-based geometry inference algorithm, which we outline below.

\subsection{General MLE formulation of hyperbolic geometry inference}
\label{sec:inference}
Given the real network of interest with randomly removed links, we aim to find the set of node coordinates $\{\mathbf{x}_{i}\} \equiv \{(r_i,\theta_i)\}$, $i=1,2,\ldots,N$, in the hyperbolic disk $\mathbb{H}^{2}$ maximizing the probability  $\mathcal{L}\left( \{\mathbf{x}_{i}\} | a_{ij}, \mathcal{P}, q \right)$  that node coordinates take particular values in the case the network is generated as the RHG with a subsequent random link removal process. Here $a_{ij}$ is the network's observed adjacency matrix, and $\mathcal{P}$ is the set of parameters of the RHG, $\mathcal{P}= \{\alpha, T, R\}$.

By the Bayes rule the thought probability is given by
\begin{equation}
\mathcal{L}\left( \{\mathbf{x}_{i}\} | a_{ij}, \mathcal{P}, q \right) = \frac{\mathcal{L}\left(  a_{ij}| \{\mathbf{x}_{i}\}, \mathcal{P}, q \right) {\rm Prob} (\mathbf{x}_i)}{\mathcal{L}\left( a_{ij}| \mathcal{P}, q \right)},
\label{eq:general_likelihood}
\end{equation}
where $\mathcal{L}\left(  a_{ij}| \{\mathbf{x}_{i}\}, \mathcal{P}, q \right)$ is the likelihood that network $a_{ij}$ is generated as RHG with subsequent random link removal, ${\rm Prob} (\mathbf{x}_i)$ is the prior probability of node coordinates generated by the RHG, and $\mathcal{L}\left(  a_{ij}|  \mathcal{P}, q \right)$ is the probability that the network has been generated as the RHG with random link removal.

In the following we assume the uniform prior probability
\begin{equation}
{\rm Prob} (\mathbf{x}_i) = \frac{1}{\left(2 \pi\right)^{N}} \prod_{i=1}^{N} \rho(r_{i}),
\end{equation}
where $\rho(r_i)$ are given by Eq.~(\ref{eq:rho_r}). Since node pairs are connected independently, the likelihood is given by
\begin{equation}
\mathcal{L}\left(  a_{ij}| \{\mathbf{x}_{i}\}, \mathcal{P}, q \right) = \prod_{i < j} \left[\tilde{p}\left(x_{ij}\right)\right]^{a_{ij}}  \left[1 -\tilde{p}\left(x_{ij}\right)\right]^{1 - a_{ij}},
\label{eq:likelihood}
\end{equation}
where $\tilde{p}\left(x_{ij}\right)$ is the effective connection probability in the RHG generation process with subsequent random link removal:
\begin{equation}
\tilde{p}\left(x\right) \equiv q p(x),
\label{eq:pijq}
\end{equation}
where $p(x)$ is the RHG connection probability function prescribed by Eq.~(\ref{eq:conn}).

The MLE inference aims to find node coordinates $\hat{\mathbf{x}}_i$ maximizing the likelihood $\mathcal{L}\left( \{\mathbf{x}_{i}\} | a_{ij}, \mathcal{P}, q \right)$, or equivalently, its logarithm
\begin{equation}
\ln \mathcal{L}\left( \{\mathbf{x}_{i}\} | a_{ij}, \mathcal{P}, q \right) = K + \sum^{N}_{i=1} \ln \rho(r_i) + \sum_{i < j} \left[a_{ij} \ln \tilde{p}\left(x_{ij}\right)  + \left(1 - a_{ij}\right) \ln \left(1 - \tilde{p}\left(x_{ij}\right)\right) \right],
\label{eq:loglikelihood}
\end{equation}
where constant $K$ absorbs all terms independent of $ \{\mathbf{x}_{i}\}$.

Our hyperbolic geometry inference procedure consists of three components: (1) finite-size effects and model parameter inference, (2) MLE-based inference of radial node coordinates, (3) MLE-based inference of angular node coordinates.

\subsection{Finite-size effects and model parameter inference}
The RHG has four parameters: the number of nodes $N$, hyperbolic disk radius $R$, node density parameter $\alpha$ and temperature $T$.

To infer $\alpha$ we first estimate the degree distribution exponent $\gamma$ through the inspection of the network degree distribution $P(k)$. Node density $\alpha$ is related to $\gamma$ through Eq.~(\ref{eq:gamma}):
\begin{equation}
\alpha = \frac{1}{2}\left(\gamma - 1\right).
\end{equation}

The estimation of $N$ and in $R$ is less straightforward due to finite-size effects. First, in a real network one normally can only observe nodes with nonzero degrees. In contrast, the RHG may generate nodes of zero degree, which are accounted for in the calculation of the network's average degree, $\overline{k}$, Eq.~(\ref{eq:avgk}).

Second, due to finite-size effects, there is a cutoff value for the smallest node radius, $R_{0}$, affecting $\langle e^{-r/2} \rangle$ and, as a result, the observable $\overline{k}(r)$ and $\overline{k}$, Eqs.~(\ref{eq:avgk}) and (\ref{eq:avgk_r}). Specifically, with the radius cutoff $R_{0}$
\begin{equation}
\langle e^{-r/2} \rangle \left( R_0 \right) = \int_{R_{0}}^{R} e^{-r/2} \rho(r) {\rm d} r = \langle e^{-r/2} \rangle \lambda\left(\alpha, R-R_{0}\right),
\end{equation}
where $\lambda(\alpha,x)$ is the finite-size correction coefficient:
\begin{equation}
\lambda\left(\alpha, x\right) \equiv 1 - e^{-(\alpha - 1/2)x}.
\end{equation}
In the thermodynamic limit $ \lambda (\alpha, \left(R-R_0\right) ) \to 1$  as
\begin{equation}
1 - \lambda (\alpha, \left(R-R_0\right) ) \sim N^\frac{1-2\alpha }{2\alpha} = N ^{\frac{2-\gamma}{\gamma-1}}.
\label{eq:finite}
\end{equation}
However, in networks with $\alpha$ close to $1/2$ ($\gamma$ close 2) the rate of $\lambda$ convergence is slow and one needs to account for nonzero $R_0$.

Third, one needs to account for missing links that affect all observable properties of the RHG. In the particular case links are missing uniformly with probability $1-q$, the connection probability function  $p(x)$ gets attenuated by the factor of $q$, Eq.~(\ref{eq:pijq}), affecting all observable network properties.

Taken together, zero degree nodes, minimum radius cutoff, and missing links affect observable network properties as follows:
\begin{eqnarray}
\tilde{N} &=& N (1-P(0)),\label{eq:tilden}\\
\tilde{k} &=& \frac{q \left[\lambda\left(\alpha, R-R_{0} \right)\right]^{2}}{1-P(0)} \overline{k},\label{eq:tildek}\\
\tilde{k}_{\rm max} & \approx & q \lambda\left(\alpha, R-R_{0} \right) \overline{k} \frac{e^{-R_{0}/2} }{\langle e^{-r/2}\rangle}\label{eq:tildekm},
\end{eqnarray}
where $\tilde{k}_{\rm max}$ is the maximum degree observed in the network and $P(0)$ is the fraction of zero degree nodes in the network. The latter can be estimated by averaging the conditional degree distribution $P(k=0|r)$ in  Eq.~(\ref{eq:cond_pkr}) over possible $r$ values:
\begin{eqnarray}
P(0) &=& 2 \alpha \tau^{2\alpha} \Gamma\left[-2\alpha, \tau \right], \label{eq:p0}\\
\tau &\equiv& q \left[\lambda\left(\alpha, R-R_{0} \right)\right] \overline{k} \frac{e^{-R/2}}{\langle e^{-r/2}\rangle},
\label{eq:tau}
\end{eqnarray}
where $\Gamma\left[s,x\right]$ is the upper incomplete gamma function.

Equations.~(\ref{eq:tilden}), (\ref{eq:tildek}), (\ref{eq:tildekm}), (\ref{eq:p0}) and (\ref{eq:tau}) allow one to infer the RHG parameters $R_{0}$, $R$, $N$ as well as resulting $\overline{k}$, and $P(0)$ by measuring observables $\tilde{N}$, $\tilde{k}$, and $\tilde{k}_{\rm max}$.

The caveat here is that parameter estimation presumes the knowledge of the missing link probability $1-q$. While this information is available in our synthetic experiments, it may not be available in real networks. In case the fraction of missing links is small, one can assume that $q=1$. The most general case of substantially incomplete networks where $q \ll 1$ is beyond the scope of this paper and will be studied elsewhere.

Finally, the temperature parameter $T$ needs to be estimated numerically by finding the solution of
\begin{equation}
\overline{c}(T) = c_{0},
\end{equation}
where $c_{0}$ is the average clustering coefficient of the network of interest and $\overline{c}(T)$ is the average clustering coefficient of the RHG generated with temperature $T$.  We utilize this approach to  infer $T$ of real networks in Sec.~\ref{sec:real_infer_coords}, while in experiments with RHGs we use actual $T$ values.

\subsection{MLE-based inference of radial node coordinates}
To infer radial node coordinates we extremize the logarithm of the  likelihood function,
\begin{equation}
\frac{\partial }{\partial r_{\ell}}  \ln \mathcal{L}\left( \{\mathbf{x}_{i}\} | a_{ij}, \mathcal{P}, q \right) = 0,
\end{equation}
obtaining
\begin{equation}
2\alpha T {\rm coth}\left(\alpha r_{\ell}\right) + \sum_{j} \left[ \frac{1 - p\left(x_{\ell j}\right)    }{1 - q p\left(x_{\ell j}\right)} \left( a_{\ell,j} -q p\left(x_{\ell, j}\right) \right) \right]\frac{\partial x_{\ell j}}{\partial r_{\ell}} = 0.
\label{eq:infer_r}
\end{equation}
In the case of sufficiently large $r$ values ${\rm coth}\left(\alpha r_{\ell}\right) \approx 1$. Further, one can approximate $x_{\ell j}$ as
\begin{equation}
x_{\ell j} = r_{\ell} + r_{j} + \ln \sin \theta_{\ell j}/2,
\end{equation}
resulting in  $\frac{\partial x_{\ell j}}{\partial r_{\ell}}\approx 1$.  Taken together, these approximations allow us to simplify Eq.(\ref{eq:infer_r}) as
\begin{equation}
2\alpha T + \sum_{j} a_{\ell,j} - q \sum_{j} p\left(x_{\ell, j}\right) = 0
\label{eq:r_infer}
\end{equation}
for $1-q \ll 1$.
Note that the first summation in Eq.~(\ref{eq:r_infer}) is the degree of node $\ell$, $\sum_{j} a_{\ell j} = k_{\ell}$, while the second summation is the expected degree of the node with $r_{\ell}$, $\tilde{k}\left(r_{\ell} \right) = q \sum_{j} p\left(x_{\ell, j}\right)$. As a result, the value of $\hat{r}_{\ell}$ extremizing the likelihood is given by
\begin{equation}
\tilde{k}\left(\hat{r}_{\ell} \right) = k_{\ell} + 2\alpha T,
\end{equation}
where $\tilde{k}\left(r \right)$ is the observable expected degree of the node with radial coordinate $r$.  Since the latter is given by
\begin{equation}
\tilde{k}\left( r \right) = q \lambda \left(\alpha, R - R_0 \right) \overline{k} \frac{e^{-r/2}}{\langle e^{-r/2} \rangle},
\end{equation}
one can estimate $\hat{r}_{\ell}$ as
\begin{equation}
\hat{r}_{\ell} = 2 \ln \left[\frac{ q  \lambda \left(\alpha, R - R_0 \right) \overline{k}}  { \left( k_{\ell} +  2 \alpha T\right) \langle e^{-r/2} \rangle }\right].
\label{eq:r_inference}
\end{equation}

\subsection{MLE inference of angular node coordinates}
\begin{algorithm}[t]
\algsetup{indent=2em}
\caption{Angular MLE Inference}
\begin{algorithmic}
\STATE organize network nodes into layers $\{s_i\}$ and cores $\{cr_i\}$, $i=0,1,..., m$.
\STATE define the sequence of subgraphs $\{G_{i}\}$ spanned by nodes in $\{cr_i\}$.
\FOR {${\rm iter} = 0$ to ${\rm max\_iter}$}
    \FOR{$\ell=0$ to $\lfloor m/2 \rfloor $ (first half)}
        \STATE assign random angle values, $\theta_{i} \leftarrow U[0, 2\pi]$,  to nodes in $s_\ell$. \\
        Other nodes in $G_\ell$ retain their previous angular positions.
        \STATE $a(\ell) \leftarrow \frac{\pi}{4}\left(1 - \frac{\ell}{m}\right)+ a_0$.
        \FOR{all nodes $i$  in $G_{\ell}$}
            \STATE $X_{i} \leftarrow U\left(- \frac{\pi}{2}, \frac{\pi}{2} \right) $.
            \STATE $\hat{\theta}_i \leftarrow \hat{\theta}_i + a(\ell) X_i$.
        \ENDFOR
        \REPEAT
            \FOR{all nodes $i$  in $G_{\ell}$}
                \STATE $\hat{\theta}_{i} \leftarrow  {\rm argmax}~{\rm ln}~\mathcal{L} \left[G_\ell\right]_{i} $, see Alg.~\ref{alg2}.
            \ENDFOR
        \UNTIL{ ($ {\rm max}_{i \in G_{\ell}} \Delta \hat{\theta}_i  < \epsilon $) or
        (\# rounds $> {\rm max\_rounds}$ )}

    \ENDFOR
    \STATE compute resulting log-likelihood ${\rm ln}~\mathcal{L} \left[G_{\lfloor m/2 \rfloor}\right]$ value and save corresponding $\{\theta_i\}$ values.
\ENDFOR
\STATE continue with $\{\theta_i\}$ values corresponding to the largest ${\rm ln}~\mathcal{L} \left[G_{\lfloor m/2 \rfloor}\right]$.
\FOR{$\ell=\lfloor m/2 \rfloor + 1$ to $m$ (second half)}
    \STATE assign random $\{\theta_{i}\}$ values to nodes in $s_\ell$. Other nodes in $G_\ell$ retain their previous angular positions.
    \REPEAT
        \FOR{all nodes $i$  in $G_{\ell}$}
            \STATE $\hat{\theta}_{i} \leftarrow  {\rm argmax}~{\rm ln}~\mathcal{L} \left[G_\ell\right]_{i} $, see Alg.~\ref{alg2}.
        \ENDFOR
    \UNTIL{ ($ {\rm max}_{i \in G_{\ell}} \Delta \hat{\theta}_i  < \epsilon $) or
    (\# rounds $> {\rm max\_rounds}$ )}
    \STATE $a(\ell) \leftarrow \frac{\pi}{4}\left(1 - \frac{\ell}{m}\right)+ a_0$.
    \FOR{all nodes $i$  in $G_{\ell}$}
        \STATE $X_{i} \leftarrow U\left(- \frac{\pi}{2}, \frac{\pi}{2} \right) $.
        \STATE $\hat{\theta}_i \leftarrow \hat{\theta}_i + a(\ell) X_i$.
    \ENDFOR
\ENDFOR
\FOR{$20$ iterations}
        \FOR{all nodes $i$  in $G$}
            \STATE $X_{i} \leftarrow U\left(- \frac{\pi}{2}, \frac{\pi}{2} \right) $.
            \STATE $\hat{\theta}_i \leftarrow \hat{\theta}_i + a_{0} X_i$.
        \ENDFOR
        \FOR{all nodes $i$  in $G$}
            \STATE $\hat{\theta}_{i} \leftarrow  {\rm argmax}~{\rm ln}~\mathcal{L} \left[G\right]_{i} $, see Alg.~\ref{alg2}.
        \ENDFOR
\ENDFOR
\end{algorithmic}
\label{alg1}
\end{algorithm}
\begin{figure}
\includegraphics[width=3in]{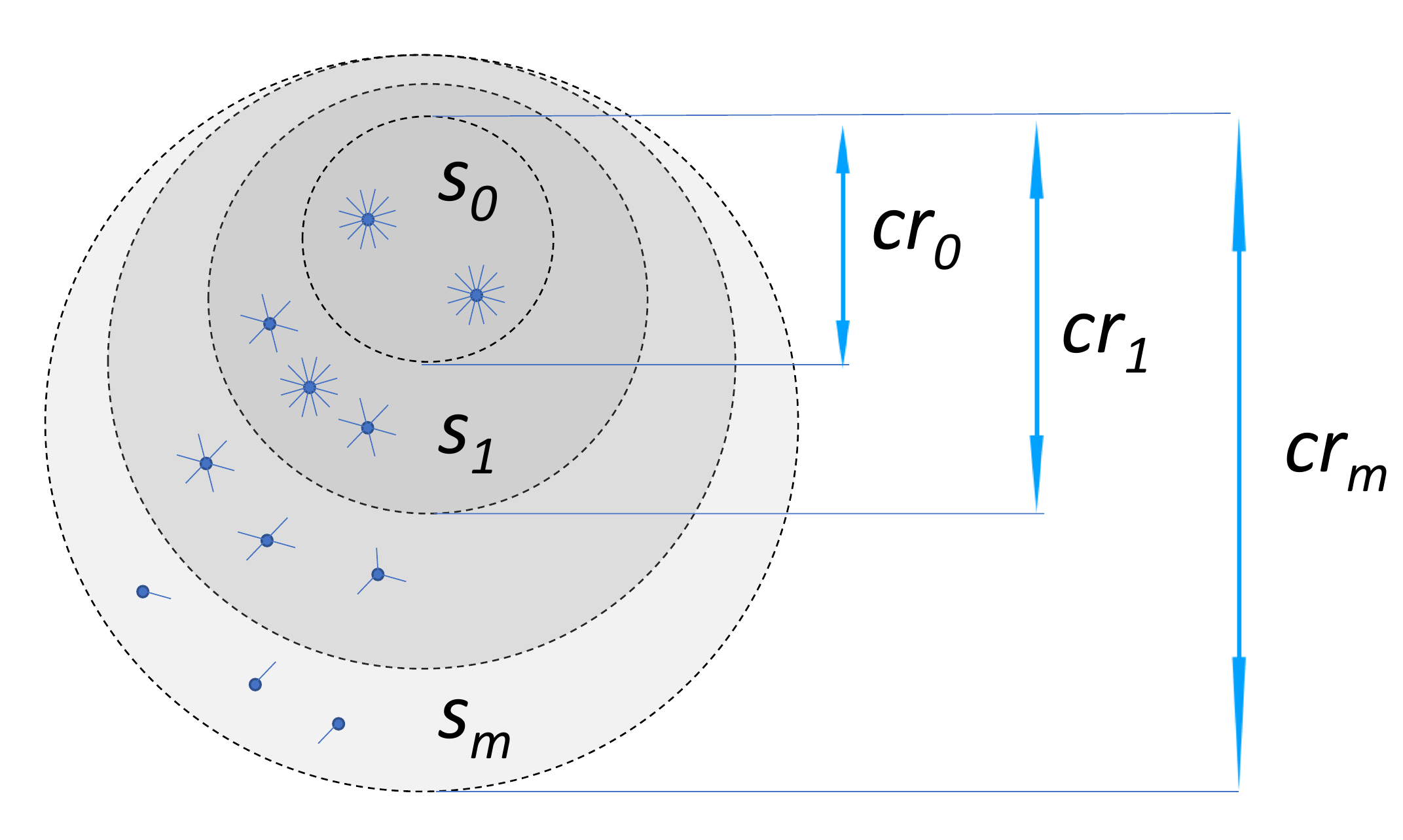}
\caption{\footnotesize Layered network structure for MLE inference. Nodes are sorted in the decreasing order of their degree and placed into logarithmically sized layers. The outer layer contains only $k=1$ nodes.}
\label{fig:a2}
\end{figure}

To infer angular node coordinates one needs to maximize the likelihood $ {\rm ln}~\mathcal{L}\left( \{\mathbf{x}_{i}\}|  a_{ij}, \mathcal{P}, q \right)$ in Eq.~(\ref{eq:loglikelihood}) with respect to angular coordinates $\{\theta_i\}$, given the MLE values for radial coordinates $\{\hat{r}_{i}\}$. Since the maximization of $ {\rm ln}~\mathcal{L}\left( \{\mathbf{x}_{i}\}|  a_{ij}, \mathcal{P}, q \right)$ with respect to $\{\theta_{i}\}$ cannot be performed analytically, we have to rely on numerical approximations. To this end, we developed an MLE-based algorithm optimized for the linked prediction problem.

Conceptually, our algorithm is similar to the one developed in Ref.~\cite{Boguna2010sustaining} but has several important differences.

Following the exposition of Ref.~\cite{Boguna2010sustaining}, we make two observations based on the link independence in RHG.  First, angular coordinates of any node subset $\mathbb{S}$ can be inferred independently (albeit, with lower accuracy) based only on the partial information contained in the graph $G_{\mathbb{S}}$ formed by these nodes. In other words, the inference of angular coordinates in  $\mathbb{S}$  is possible by maximizing the $\mathbb{S}$-specific log likelihood:
\begin{equation}
{\rm ln}~\mathcal{L}\left[G_{\mathbb{S}}\right] = \frac{1}{2}\sum_{\{i,j\}\in G_{\mathbb{S}}} \left[a_{ij} \ln \tilde{p}\left(x_{ij}\right)  + \left(1 - a_{ij}\right) \ln\left(1 - \tilde{p}\left(x_{ij}\right)\right) \right].
\label{eq:log_likelihood}
\end{equation}
Second, any log likelihood $\mathcal{L}\left[G_{\mathbb{S}}\right]$ can be represented as a sum of local contributions $\mathcal{L}\left[G_{\mathbb{S}}\right]_{i}$:
\begin{equation}
{\rm ln}~\mathcal{L}\left[G_{\mathbb{S}}\right] = \frac{1}{2} \sum_{i} {\rm ln}~\mathcal{L}\left[G_{\mathbb{S}}\right]_{i},
\end{equation}
where
\begin{equation}
{\rm ln}~\mathcal{L}\left[G_{\mathbb{S}}\right]_{i} = \sum_{j \neq i \in G_\mathbb{S}} \left[a_{ij} \ln \tilde{p}\left(x_{ij}\right)  + \left(1 - a_{ij}\right) \ln\left(1 - \tilde{p}\left(x_{ij}\right)\right) \right].
\end{equation}

Since the log-likelihood profile $ {\rm ln}~\mathcal{L}\left( \{\mathbf{x}_{i}\}|  a_{ij}, \mathcal{P}, q \right)$  is nonconvex with abundant local maxima, we do not intend to find its global maximum by optimizing all angles at once. Instead, we proceed in a nested fashion by organizing network nodes into logarithmically sized layers with nodes of larger degree belonging to inner layers. To this end, we define the set $\mathbb{C}$ of all nodes with degrees $k > 1$. We then rank all nodes in $\mathbb{C}$  in the decreasing order of their degree value, and split the resulting node list into $m$ layers with logarithmically growing sizes $s_i$, $i = 0,.., m-1$:
\begin{eqnarray}
s_{i+1} &=& \lfloor w \times s_{i} \rfloor,\\
w &=& \left[N(k>1)\right]^{ 1 / m },
\end{eqnarray}
where $N(k>1)$ is the number of nodes with degree $ k>1 $, and $s_0 \ll N$. Unless otherwise noted, we set $s_0=20$. Finally, all $ k = 1 $  nodes are assigned to the outer layer $s_{m}$.

Complementary to layers $\{ s_i \}$,  we also define self-enclosed cores $\{cr_{i}\}$, $i = 0,.., m$, such that core $cr_{i}$ contains all layer with indices $j \leq i$, $cr_{i} = \prod_{j =0}^{i} \bigcup s_j$, as well as the sequence of nested subgraphs $\{G_{i}\}$, $i=0,..,m$, spanned by the nodes in corresponding cores, see Fig.~\ref{fig:a2}.

 We start by inferring node angular coordinates $i \in cr_{0}$ by maximizing $G_{0}$-specific likelihood ${\rm ln} \mathcal{L} \left[G_0\right]$. We then utilize the inferred angles $\{\theta_{i}\} \in cr_0$ as initial approximation to  maximize $ {\rm ln} \mathcal{L}\left[G_1\right]$. We continue the angular coordinate inference procedure in the nested fashion to find angular values  maximizing $ {\rm ln} \mathcal{L} \left[G_m\right]$:
\begin{equation}
 {\rm ln} \mathcal{L} \left[G_0\right] \to {\rm ln} \mathcal{L} \left[G_1\right] \to ... \to {\rm ln} \mathcal{L} \left[G_m\right].
\end{equation}

We maximize each log likelihood ${\rm ln} \mathcal{L} \left[G_\ell\right]$ iteratively by visiting $G_\ell$ nodes in rounds. At each round every node $i$ in $G_\ell$ is visited once and placed at $\hat{\theta}_{i}$ maximizing its local log likelihood $\mathcal{L}\left[G_{\mathbb{S}}\right]_{i}$ with respect to the current angular values of other nodes in $G_\ell$. The procedure is continued until we arrive at the stable angular configuration:
\begin{equation}
{\rm max}_{i \in G_{\ell}} \Delta \hat{\theta}_i  < \epsilon,
\end{equation}
where $0 <\epsilon \ll 1$ is the precision parameter and $ \Delta \hat{\theta}_i$ is the angular difference between angular positions of node $i$ in two consecutive rounds.
In our experiments we set $\epsilon = 10^{-4}$ radians.

The required total number of all-node visit rounds is typically small, of the order of the network average degree. In certain circumstances, e.g., in the case of the global ${\rm ln}~\mathcal{L} \left[G_\ell\right]$ maximum close to the second largest maximum, the procedure may require a large number of rounds to converge. To avoid these scenarios we limit the maximum number of rounds to $10$  per $G_{\ell}$.

Our experiments indicate that the resulting \textsc{hyperlink} link prediction accuracy is highly sensitive to the correct placement of highest degree nodes. Thus, to further improve angular inference of the most connected nodes, we split the procedure into two parts, $\ell = 0,1,..,\lfloor m/2 \rfloor$ and $\lfloor m/2 \rfloor+1,.., m$, respectively. The first part is repeated independently for ${\rm max\_iter = 20}$ times, starting from different initial angle values. For each repetition the resulting  ${\rm ln}~\mathcal{L} \left[G_{\lfloor m/2 \rfloor}\right]$ value is computed. The second part is carried out only once using $\{\theta_i\}$ values corresponding to the iteration with largest  ${\rm ln}~\mathcal{L} \left[G_{\lfloor m/2 \rfloor}\right]$ value. Since $\ell = 0,1,..,\lfloor m/2 \rfloor$ cores are significantly smaller than $\ell = \lfloor m/2 \rfloor+1,.., m/2$ cores, the first part is carried out much faster than the second, despite the large number of repetitions.

After each round $\ell$ we perturb the angular coordinates  $\hat{\theta}_{i}$, $i \in {\rm cr_{\ell}}$, by adding random noise:
\begin{eqnarray}
\hat{\theta}_{i} &\leftarrow& \hat{\theta}_{i} + a(\ell) X_{i},\\
\label{eq:noise}
X_{i}  &\leftarrow& U\left(-\frac{\pi} {2}, \frac{\pi}{2}\right),
\end{eqnarray}
with amplitude $a(\ell)$, which we decrease linearly as  $a(\ell) = \frac{\pi}{4}\left(1 - \frac{\ell}{m}\right)+ a_0$.  These coordinate perturbations allow us to avoid getting trapped in local maxima of the log-likelihood function and to arrive to the optimal angles $\{\theta_i\}$ faster. We also stress the importance of the nonzero residual noise amplitude of $a_0$. In the final $\ell = m$ stage residual noise allows us to effectively "repel" $k=1$ nodes connected to the same node. Without residual noise at the $\ell=m$ step, all $k=1$ nodes connected to the same node are likely to be placed very close to each other and their common neighbor. As a result, pairs of these $k=1$ nodes will be ranked as the most likely candidates for link prediction, and will adversely affect the \textsc{hyperlink} accuracy. Our experiments indicate that the \textsc{hyperlink}  accuracy is not sensitive to specific $a_0$ values, as long as $a_0 \in \left[10^{-6}, 10^{-3}\right]$. In all our experiments we set $a_0= 10^{-4}$ radians. 

The final part of the embedder algorithm is the series of $20$ coordinate perturbations, following local coordinate inferences in the entire network $G$. This last step often helps to further improve coordinate inference accuracy and, consequently, the accuracy of link prediction. The angular inference procedure is summarized in Alg.~\ref{alg1}.

\begin{figure}
\includegraphics[width=3in]{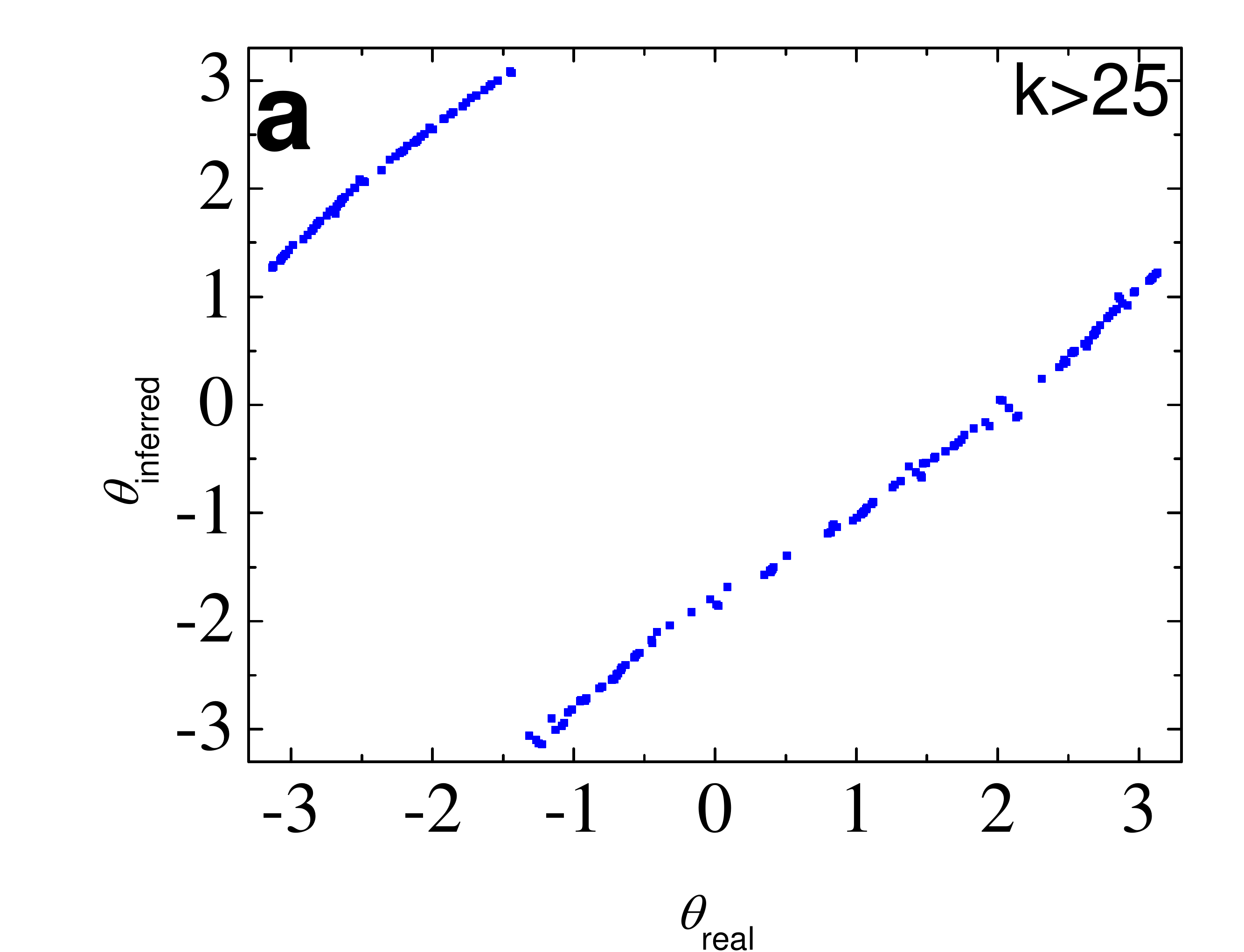}
\includegraphics[width=3in]{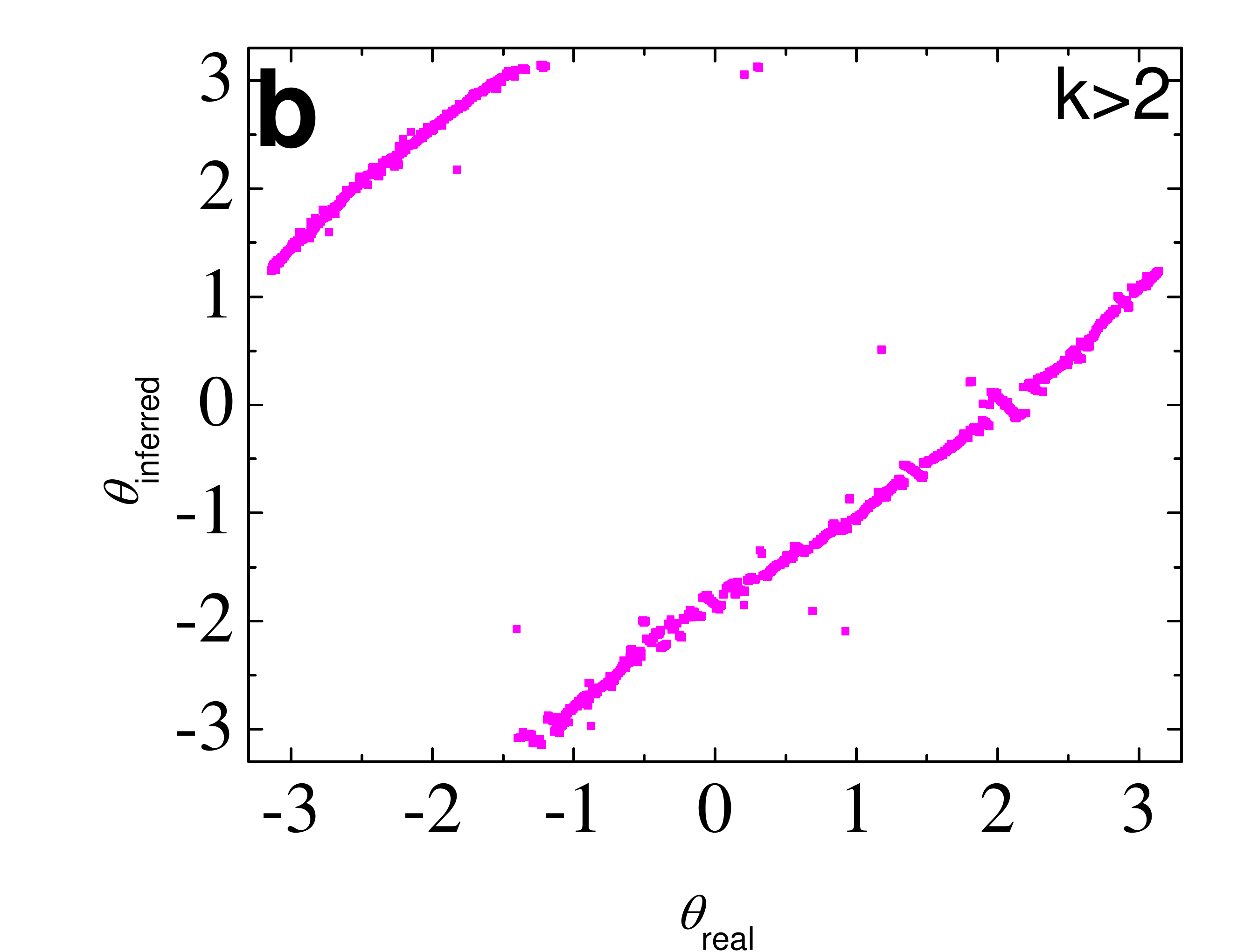}
\includegraphics[width=3in]{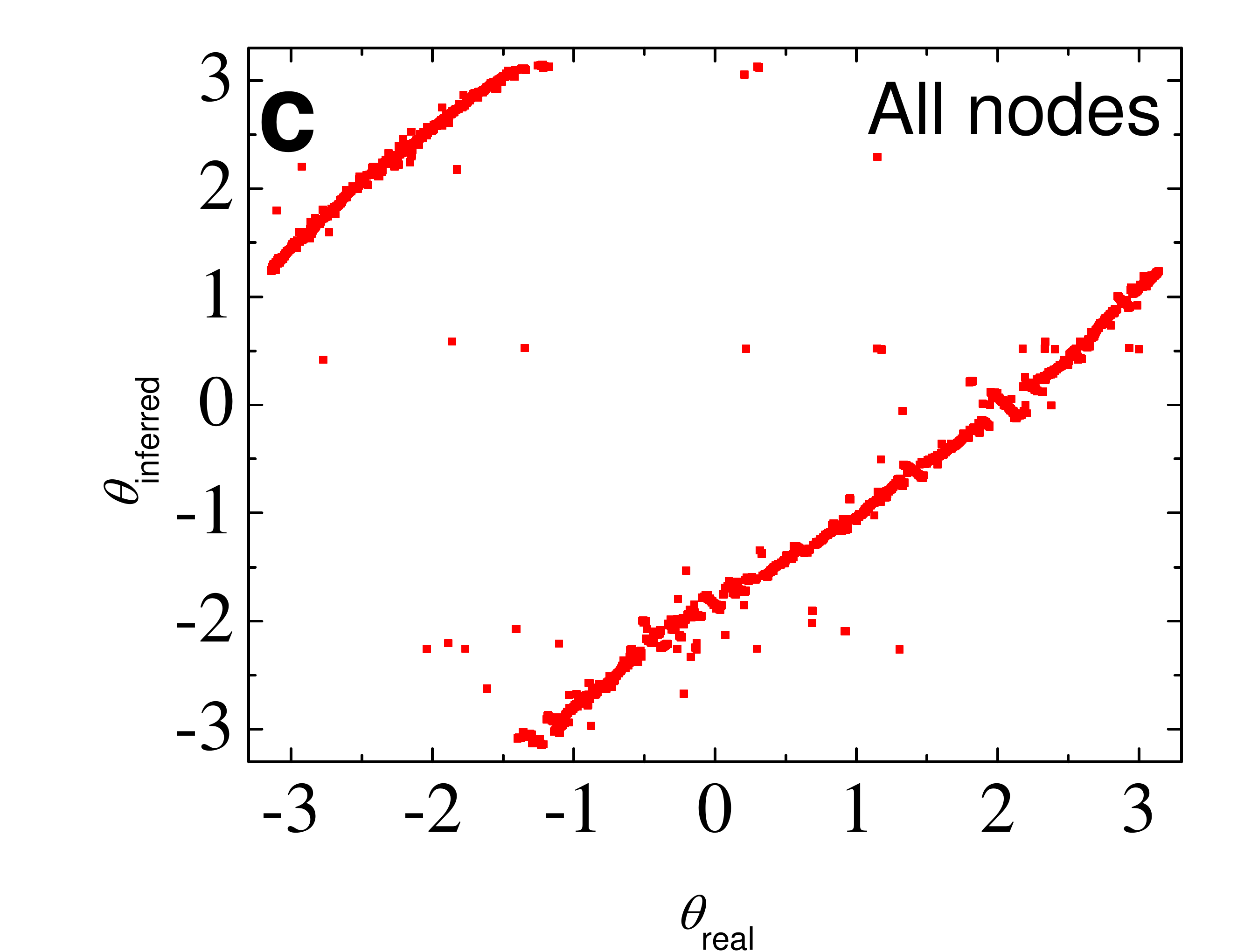}
\includegraphics[width=3in]{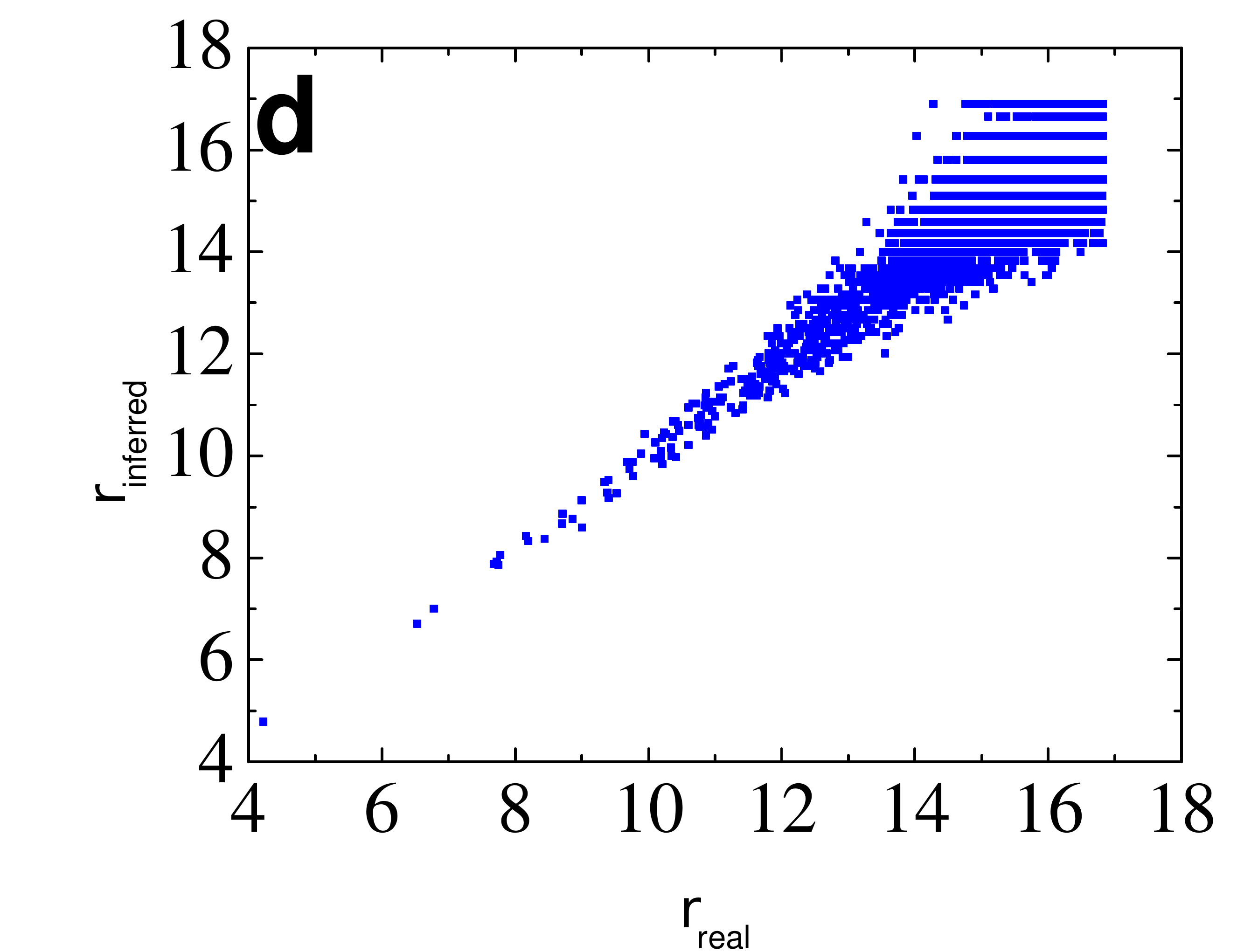}
\caption{\footnotesize {\bf Testing the hyperbolic geometry inference algorithm.} Here we plot inferred vs original node coordinates for the RHG that we map to the hyperbolic space. All plots correspond to the same RHG of $N=5,000$, $\langle k \rangle = 10$, $T=0.5$, and $\gamma = 2.5$. Panels {\bf a} and {\bf b} display angular coordinates for nodes with degrees $k>25$ and $2$, respectively. Panel {\bf c} displays angular coordinates of all nodes. Panel {\bf d} displays radial coordinates of all nodes in the graph.}
\label{fig:validation}
\end{figure}

Having sketched the angular inference procedure, we now focus on the individual node placement subroutine. We determine $\hat{\theta}_i$ for each node by maximizing the corresponding local log likelihood ${\rm ln}~\mathcal{L} \left[G_\ell\right]_{i}$. To this end, we split the angular space $\left[-\pi,\pi\right]$ evenly into $\mathcal{O}(N_{\ell})$ regions, where $N_{\ell}$ is the number of nodes in $G_\ell$. By placing node $i$ into each of these regions we then identify $\hat{\theta}_i$ maximizing its local likelihood. Since ${\rm ln}~\mathcal{L} \left[G_\ell\right]_{i}$ calculation takes $\mathcal{O}(N_{\ell})$ steps for each $\theta_i$ value, it takes  $\mathcal{O}(N_{\ell}^{3})$ steps to execute each round $\ell$. As a result, the overall running-time complexity for $m$ layers, $\mathcal{O}(m N^{3})$, is prohibitive for large networks.

To reduce the running time complexity to $\mathcal{O}(m \langle k \rangle N^{2})$, where $\langle k \rangle$ is the average degree of the entire network, we utilize the following approximation, first offered in Ref.~\cite{Boguna2010sustaining}. If the number of nodes in $G_{\ell}$ is larger than or equal to $500$, for each node we first obtain the rough  estimate of $\hat{\theta}_i$ by taking into account only its neighboring nodes in $G_{\ell}$. To this end we find the nearly optimal placement $\tilde{\theta}_i$ by maximizing
\begin{equation}
{\rm ln}~\mathcal{\tilde{L}}\left[G_{\mathbb{S}}\right]_{i} = \sum_{j \neq i \in G_\mathbb{S}} a_{ij} \ln \tilde{p}\left(x_{ij}\right).
\label{eq:local_log_likelihood}
\end{equation}
Since the summation in Eq.~(\ref{eq:local_log_likelihood}) goes only through node $i$ neighbors, it now takes $\mathcal{O}( k_i N)$ steps to find $\tilde{\theta}_i$. Having obtained the initial approximation, we then look for the optimal angle $\hat{\theta}_i$ in the neighborhood of $\tilde{\theta}_i$ maximizing the full local likelihood ${\rm ln}~\mathcal{L}\left[G_{\mathbb{S}}\right]_{i}$, which takes $\mathcal{O}( L N)$ steps, where $L$ is the neighborhood  centered at $\tilde{\theta}_i$. Specifically, we search for $\hat{\theta}_i$ within $L=300 \frac{N_{\ell}}{N}$ regions on both sides of $\tilde{\theta}_i$, which takes  $\mathcal{O}\left( \frac{N_{\ell}^{2}}{N} \right)$ steps, leading to the overall running time complexity of $\mathcal{O}(m \langle k \rangle N^{2})$ steps. The individual node placement subroutine is summarized in Alg.~\ref{alg2}.

\begin{algorithm}[t]
\algsetup{indent=2em}
\caption{Individual node placement subroutine}
\begin{algorithmic}
\IF {$N_{\ell} < 500$}
\STATE split the angular space $\left[-\pi,\pi\right]$ evenly into $\mathcal{O}(N_{\ell})$ regions.
\FOR{each region $r$ in $\left[-\pi,\pi\right]$ }
    \STATE assign $\theta_{i}(r)$ values to lower boundaries of each region $r$.
    \STATE compute ${\rm ln}~\mathcal{L}\left[G_{\ell}\right]_{i}$ for $\theta_{i}(r)$, as defined in  Eq.~(\ref{eq:log_likelihood}).
\ENDFOR
\STATE  $\hat{\theta}_i \leftarrow {\rm argmax}_{r \in [-\pi, \pi]}~{\rm ln}~\mathcal{L}\left[G_{\ell}\right]_{i}$

\ELSE
    \STATE split the angular space $\left[-\pi,\pi\right]$ evenly into $\mathcal{O}(N_{\ell})$ regions.
    \FOR{each region $r$ in $\left[-\pi,\pi\right]$ }
         \STATE sample $\theta_{i}(r)$ uniformly at random from region $r$.
         \STATE compute ${\rm ln}~\mathcal{\tilde{L}}\left[G_{\ell}\right]_{i}$ for $\theta_{i}(r)$, as defined in  Eq.~(\ref{eq:local_log_likelihood}).
    \ENDFOR
    \STATE  $\tilde{\theta}_i \leftarrow {\rm argmax}_{r \in [-\pi, \pi]}~{\rm ln}~\mathcal{\tilde{L}}\left[G_{\ell}\right]_{i}$
    \FOR{each region $r$ in $\left[\|\tilde{\theta}_{i} -L\|, \|\tilde{\theta}_{i} +L \| \right]$ }
        \STATE assign $\theta_{i}(r)$ values to lower boundaries of each region $r$.
        \STATE compute ${\rm ln}~\mathcal{L}\left[G_{\ell}\right]_{i}$ for $\theta_{i}(r)$.
    \ENDFOR
    \STATE Identify $\hat{r}$ maximizing ${\rm ln}~\mathcal{L}\left[G_\ell\right]_{i}$. $\hat{\theta}_{i} \leftarrow \theta_{i}(\hat{r})$
    \STATE  $\hat{\theta}_i \leftarrow {\rm argmax}_{r \in \left[\|\tilde{\theta}_{i} -L\|, \|\tilde{\theta}_{i} +L \| \right]}~{\rm ln}~\mathcal{L}\left[G_{\ell}\right]_{i}$
\ENDIF
\end{algorithmic}
\label{alg2}
\end{algorithm}

The outline of the HYPERLINK embedder above is its simplified description omitting a number of important details and presenting some of them slightly differently. The full detailed description of the algorithm exactly as used in this paper is included in its Bitbucket repository~\cite{codeHLembedder}.

To validate the hyperbolic geometry inference algorithm we compare inferred  coordinates in the RHG to its true coordinates.
Parameters of the RHG are taken to be $N=5,000$, $\langle k \rangle = 10$, $T=0.5$, and $\gamma = 2.5$. As seen from Figs.~\ref{fig:validation} {\bf a-c}, the accuracy of the  angular coordinate inference does not decline significantly for small degree nodes. This is the case, mainly, due to the nested inference with inference cores $cr_{i}$ covering all network nodes, in contrast to the original algorithm of Ref.~\cite{Boguna2010sustaining}, where cores only cover the most connected nodes.

As seen from Fig.~\ref{fig:validation} {\bf d}, Eq.~(\ref{eq:r_inference}) allows for accurate inference of small radial coordinates. At the same time, radial coordinates inference is less accurate for large radial coordinates. To explain this observation we recall that the key assumption in  Eq.~(\ref{eq:r_inference}) is that the node degree in the RHG is fully determined by its radial coordinate. In other words, we assume that possible node degree values are narrowly distributed around its expected value,  which is given by Eq.~(\ref{eq:r_inference}). This is indeed the case since node degrees are distributed according to the Poisson distribution,  Eq.~(\ref{eq:cond_pkr}).
The coefficient of variation of the Poisson distribution, however, is large for small mean values. This leads to significant variation in node degree values in the case of nodes with large radial coordinates, making Eq.~(\ref{eq:r_inference}) inaccurate.

\begin{figure}[H]
\includegraphics[width=7in]{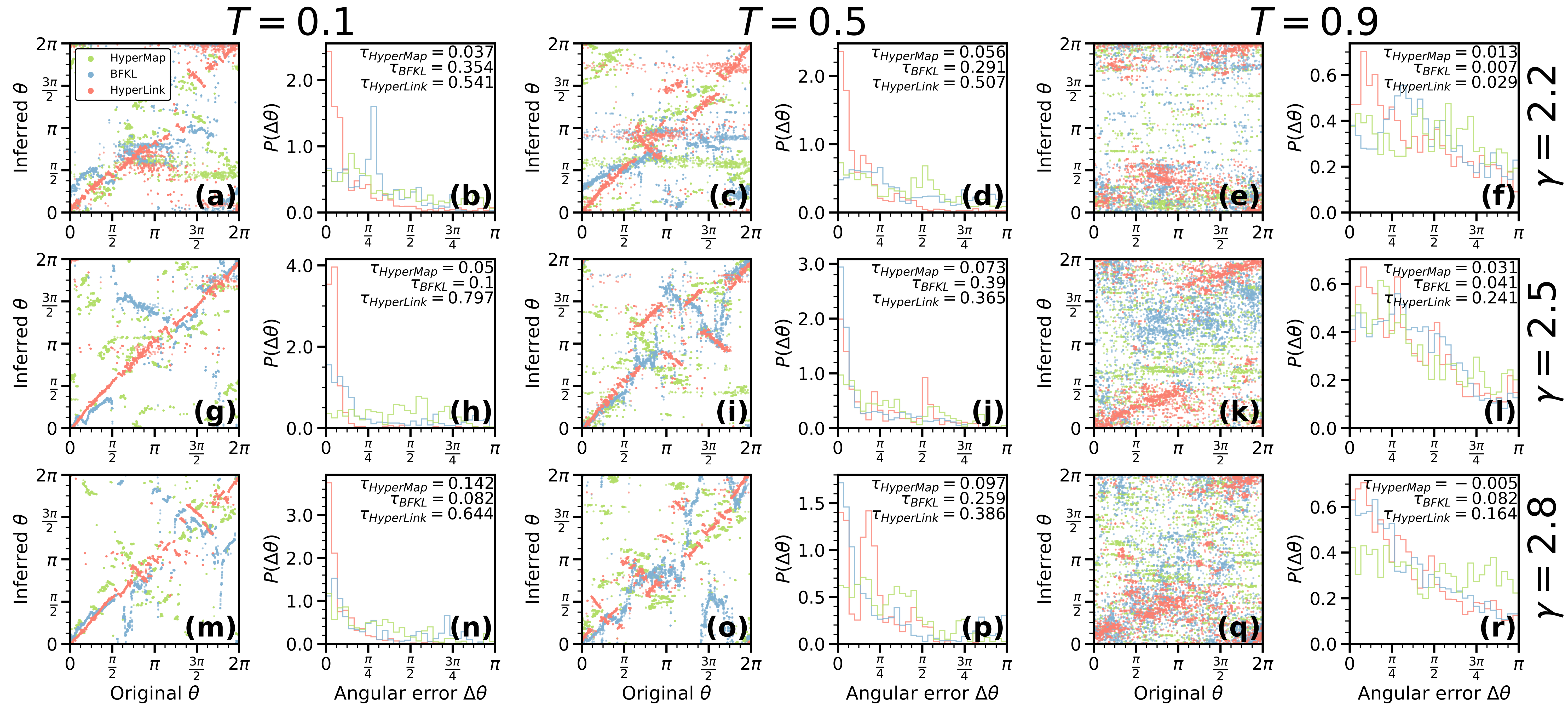}
\caption{\footnotesize {\bf HYPERLINK embedder accuracy compared to other embedding algorithms.}   RHGs are embedded to the hyperbolic disk by (red) \textsc{hyperlink} embedder, (blue) the algorithm by Bl\"{a}sius et al.~\cite{Blasius2016efficient} (BFKL), and (green) the \textsc{hypermap}~\cite{Papadopoulos2015network} algorithm. All comparisons correspond to RHGs consisting of $N=5,000$ nodes, $\overline{k}=10$, $1-q=0.5$ missing links, and various $T$ and $\gamma$ parameters. Panels are arranged according to $T$ and $\gamma$ parameters. Panels {\bf a, c, e, g, i, k, m, o, q} correspond to the scatter plots displaying inferred angular coordinates as a function of true angular coordinates. To quantify the embedding accuracy, we plot the distributions of embedding errors, $P(\Delta \theta)$, where $ \Delta \theta \equiv \pi - |\pi -  |\theta_{\rm inferred} - \theta_{original}||$ in panels {\bf b, d, f, h, j, l, n, p, r}, respectively. To quantify the association between the inferred and the original angular coordinates for each embedding we employ the U-statistic $\tau \in [-1,1]$, Ref~\cite{fisher1981nonparametric}. The U-statistic $\tau$ quantifies the correlation between the ordering of the inferred and original and angular coordinates and ranges from $\tau = 1$, in the case the two orderings are the same, to $\tau = -1$ in the case the two orderings are inverted with respect to one another.  The U-statistic $\tau$ is invariant under global shifts of the inferred coordinates. Our results indicate that the \textsc{hyperlink} accuracy is higher than that of the considered two algorithms in all cases, with the only exception of the $T=0.5$, $\gamma=2.5$ case, where BFKL is slightly better.}
\label{fig:HL_comparion}
\end{figure}

The \textsc{hyperlink} embedder allows for accurate node coordinate inference even in substantially incomplete networks in contrast to other mapping methods, e.g., \textsc{hypermap}~\cite{Papadopoulos2015network1} and the algorithm by Bl\"{a}sius et al.~\cite{Blasius2016efficient}, which become less accurate in the case of large $T$ values, Fig.~\ref{fig:HL_comparion}.

As evidenced by Fig.~\ref{fig:validation} and, indirectly, by our link prediction results in Secs.~\ref{sec:model_true_coords}~and~\ref{sec:inferred_coords}, our hyperbolic inference algorithm is sufficiently accurate for the prediction of missing links on both synthetic and real networks. At the same time, the algorithm does have limitations. First, it is designed to map networks with links removed uniformly at random. The link presence rate $q$ is the required parameter of the algorithm. In cases when the fraction of missing links is unknown, $q$ needs to be estimated and this may lead to less accurate mapping. The second limitation is the algorithm's running time complexity of $\mathcal{O}\left(N^{2}\right)$ restricting its utility to networks of smaller size. Finally, the third limitation is the analytic estimation of radial coordinates, which is not accurate for small degree nodes. Addressing these limitations is the subject of future work that is expected to further improve the accuracy and the utility of link prediction with hyperbolic geometry.

\twocolumngrid


\begin{thebibliography}{74}%
\makeatletter
\providecommand \@ifxundefined [1]{%
 \@ifx{#1\undefined}
}%
\providecommand \@ifnum [1]{%
 \ifnum #1\expandafter \@firstoftwo
 \else \expandafter \@secondoftwo
 \fi
}%
\providecommand \@ifx [1]{%
 \ifx #1\expandafter \@firstoftwo
 \else \expandafter \@secondoftwo
 \fi
}%
\providecommand \natexlab [1]{#1}%
\providecommand \enquote  [1]{``#1''}%
\providecommand \bibnamefont  [1]{#1}%
\providecommand \bibfnamefont [1]{#1}%
\providecommand \citenamefont [1]{#1}%
\providecommand \href@noop [0]{\@secondoftwo}%
\providecommand \href [0]{\begingroup \@sanitize@url \@href}%
\providecommand \@href[1]{\@@startlink{#1}\@@href}%
\providecommand \@@href[1]{\endgroup#1\@@endlink}%
\providecommand \@sanitize@url [0]{\catcode `\\12\catcode `\$12\catcode
  `\&12\catcode `\#12\catcode `\^12\catcode `\_12\catcode `\%12\relax}%
\providecommand \@@startlink[1]{}%
\providecommand \@@endlink[0]{}%
\providecommand \url  [0]{\begingroup\@sanitize@url \@url }%
\providecommand \@url [1]{\endgroup\@href {#1}{\urlprefix }}%
\providecommand \urlprefix  [0]{URL }%
\providecommand \Eprint [0]{\href }%
\providecommand \doibase [0]{http://dx.doi.org/}%
\providecommand \selectlanguage [0]{\@gobble}%
\providecommand \bibinfo  [0]{\@secondoftwo}%
\providecommand \bibfield  [0]{\@secondoftwo}%
\providecommand \translation [1]{[#1]}%
\providecommand \BibitemOpen [0]{}%
\providecommand \bibitemStop [0]{}%
\providecommand \bibitemNoStop [0]{.\EOS\space}%
\providecommand \EOS [0]{\spacefactor3000\relax}%
\providecommand \BibitemShut  [1]{\csname bibitem#1\endcsname}%
\let\auto@bib@innerbib\@empty
\bibitem [{\citenamefont {Peng}\ \emph {et~al.}(2015)\citenamefont {Peng},
  \citenamefont {Baowen}, \citenamefont {Yurong},\ and\ \citenamefont
  {Xiaoyu}}]{Peng2015}%
  \BibitemOpen
  \bibfield  {author} {\bibinfo {author} {\bibfnamefont {W.}~\bibnamefont
  {Peng}}, \bibinfo {author} {\bibfnamefont {X.}~\bibnamefont {Baowen}},
  \bibinfo {author} {\bibfnamefont {W.}~\bibnamefont {Yurong}}, \ and\ \bibinfo
  {author} {\bibfnamefont {Z.}~\bibnamefont {Xiaoyu}},\ }\bibfield  {title}
  {\emph {\bibinfo {title} {{Link prediction in social networks : the
  state-of-the-art}},\ }}\href {\doibase 10.1007/s11432-014-5237-y} {\bibfield
  {journal} {\bibinfo  {journal} {Sci. China Inf. Sci.}\ }\textbf {\bibinfo
  {volume} {58}},\ \bibinfo {pages} {011101} (\bibinfo {year}
  {2015})}\BibitemShut {NoStop}%
\bibitem [{\citenamefont {L{\"{u}}}\ and\ \citenamefont
  {Zhou}(2011)}]{lu2011link}%
  \BibitemOpen
  \bibfield  {author} {\bibinfo {author} {\bibfnamefont {L.}~\bibnamefont
  {L{\"{u}}}}\ and\ \bibinfo {author} {\bibfnamefont {T.}~\bibnamefont
  {Zhou}},\ }\bibfield  {title} {\emph {\bibinfo {title} {{Link prediction in
  complex networks: A survey}},\ }}\href {\doibase 10.1016/j.physa.2010.11.027}
  {\bibfield  {journal} {\bibinfo  {journal} {Phys. A Stat. Mech. its Appl.}\
  }\textbf {\bibinfo {volume} {390}},\ \bibinfo {pages} {1150} (\bibinfo {year}
  {2011})}\BibitemShut {NoStop}%
\bibitem [{\citenamefont {Menon}\ and\ \citenamefont
  {Elkan}(2011)}]{Menon2011}%
  \BibitemOpen
  \bibfield  {author} {\bibinfo {author} {\bibfnamefont {A.~K.}\ \bibnamefont
  {Menon}}\ and\ \bibinfo {author} {\bibfnamefont {C.}~\bibnamefont {Elkan}},\
  }in\ \href {\doibase 10.1007/978-3-642-23783-6_28} {\emph {\bibinfo
  {booktitle} {ECML PKDD 2011}}},\ Vol.\ \bibinfo {volume} {6912}\ (\bibinfo
  {year} {2011})\ pp.\ \bibinfo {pages} {437--452}\BibitemShut {NoStop}%
\bibitem [{\citenamefont {Peixoto}(2018)}]{peixoto2018reconstructing}%
  \BibitemOpen
  \bibfield  {author} {\bibinfo {author} {\bibfnamefont {T.~P.}\ \bibnamefont
  {Peixoto}},\ }\bibfield  {title} {\emph {\bibinfo {title} {{Reconstructing
  Networks with Unknown and Heterogeneous Errors}},\ }}\href {\doibase
  10.1103/PhysRevX.8.041011} {\bibfield  {journal} {\bibinfo  {journal} {Phys.
  Rev. X}\ }\textbf {\bibinfo {volume} {8}},\ \bibinfo {pages} {041011}
  (\bibinfo {year} {2018})}\BibitemShut {NoStop}%
\bibitem [{\citenamefont {Marchette}\ and\ \citenamefont
  {Priebe}(2008)}]{Marchette2008}%
  \BibitemOpen
  \bibfield  {author} {\bibinfo {author} {\bibfnamefont {D.~J.}\ \bibnamefont
  {Marchette}}\ and\ \bibinfo {author} {\bibfnamefont {C.~E.}\ \bibnamefont
  {Priebe}},\ }\bibfield  {title} {\emph {\bibinfo {title} {{Predicting
  unobserved links in incompletely observed networks}},\ }}\href {\doibase
  10.1016/j.csda.2007.03.016} {\bibfield  {journal} {\bibinfo  {journal}
  {Comput. Stat. Data Anal.}\ }\textbf {\bibinfo {volume} {52}},\ \bibinfo
  {pages} {1373} (\bibinfo {year} {2008})}\BibitemShut {NoStop}%
\bibitem [{\citenamefont {Guimer{\`{a}}}\ and\ \citenamefont
  {Sales-Pardo}(2009)}]{guimera2009missing}%
  \BibitemOpen
  \bibfield  {author} {\bibinfo {author} {\bibfnamefont {R.}~\bibnamefont
  {Guimer{\`{a}}}}\ and\ \bibinfo {author} {\bibfnamefont {M.}~\bibnamefont
  {Sales-Pardo}},\ }\bibfield  {title} {\emph {\bibinfo {title} {{Missing and
  spurious interactions and the reconstruction of complex networks}},\ }}\href
  {\doibase 10.1073/pnas.0908366106} {\bibfield  {journal} {\bibinfo  {journal}
  {Proc. Natl. Acad. Sci.}\ }\textbf {\bibinfo {volume} {106}},\ \bibinfo
  {pages} {22073} (\bibinfo {year} {2009})}\BibitemShut {NoStop}%
\bibitem [{\citenamefont {Kim}\ and\ \citenamefont {Leskovec}(2011)}]{Kim2011}%
  \BibitemOpen
  \bibfield  {author} {\bibinfo {author} {\bibfnamefont {M.}~\bibnamefont
  {Kim}}\ and\ \bibinfo {author} {\bibfnamefont {J.}~\bibnamefont {Leskovec}},\
  }\bibfield  {title} {\emph {\bibinfo {title} {{The Network completion
  problem: Inferring missing nodes and edges in networks}},\ }}\href {\doibase
  10.1137/1.9781611972818.5} {\bibfield  {journal} {\bibinfo  {journal} {SIAM
  Int. Conf. Data Min.}\ ,\ \bibinfo {pages} {47}} (\bibinfo {year}
  {2011})}\BibitemShut {NoStop}%
\bibitem [{\citenamefont {Adamic}\ and\ \citenamefont
  {Adar}(2003)}]{Adamic2003friends}%
  \BibitemOpen
  \bibfield  {author} {\bibinfo {author} {\bibfnamefont {L.~A.}\ \bibnamefont
  {Adamic}}\ and\ \bibinfo {author} {\bibfnamefont {E.}~\bibnamefont {Adar}},\
  }\bibfield  {title} {\emph {\bibinfo {title} {{Friends and neighbors on the
  Web}},\ }}\href {\doibase 10.1016/S0378-8733(03)00009-1} {\bibfield
  {journal} {\bibinfo  {journal} {Soc. Networks}\ }\textbf {\bibinfo {volume}
  {25}},\ \bibinfo {pages} {211} (\bibinfo {year} {2003})}\BibitemShut
  {NoStop}%
\bibitem [{\citenamefont {Newman}\ and\ \citenamefont
  {Clauset}(2016)}]{Newman2016}%
  \BibitemOpen
  \bibfield  {author} {\bibinfo {author} {\bibfnamefont {M.~E.~J.}\
  \bibnamefont {Newman}}\ and\ \bibinfo {author} {\bibfnamefont
  {A.}~\bibnamefont {Clauset}},\ }\bibfield  {title} {\emph {\bibinfo {title}
  {{Structure and inference in annotated networks}},\ }}\href {\doibase
  10.1038/ncomms11863} {\bibfield  {journal} {\bibinfo  {journal} {Nat.
  Commun.}\ }\textbf {\bibinfo {volume} {7}},\ \bibinfo {pages} {1} (\bibinfo
  {year} {2016})}\BibitemShut {NoStop}%
\bibitem [{\citenamefont {von Mering}\ \emph {et~al.}(2002)\citenamefont {von
  Mering}, \citenamefont {Krause}, \citenamefont {Snel}, \citenamefont
  {Cornell}, \citenamefont {Oliver}, \citenamefont {Fields},\ and\
  \citenamefont {Bork}}]{VonMering2002a}%
  \BibitemOpen
  \bibfield  {author} {\bibinfo {author} {\bibfnamefont {C.}~\bibnamefont {von
  Mering}}, \bibinfo {author} {\bibfnamefont {R.}~\bibnamefont {Krause}},
  \bibinfo {author} {\bibfnamefont {B.}~\bibnamefont {Snel}}, \bibinfo {author}
  {\bibfnamefont {M.}~\bibnamefont {Cornell}}, \bibinfo {author} {\bibfnamefont
  {S.~G.}\ \bibnamefont {Oliver}}, \bibinfo {author} {\bibfnamefont
  {S.}~\bibnamefont {Fields}}, \ and\ \bibinfo {author} {\bibfnamefont
  {P.}~\bibnamefont {Bork}},\ }\bibfield  {title} {\emph {\bibinfo {title}
  {{Comparative assessment of large-scale data sets of protein–protein
  interactions}},\ }}\href {\doibase 10.1038/nature750} {\bibfield  {journal}
  {\bibinfo  {journal} {Nature}\ }\textbf {\bibinfo {volume} {217}},\ \bibinfo
  {pages} {399} (\bibinfo {year} {2002})}\BibitemShut {NoStop}%
\bibitem [{\citenamefont {Yu}\ \emph {et~al.}(2008)\citenamefont {Yu},
  \citenamefont {Braun}, \citenamefont {Yildirim}, \citenamefont {Lemmens},
  \citenamefont {Venkatesan}, \citenamefont {Sahalie}, \citenamefont
  {Hirozane-Kishikawa}, \citenamefont {Gebreab}, \citenamefont {Li},
  \citenamefont {Simonis}, \citenamefont {Hao}, \citenamefont {Rual},
  \citenamefont {Dricot}, \citenamefont {Vazquez}, \citenamefont {Murray},
  \citenamefont {Simon}, \citenamefont {Tardivo}, \citenamefont {Tam},
  \citenamefont {Svrzikapa}, \citenamefont {Fan}, \citenamefont {de~Smet},
  \citenamefont {Motyl}, \citenamefont {Hudson}, \citenamefont {Park},
  \citenamefont {Xin}, \citenamefont {Cusick}, \citenamefont {Moore},
  \citenamefont {Boone}, \citenamefont {Snyder}, \citenamefont {Roth},
  \citenamefont {Barab{\'{a}}si}, \citenamefont {Tavernier}, \citenamefont
  {Hill}, \citenamefont {Vidal},\ and\ \citenamefont {Yıldırım}}]{Yu2008}%
  \BibitemOpen
  \bibfield  {author} {\bibinfo {author} {\bibfnamefont {H.}~\bibnamefont
  {Yu}}, \bibinfo {author} {\bibfnamefont {P.}~\bibnamefont {Braun}}, \bibinfo
  {author} {\bibfnamefont {M.~A.}\ \bibnamefont {Yildirim}}, \bibinfo {author}
  {\bibfnamefont {I.}~\bibnamefont {Lemmens}}, \bibinfo {author} {\bibfnamefont
  {K.}~\bibnamefont {Venkatesan}}, \bibinfo {author} {\bibfnamefont
  {J.}~\bibnamefont {Sahalie}}, \bibinfo {author} {\bibfnamefont
  {T.}~\bibnamefont {Hirozane-Kishikawa}}, \bibinfo {author} {\bibfnamefont
  {F.}~\bibnamefont {Gebreab}}, \bibinfo {author} {\bibfnamefont
  {N.}~\bibnamefont {Li}}, \bibinfo {author} {\bibfnamefont {N.}~\bibnamefont
  {Simonis}}, \bibinfo {author} {\bibfnamefont {T.}~\bibnamefont {Hao}},
  \bibinfo {author} {\bibfnamefont {J.-F.}\ \bibnamefont {Rual}}, \bibinfo
  {author} {\bibfnamefont {A.}~\bibnamefont {Dricot}}, \bibinfo {author}
  {\bibfnamefont {A.}~\bibnamefont {Vazquez}}, \bibinfo {author} {\bibfnamefont
  {R.~R.}\ \bibnamefont {Murray}}, \bibinfo {author} {\bibfnamefont
  {C.}~\bibnamefont {Simon}}, \bibinfo {author} {\bibfnamefont
  {L.}~\bibnamefont {Tardivo}}, \bibinfo {author} {\bibfnamefont
  {S.}~\bibnamefont {Tam}}, \bibinfo {author} {\bibfnamefont {N.}~\bibnamefont
  {Svrzikapa}}, \bibinfo {author} {\bibfnamefont {C.}~\bibnamefont {Fan}},
  \bibinfo {author} {\bibfnamefont {A.-S.}\ \bibnamefont {de~Smet}}, \bibinfo
  {author} {\bibfnamefont {A.}~\bibnamefont {Motyl}}, \bibinfo {author}
  {\bibfnamefont {M.~E.}\ \bibnamefont {Hudson}}, \bibinfo {author}
  {\bibfnamefont {J.}~\bibnamefont {Park}}, \bibinfo {author} {\bibfnamefont
  {X.}~\bibnamefont {Xin}}, \bibinfo {author} {\bibfnamefont {M.~E.}\
  \bibnamefont {Cusick}}, \bibinfo {author} {\bibfnamefont {T.}~\bibnamefont
  {Moore}}, \bibinfo {author} {\bibfnamefont {C.}~\bibnamefont {Boone}},
  \bibinfo {author} {\bibfnamefont {M.}~\bibnamefont {Snyder}}, \bibinfo
  {author} {\bibfnamefont {F.~P.}\ \bibnamefont {Roth}}, \bibinfo {author}
  {\bibfnamefont {A.-L.}\ \bibnamefont {Barab{\'{a}}si}}, \bibinfo {author}
  {\bibfnamefont {J.}~\bibnamefont {Tavernier}}, \bibinfo {author}
  {\bibfnamefont {D.~E.}\ \bibnamefont {Hill}}, \bibinfo {author}
  {\bibfnamefont {M.}~\bibnamefont {Vidal}}, \ and\ \bibinfo {author}
  {\bibfnamefont {M.}~\bibnamefont {Yıldırım}},\ }\bibfield  {title} {\emph
  {\bibinfo {title} {{High-quality binary protein interaction map of the yeast
  interactome network.}}\ }}\href {\doibase 10.1126/science.1158684} {\bibfield
   {journal} {\bibinfo  {journal} {Science}\ }\textbf {\bibinfo {volume}
  {322}},\ \bibinfo {pages} {104} (\bibinfo {year} {2008})}\BibitemShut
  {NoStop}%
\bibitem [{\citenamefont {Kov{\'{a}}cs}\ \emph {et~al.}(2019)\citenamefont
  {Kov{\'{a}}cs}, \citenamefont {Luck}, \citenamefont {Spirohn}, \citenamefont
  {Wang}, \citenamefont {Pollis}, \citenamefont {Schlabach}, \citenamefont
  {Bian}, \citenamefont {Kim}, \citenamefont {Kishore}, \citenamefont {Hao},
  \citenamefont {Calderwood}, \citenamefont {Vidal},\ and\ \citenamefont
  {Barab{\'{a}}si}}]{kovacs2019network}%
  \BibitemOpen
  \bibfield  {author} {\bibinfo {author} {\bibfnamefont {I.~A.}\ \bibnamefont
  {Kov{\'{a}}cs}}, \bibinfo {author} {\bibfnamefont {K.}~\bibnamefont {Luck}},
  \bibinfo {author} {\bibfnamefont {K.}~\bibnamefont {Spirohn}}, \bibinfo
  {author} {\bibfnamefont {Y.}~\bibnamefont {Wang}}, \bibinfo {author}
  {\bibfnamefont {C.}~\bibnamefont {Pollis}}, \bibinfo {author} {\bibfnamefont
  {S.}~\bibnamefont {Schlabach}}, \bibinfo {author} {\bibfnamefont
  {W.}~\bibnamefont {Bian}}, \bibinfo {author} {\bibfnamefont {D.-K.}\
  \bibnamefont {Kim}}, \bibinfo {author} {\bibfnamefont {N.}~\bibnamefont
  {Kishore}}, \bibinfo {author} {\bibfnamefont {T.}~\bibnamefont {Hao}},
  \bibinfo {author} {\bibfnamefont {M.~A.}\ \bibnamefont {Calderwood}},
  \bibinfo {author} {\bibfnamefont {M.}~\bibnamefont {Vidal}}, \ and\ \bibinfo
  {author} {\bibfnamefont {A.-L.}\ \bibnamefont {Barab{\'{a}}si}},\ }\bibfield
  {title} {\emph {\bibinfo {title} {{Network-based prediction of protein
  interactions}},\ }}\href {\doibase 10.1038/s41467-019-09177-y} {\bibfield
  {journal} {\bibinfo  {journal} {Nat. Commun.}\ }\textbf {\bibinfo {volume}
  {10}},\ \bibinfo {pages} {1240} (\bibinfo {year} {2019})}\BibitemShut
  {NoStop}%
\bibitem [{\citenamefont {Zhou}\ \emph {et~al.}(2007)\citenamefont {Zhou},
  \citenamefont {Ren}, \citenamefont {Medo},\ and\ \citenamefont
  {Zhang}}]{zhou2007bipartite}%
  \BibitemOpen
  \bibfield  {author} {\bibinfo {author} {\bibfnamefont {T.}~\bibnamefont
  {Zhou}}, \bibinfo {author} {\bibfnamefont {J.}~\bibnamefont {Ren}}, \bibinfo
  {author} {\bibfnamefont {M.}~\bibnamefont {Medo}}, \ and\ \bibinfo {author}
  {\bibfnamefont {Y.-C.}\ \bibnamefont {Zhang}},\ }\bibfield  {title} {\emph
  {\bibinfo {title} {{Bipartite network projection and personal
  recommendation}},\ }}\href {\doibase 10.1103/PhysRevE.76.046115} {\bibfield
  {journal} {\bibinfo  {journal} {Phys. Rev. E}\ }\textbf {\bibinfo {volume}
  {76}},\ \bibinfo {pages} {046115} (\bibinfo {year} {2007})}\BibitemShut
  {NoStop}%
\bibitem [{\citenamefont {L{\"{u}}}\ \emph {et~al.}(2012)\citenamefont
  {L{\"{u}}}, \citenamefont {Medo}, \citenamefont {Yeung}, \citenamefont
  {Zhang}, \citenamefont {Zhang},\ and\ \citenamefont
  {Zhou}}]{lu2012recommender}%
  \BibitemOpen
  \bibfield  {author} {\bibinfo {author} {\bibfnamefont {L.}~\bibnamefont
  {L{\"{u}}}}, \bibinfo {author} {\bibfnamefont {M.}~\bibnamefont {Medo}},
  \bibinfo {author} {\bibfnamefont {C.~H.}\ \bibnamefont {Yeung}}, \bibinfo
  {author} {\bibfnamefont {Y.-C.}\ \bibnamefont {Zhang}}, \bibinfo {author}
  {\bibfnamefont {Z.-K.}\ \bibnamefont {Zhang}}, \ and\ \bibinfo {author}
  {\bibfnamefont {T.}~\bibnamefont {Zhou}},\ }\bibfield  {title} {\emph
  {\bibinfo {title} {{Recommender systems}},\ }}\href {\doibase
  10.1016/j.physrep.2012.02.006} {\bibfield  {journal} {\bibinfo  {journal}
  {Phys. Rep.}\ }\textbf {\bibinfo {volume} {519}},\ \bibinfo {pages} {1}
  (\bibinfo {year} {2012})}\BibitemShut {NoStop}%
\bibitem [{\citenamefont {Bobadilla}\ \emph {et~al.}(2013)\citenamefont
  {Bobadilla}, \citenamefont {Ortega}, \citenamefont {Hernando},\ and\
  \citenamefont {Guti{\'{e}}rrez}}]{bobadilla2013recommender}%
  \BibitemOpen
  \bibfield  {author} {\bibinfo {author} {\bibfnamefont {J.}~\bibnamefont
  {Bobadilla}}, \bibinfo {author} {\bibfnamefont {F.}~\bibnamefont {Ortega}},
  \bibinfo {author} {\bibfnamefont {A.}~\bibnamefont {Hernando}}, \ and\
  \bibinfo {author} {\bibfnamefont {A.}~\bibnamefont {Guti{\'{e}}rrez}},\
  }\bibfield  {title} {\emph {\bibinfo {title} {{Recommender systems survey}},\
  }}\href {\doibase 10.1016/j.knosys.2013.03.012} {\bibfield  {journal}
  {\bibinfo  {journal} {Knowledge-Based Syst.}\ }\textbf {\bibinfo {volume}
  {46}},\ \bibinfo {pages} {109} (\bibinfo {year} {2013})}\BibitemShut
  {NoStop}%
\bibitem [{\citenamefont {Schafer}\ \emph {et~al.}(1999)\citenamefont
  {Schafer}, \citenamefont {Konstan},\ and\ \citenamefont
  {Riedl}}]{schafer1999recommender}%
  \BibitemOpen
  \bibfield  {author} {\bibinfo {author} {\bibfnamefont {J.~B.}\ \bibnamefont
  {Schafer}}, \bibinfo {author} {\bibfnamefont {J.}~\bibnamefont {Konstan}}, \
  and\ \bibinfo {author} {\bibfnamefont {J.}~\bibnamefont {Riedl}},\ }in\ \href
  {\doibase 10.1145/336992.337035} {\emph {\bibinfo {booktitle} {Proc. 1st ACM
  Conf. Electron. Commer.}}}\ (\bibinfo {organization} {ACM},\ \bibinfo {year}
  {1999})\ pp.\ \bibinfo {pages} {158--166}\BibitemShut {NoStop}%
\bibitem [{\citenamefont {Gilbert}(1961)}]{gilbert1961random}%
  \BibitemOpen
  \bibfield  {author} {\bibinfo {author} {\bibfnamefont {E.~N.}\ \bibnamefont
  {Gilbert}},\ }\bibfield  {title} {\emph {\bibinfo {title} {{Random plane
  networks}},\ }}\href@noop {} {\bibfield  {journal} {\bibinfo  {journal} {J.
  Soc. Ind. Appl. Math.}\ }\textbf {\bibinfo {volume} {9}},\ \bibinfo {pages}
  {533} (\bibinfo {year} {1961})}\BibitemShut {NoStop}%
\bibitem [{\citenamefont {McFarland}\ and\ \citenamefont
  {Brown}(1973)}]{mcfarland1973social}%
  \BibitemOpen
  \bibfield  {author} {\bibinfo {author} {\bibfnamefont {D.~D.}\ \bibnamefont
  {McFarland}}\ and\ \bibinfo {author} {\bibfnamefont {D.~J.}\ \bibnamefont
  {Brown}},\ }\bibfield  {title} {\emph {\bibinfo {title} {{Social distance as
  a metric: a systematic introduction to smallest space analysis}},\ }}\href
  {https://www.semanticscholar.org/paper/Social-distance-as-a-metric%3A-a-systematic-to-space-McFarland-Brown/d71c2bb6c72fac36a09e5a5acb9c306c560d6d2e}
  {\bibfield  {journal} {\bibinfo  {journal} {Bond. Plur. Form Subst. Urban
  Soc. Networks}\ }\textbf {\bibinfo {volume} {6}},\ \bibinfo {pages} {213}
  (\bibinfo {year} {1973})}\BibitemShut {NoStop}%
\bibitem [{\citenamefont {McPherson}\ \emph {et~al.}(2001)\citenamefont
  {McPherson}, \citenamefont {Smith-Lovin},\ and\ \citenamefont
  {Cook}}]{McPherson2001}%
  \BibitemOpen
  \bibfield  {author} {\bibinfo {author} {\bibfnamefont {M.}~\bibnamefont
  {McPherson}}, \bibinfo {author} {\bibfnamefont {L.}~\bibnamefont
  {Smith-Lovin}}, \ and\ \bibinfo {author} {\bibfnamefont {J.~M.}\ \bibnamefont
  {Cook}},\ }\bibfield  {title} {\emph {\bibinfo {title} {{Birds of a feather:
  Homophily in social networks}},\ }}\href {\doibase
  10.1146/annurev.soc.27.1.415} {\bibfield  {journal} {\bibinfo  {journal}
  {Annu. Rev. Sociol.}\ }\textbf {\bibinfo {volume} {27}},\ \bibinfo {pages}
  {415} (\bibinfo {year} {2001})}\BibitemShut {NoStop}%
\bibitem [{\citenamefont {Krioukov}\ \emph {et~al.}(2010)\citenamefont
  {Krioukov}, \citenamefont {Papadopoulos}, \citenamefont {Kitsak},
  \citenamefont {Vahdat},\ and\ \citenamefont
  {Bogu{\~{n}}{\'{a}}}}]{Krioukov2010hyperbolic}%
  \BibitemOpen
  \bibfield  {author} {\bibinfo {author} {\bibfnamefont {D.}~\bibnamefont
  {Krioukov}}, \bibinfo {author} {\bibfnamefont {F.}~\bibnamefont
  {Papadopoulos}}, \bibinfo {author} {\bibfnamefont {M.}~\bibnamefont
  {Kitsak}}, \bibinfo {author} {\bibfnamefont {A.}~\bibnamefont {Vahdat}}, \
  and\ \bibinfo {author} {\bibfnamefont {M.}~\bibnamefont
  {Bogu{\~{n}}{\'{a}}}},\ }\bibfield  {title} {\emph {\bibinfo {title}
  {{Hyperbolic geometry of complex networks}},\ }}\href {\doibase
  10.1103/PhysRevE.82.036106} {\bibfield  {journal} {\bibinfo  {journal} {Phys.
  Rev. E}\ }\textbf {\bibinfo {volume} {82}},\ \bibinfo {pages} {036106}
  (\bibinfo {year} {2010})}\BibitemShut {NoStop}%
\bibitem [{\citenamefont {Newman}\ and\ \citenamefont
  {Peixoto}(2015)}]{newman2015generalized}%
  \BibitemOpen
  \bibfield  {author} {\bibinfo {author} {\bibfnamefont {M.~E.~J.}\
  \bibnamefont {Newman}}\ and\ \bibinfo {author} {\bibfnamefont {T.~P.}\
  \bibnamefont {Peixoto}},\ }\bibfield  {title} {\emph {\bibinfo {title}
  {{Generalized communities in networks}},\ }}\href {\doibase
  10.1103/PhysRevLett.115.088701} {\bibfield  {journal} {\bibinfo  {journal}
  {Phys. Rev. Lett.}\ }\textbf {\bibinfo {volume} {115}},\ \bibinfo {pages}
  {088701} (\bibinfo {year} {2015})}\BibitemShut {NoStop}%
\bibitem [{\citenamefont {Brew}\ and\ \citenamefont
  {Salter-Townshend}(2010)}]{brew2010latent}%
  \BibitemOpen
  \bibfield  {author} {\bibinfo {author} {\bibfnamefont {A.}~\bibnamefont
  {Brew}}\ and\ \bibinfo {author} {\bibfnamefont {M.}~\bibnamefont
  {Salter-Townshend}},\ }in\ \href@noop {} {\emph {\bibinfo {booktitle}
  {{Workshop on Networks Across Disciplines: Theory and Applications}}}}\
  (\bibinfo {year} {2010})\BibitemShut {NoStop}%
\bibitem [{\citenamefont {Zhu}\ \emph {et~al.}(2016)\citenamefont {Zhu},
  \citenamefont {Guo}, \citenamefont {Yin}, \citenamefont {Steeg},\ and\
  \citenamefont {Galstyan}}]{zhu2016scalable}%
  \BibitemOpen
  \bibfield  {author} {\bibinfo {author} {\bibfnamefont {L.}~\bibnamefont
  {Zhu}}, \bibinfo {author} {\bibfnamefont {D.}~\bibnamefont {Guo}}, \bibinfo
  {author} {\bibfnamefont {J.}~\bibnamefont {Yin}}, \bibinfo {author}
  {\bibfnamefont {G.~V.}\ \bibnamefont {Steeg}}, \ and\ \bibinfo {author}
  {\bibfnamefont {A.}~\bibnamefont {Galstyan}},\ }\bibfield  {title} {\emph
  {\bibinfo {title} {{Scalable temporal latent space inference for link
  prediction in dynamic social networks}},\ }}\href {\doibase
  10.1109/TKDE.2016.2591009} {\bibfield  {journal} {\bibinfo  {journal} {IEEE
  Trans. Knowl. Data Eng.}\ }\textbf {\bibinfo {volume} {28}},\ \bibinfo
  {pages} {2765 } (\bibinfo {year} {2016})}\BibitemShut {NoStop}%
\bibitem [{\citenamefont {Garc{\'{i}}a-P{\'{e}}rez}\ \emph
  {et~al.}(2020)\citenamefont {Garc{\'{i}}a-P{\'{e}}rez}, \citenamefont
  {Aliakbarisani}, \citenamefont {Ghasemi},\ and\ \citenamefont
  {Serrano}}]{garcia-perez2020precision}%
  \BibitemOpen
  \bibfield  {author} {\bibinfo {author} {\bibfnamefont {G.}~\bibnamefont
  {Garc{\'{i}}a-P{\'{e}}rez}}, \bibinfo {author} {\bibfnamefont
  {R.}~\bibnamefont {Aliakbarisani}}, \bibinfo {author} {\bibfnamefont
  {A.}~\bibnamefont {Ghasemi}}, \ and\ \bibinfo {author} {\bibfnamefont
  {M.~{\'{A}}.}\ \bibnamefont {Serrano}},\ }\bibfield  {title} {\emph {\bibinfo
  {title} {{Precision as a measure of predictability of missing links in real
  networks}},\ }}\href {\doibase 10.1103/PhysRevE.101.052318} {\bibfield
  {journal} {\bibinfo  {journal} {Phys. Rev. E}\ }\textbf {\bibinfo {volume}
  {101}},\ \bibinfo {pages} {052318} (\bibinfo {year} {2020})}\BibitemShut
  {NoStop}%
\bibitem [{\citenamefont {Serrano}\ \emph {et~al.}(2008)\citenamefont
  {Serrano}, \citenamefont {Krioukov},\ and\ \citenamefont
  {Bogu{\~{n}}{\'{a}}}}]{Serrano2008}%
  \BibitemOpen
  \bibfield  {author} {\bibinfo {author} {\bibfnamefont {M.}~\bibnamefont
  {Serrano}}, \bibinfo {author} {\bibfnamefont {D.}~\bibnamefont {Krioukov}}, \
  and\ \bibinfo {author} {\bibfnamefont {M.}~\bibnamefont
  {Bogu{\~{n}}{\'{a}}}},\ }\bibfield  {title} {\emph {\bibinfo {title}
  {{Self-similarity of complex networks and hidden metric spaces}},\ }}\href
  {\doibase 10.1103/PhysRevLett.100.078701} {\bibfield  {journal} {\bibinfo
  {journal} {Phys. Rev. Lett.}\ }\textbf {\bibinfo {volume} {100}},\ \bibinfo
  {pages} {078701} (\bibinfo {year} {2008})}\BibitemShut {NoStop}%
\bibitem [{\citenamefont {Papadopoulos}\ \emph {et~al.}(2012)\citenamefont
  {Papadopoulos}, \citenamefont {Kitsak}, \citenamefont {Serrano},
  \citenamefont {Bogu{\~{n}}{\'{a}}},\ and\ \citenamefont
  {Krioukov}}]{Papadopoulos2012popularity}%
  \BibitemOpen
  \bibfield  {author} {\bibinfo {author} {\bibfnamefont {F.}~\bibnamefont
  {Papadopoulos}}, \bibinfo {author} {\bibfnamefont {M.}~\bibnamefont
  {Kitsak}}, \bibinfo {author} {\bibfnamefont {M.~{\'{A}}.}\ \bibnamefont
  {Serrano}}, \bibinfo {author} {\bibfnamefont {M.}~\bibnamefont
  {Bogu{\~{n}}{\'{a}}}}, \ and\ \bibinfo {author} {\bibfnamefont
  {D.}~\bibnamefont {Krioukov}},\ }\bibfield  {title} {\emph {\bibinfo {title}
  {{Popularity versus similarity in growing networks}},\ }}\href {\doibase
  10.1038/nature11459} {\bibfield  {journal} {\bibinfo  {journal} {Nature}\
  }\textbf {\bibinfo {volume} {489}},\ \bibinfo {pages} {537} (\bibinfo {year}
  {2012})}\BibitemShut {NoStop}%
\bibitem [{\citenamefont {Zuev}\ \emph {et~al.}(2015)\citenamefont {Zuev},
  \citenamefont {Bogu{\~{n}}{\'{a}}}, \citenamefont {Bianconi},\ and\
  \citenamefont {Krioukov}}]{zuev2015emergence}%
  \BibitemOpen
  \bibfield  {author} {\bibinfo {author} {\bibfnamefont {K.}~\bibnamefont
  {Zuev}}, \bibinfo {author} {\bibfnamefont {M.}~\bibnamefont
  {Bogu{\~{n}}{\'{a}}}}, \bibinfo {author} {\bibfnamefont {G.}~\bibnamefont
  {Bianconi}}, \ and\ \bibinfo {author} {\bibfnamefont {D.}~\bibnamefont
  {Krioukov}},\ }\bibfield  {title} {\emph {\bibinfo {title} {{Emergence of
  soft communities from geometric preferential attachment}},\ }}\href {\doibase
  10.1038/srep09421} {\bibfield  {journal} {\bibinfo  {journal} {Sci. Rep.}\
  }\textbf {\bibinfo {volume} {5}},\ \bibinfo {pages} {9421} (\bibinfo {year}
  {2015})}\BibitemShut {NoStop}%
\bibitem [{\citenamefont {Lazega}\ \emph {et~al.}(2006)\citenamefont {Lazega},
  \citenamefont {Wasserman},\ and\ \citenamefont {Faust}}]{Lazega2006social}%
  \BibitemOpen
  \bibfield  {author} {\bibinfo {author} {\bibfnamefont {E.}~\bibnamefont
  {Lazega}}, \bibinfo {author} {\bibfnamefont {S.}~\bibnamefont {Wasserman}}, \
  and\ \bibinfo {author} {\bibfnamefont {K.}~\bibnamefont {Faust}},\ }\bibfield
   {title} {\emph {\bibinfo {title} {{Social Network Analysis: Methods and
  Applications}},\ }}\href {\doibase 10.2307/3322457} {\bibfield  {journal}
  {\bibinfo  {journal} {Rev. Fran{\c{c}}aise Sociol.}\ }\textbf {\bibinfo
  {volume} {36}},\ \bibinfo {pages} {781} (\bibinfo {year} {2006})}\BibitemShut
  {NoStop}%
\bibitem [{\citenamefont {Newman}(2010)}]{NEWMAN2010}%
  \BibitemOpen
  \bibfield  {author} {\bibinfo {author} {\bibfnamefont {M.}~\bibnamefont
  {Newman}},\ }\href {\doibase 10.1093/acprof:oso/9780199206650.001.0001}
  {\emph {\bibinfo {title} {{Networks: An Introduction}}}}\ (\bibinfo
  {publisher} {Oxford University Press},\ \bibinfo {year} {2010})\BibitemShut
  {NoStop}%
\bibitem [{\citenamefont {Barab{\'{a}}si}\ and\ \citenamefont
  {P{\'{o}}sfai}(2016)}]{barabasi2016network}%
  \BibitemOpen
  \bibfield  {author} {\bibinfo {author} {\bibfnamefont {A.-L.}\ \bibnamefont
  {Barab{\'{a}}si}}\ and\ \bibinfo {author} {\bibfnamefont {M.}~\bibnamefont
  {P{\'{o}}sfai}},\ }\href@noop {} {\emph {\bibinfo {title} {{Network
  science}}}}\ (\bibinfo  {publisher} {Cambridge University Press},\ \bibinfo
  {year} {2016})\ p.\ \bibinfo {pages} {456}\BibitemShut {NoStop}%
\bibitem [{\citenamefont {van~der Hoorn}\ \emph {et~al.}(2018)\citenamefont
  {van~der Hoorn}, \citenamefont {Lippner},\ and\ \citenamefont
  {Krioukov}}]{VanderHoorn2018sparse}%
  \BibitemOpen
  \bibfield  {author} {\bibinfo {author} {\bibfnamefont {P.}~\bibnamefont
  {van~der Hoorn}}, \bibinfo {author} {\bibfnamefont {G.}~\bibnamefont
  {Lippner}}, \ and\ \bibinfo {author} {\bibfnamefont {D.}~\bibnamefont
  {Krioukov}},\ }\bibfield  {title} {\emph {\bibinfo {title} {{Sparse
  Maximum-Entropy Random Graphs with a Given Power-Law Degree Distribution}},\
  }}\href {\doibase 10.1007/s10955-017-1887-7} {\bibfield  {journal} {\bibinfo
  {journal} {J. Stat. Phys.}\ }\textbf {\bibinfo {volume} {173}},\ \bibinfo
  {pages} {806} (\bibinfo {year} {2018})}\BibitemShut {NoStop}%
\bibitem [{\citenamefont {Krioukov}(2016)}]{Krioukov2016Clustering}%
  \BibitemOpen
  \bibfield  {author} {\bibinfo {author} {\bibfnamefont {D.}~\bibnamefont
  {Krioukov}},\ }\bibfield  {title} {\emph {\bibinfo {title} {{Clustering
  Implies Geometry in Networks}},\ }}\href {\doibase
  10.1103/PhysRevLett.116.208302} {\bibfield  {journal} {\bibinfo  {journal}
  {Phys. Rev. Lett.}\ }\textbf {\bibinfo {volume} {116}},\ \bibinfo {pages}
  {208302} (\bibinfo {year} {2016})}\BibitemShut {NoStop}%
\bibitem [{\citenamefont {Serrano}\ \emph {et~al.}(2012)\citenamefont
  {Serrano}, \citenamefont {Bogu{\~{n}}{\'{a}}},\ and\ \citenamefont
  {Sagu{\'{e}}s}}]{Serrano2012uncovering}%
  \BibitemOpen
  \bibfield  {author} {\bibinfo {author} {\bibfnamefont {M.~{\'{A}}.}\
  \bibnamefont {Serrano}}, \bibinfo {author} {\bibfnamefont {M.}~\bibnamefont
  {Bogu{\~{n}}{\'{a}}}}, \ and\ \bibinfo {author} {\bibfnamefont
  {F.}~\bibnamefont {Sagu{\'{e}}s}},\ }\bibfield  {title} {\emph {\bibinfo
  {title} {{Uncovering the hidden geometry behind metabolic networks}},\
  }}\href {\doibase 10.1039/c2mb05306c} {\bibfield  {journal} {\bibinfo
  {journal} {Mol. Biosyst.}\ }\textbf {\bibinfo {volume} {8}},\ \bibinfo
  {pages} {843} (\bibinfo {year} {2012})}\BibitemShut {NoStop}%
\bibitem [{\citenamefont {Papadopoulos}\ \emph
  {et~al.}(2015{\natexlab{a}})\citenamefont {Papadopoulos}, \citenamefont
  {Psomas},\ and\ \citenamefont {Krioukov}}]{Papadopoulos2015network1}%
  \BibitemOpen
  \bibfield  {author} {\bibinfo {author} {\bibfnamefont {F.}~\bibnamefont
  {Papadopoulos}}, \bibinfo {author} {\bibfnamefont {C.}~\bibnamefont
  {Psomas}}, \ and\ \bibinfo {author} {\bibfnamefont {D.}~\bibnamefont
  {Krioukov}},\ }\bibfield  {title} {\emph {\bibinfo {title} {{Network mapping
  by replaying hyperbolic growth}},\ }}\href {\doibase
  10.1109/TNET.2013.2294052} {\bibfield  {journal} {\bibinfo  {journal}
  {IEEE/ACM Trans. Netw.}\ }\textbf {\bibinfo {volume} {23}},\ \bibinfo {pages}
  {198} (\bibinfo {year} {2015}{\natexlab{a}})}\BibitemShut {NoStop}%
\bibitem [{\citenamefont {Papadopoulos}\ \emph
  {et~al.}(2015{\natexlab{b}})\citenamefont {Papadopoulos}, \citenamefont
  {Aldecoa},\ and\ \citenamefont {Krioukov}}]{Papadopoulos2015network}%
  \BibitemOpen
  \bibfield  {author} {\bibinfo {author} {\bibfnamefont {F.}~\bibnamefont
  {Papadopoulos}}, \bibinfo {author} {\bibfnamefont {R.}~\bibnamefont
  {Aldecoa}}, \ and\ \bibinfo {author} {\bibfnamefont {D.}~\bibnamefont
  {Krioukov}},\ }\bibfield  {title} {\emph {\bibinfo {title} {{Network geometry
  inference using common neighbors}},\ }}\href {\doibase
  10.1103/PhysRevE.92.022807} {\bibfield  {journal} {\bibinfo  {journal} {Phys.
  Rev. E}\ }\textbf {\bibinfo {volume} {92}},\ \bibinfo {pages} {022807}
  (\bibinfo {year} {2015}{\natexlab{b}})}\BibitemShut {NoStop}%
\bibitem [{\citenamefont {Muscoloni}\ and\ \citenamefont
  {Cannistraci}(2018{\natexlab{a}})}]{Alessandro2018leveraging}%
  \BibitemOpen
  \bibfield  {author} {\bibinfo {author} {\bibfnamefont {A.}~\bibnamefont
  {Muscoloni}}\ and\ \bibinfo {author} {\bibfnamefont {C.~V.}\ \bibnamefont
  {Cannistraci}},\ }\bibfield  {title} {\emph {\bibinfo {title} {{Leveraging
  the nonuniform PSO network model as a benchmark for performance evaluation in
  community detection and link prediction}},\ }}\href {\doibase
  10.1088/1367-2630/aac6f9} {\bibfield  {journal} {\bibinfo  {journal} {New J.
  Phys.}\ }\textbf {\bibinfo {volume} {20}} (\bibinfo {year}
  {2018}{\natexlab{a}})}\BibitemShut {NoStop}%
\bibitem [{\citenamefont {Muscoloni}\ and\ \citenamefont
  {Cannistraci}(2018{\natexlab{b}})}]{Muscoloni2018minimum}%
  \BibitemOpen
  \bibfield  {author} {\bibinfo {author} {\bibfnamefont {A.}~\bibnamefont
  {Muscoloni}}\ and\ \bibinfo {author} {\bibfnamefont {C.~V.}\ \bibnamefont
  {Cannistraci}},\ }\bibfield  {title} {\emph {\bibinfo {title} {{Minimum
  curvilinear automata with similarity attachment for network embedding and
  link prediction in the hyperbolic space}},\ }}\href@noop {} {\  (\bibinfo
  {year} {2018}{\natexlab{b}})},\ \Eprint {http://arxiv.org/abs/1802.01183}
  {arXiv:1802.01183} \BibitemShut {NoStop}%
\bibitem [{cod()}]{codeHLembedder}%
  \BibitemOpen
  \href@noop {} {\bibfield  {title} {\emph {\bibinfo {title} {{HyperLink
  embedder }},\ }}}\bibinfo {note}
  {\url{https://bitbucket.org/dk-lab/2020_code_hyperlink}}\BibitemShut
  {NoStop}%
\bibitem [{\citenamefont {Vall{\`{e}}s-Catal{\`{a}}}\ \emph
  {et~al.}(2018)\citenamefont {Vall{\`{e}}s-Catal{\`{a}}}, \citenamefont
  {Peixoto}, \citenamefont {Sales-Pardo},\ and\ \citenamefont
  {Guimer{\`{a}}}}]{valles2018consistencies}%
  \BibitemOpen
  \bibfield  {author} {\bibinfo {author} {\bibfnamefont {T.}~\bibnamefont
  {Vall{\`{e}}s-Catal{\`{a}}}}, \bibinfo {author} {\bibfnamefont {T.~P.}\
  \bibnamefont {Peixoto}}, \bibinfo {author} {\bibfnamefont {M.}~\bibnamefont
  {Sales-Pardo}}, \ and\ \bibinfo {author} {\bibfnamefont {R.}~\bibnamefont
  {Guimer{\`{a}}}},\ }\bibfield  {title} {\emph {\bibinfo {title}
  {{Consistencies and inconsistencies between model selection and link
  prediction in networks}},\ }}\href {\doibase 10.1103/PhysRevE.97.062316}
  {\bibfield  {journal} {\bibinfo  {journal} {Phys. Rev. E}\ }\textbf {\bibinfo
  {volume} {97}},\ \bibinfo {pages} {062316} (\bibinfo {year}
  {2018})}\BibitemShut {NoStop}%
\bibitem [{\citenamefont {Ghasemian}\ \emph {et~al.}(2019)\citenamefont
  {Ghasemian}, \citenamefont {Hosseinmardi},\ and\ \citenamefont
  {Clauset}}]{Ghasemian2019evaluating}%
  \BibitemOpen
  \bibfield  {author} {\bibinfo {author} {\bibfnamefont {A.}~\bibnamefont
  {Ghasemian}}, \bibinfo {author} {\bibfnamefont {H.}~\bibnamefont
  {Hosseinmardi}}, \ and\ \bibinfo {author} {\bibfnamefont {A.}~\bibnamefont
  {Clauset}},\ }\bibfield  {title} {\emph {\bibinfo {title} {{Evaluating
  Overfit and Underfit in Models of Network Community Structure}},\ }}\href
  {\doibase 10.1109/TKDE.2019.2911585} {\bibfield  {journal} {\bibinfo
  {journal} {IEEE Trans. Knowl. Data Eng.}\ }\textbf {\bibinfo {volume} {32}},\
  \bibinfo {pages} {1} (\bibinfo {year} {2019})}\BibitemShut {NoStop}%
\bibitem [{\citenamefont {Ghasemian}\ \emph {et~al.}(2020)\citenamefont
  {Ghasemian}, \citenamefont {Hosseinmardi}, \citenamefont {Galstyan},
  \citenamefont {Airoldi},\ and\ \citenamefont
  {Clauset}}]{ghasemian2019stacking}%
  \BibitemOpen
  \bibfield  {author} {\bibinfo {author} {\bibfnamefont {A.}~\bibnamefont
  {Ghasemian}}, \bibinfo {author} {\bibfnamefont {H.}~\bibnamefont
  {Hosseinmardi}}, \bibinfo {author} {\bibfnamefont {A.}~\bibnamefont
  {Galstyan}}, \bibinfo {author} {\bibfnamefont {E.~M.}\ \bibnamefont
  {Airoldi}}, \ and\ \bibinfo {author} {\bibfnamefont {A.}~\bibnamefont
  {Clauset}},\ }\bibfield  {title} {\emph {\bibinfo {title} {Stacking models
  for nearly optimal link prediction in complex networks},\ }}\href {\doibase
  10.1073/pnas.1914950117} {\bibfield  {journal} {\bibinfo  {journal} {Proc.
  Natl. Acad. Sci.}\ }\textbf {\bibinfo {volume} {117}},\ \bibinfo {pages}
  {23393} (\bibinfo {year} {2020})}\BibitemShut {NoStop}%
\bibitem [{\citenamefont {Krioukov}\ \emph {et~al.}(2009)\citenamefont
  {Krioukov}, \citenamefont {Papadopoulos}, \citenamefont {Vahdat},\ and\
  \citenamefont {Bogu{\~{n}}{\'{a}}}}]{Krioukov2009}%
  \BibitemOpen
  \bibfield  {author} {\bibinfo {author} {\bibfnamefont {D.}~\bibnamefont
  {Krioukov}}, \bibinfo {author} {\bibfnamefont {F.}~\bibnamefont
  {Papadopoulos}}, \bibinfo {author} {\bibfnamefont {A.}~\bibnamefont
  {Vahdat}}, \ and\ \bibinfo {author} {\bibfnamefont {M.}~\bibnamefont
  {Bogu{\~{n}}{\'{a}}}},\ }\bibfield  {title} {\emph {\bibinfo {title}
  {{Curvature and temperature of complex networks}},\ }}\href {\doibase
  10.1103/PhysRevE.80.035101} {\bibfield  {journal} {\bibinfo  {journal} {Phys.
  Rev. E}\ }\textbf {\bibinfo {volume} {80}},\ \bibinfo {pages} {35101}
  (\bibinfo {year} {2009})}\BibitemShut {NoStop}%
\bibitem [{\citenamefont {Bogu{\~{n}}{\'{a}}}\ \emph
  {et~al.}(2010)\citenamefont {Bogu{\~{n}}{\'{a}}}, \citenamefont
  {Papadopoulos},\ and\ \citenamefont {Krioukov}}]{Boguna2010sustaining}%
  \BibitemOpen
  \bibfield  {author} {\bibinfo {author} {\bibfnamefont {M.}~\bibnamefont
  {Bogu{\~{n}}{\'{a}}}}, \bibinfo {author} {\bibfnamefont {F.}~\bibnamefont
  {Papadopoulos}}, \ and\ \bibinfo {author} {\bibfnamefont {D.}~\bibnamefont
  {Krioukov}},\ }\bibfield  {title} {\emph {\bibinfo {title} {{Sustaining the
  Internet with hyperbolic mapping.}}\ }}\href {\doibase 10.1038/ncomms1063}
  {\bibfield  {journal} {\bibinfo  {journal} {Nat. Commun.}\ }\textbf {\bibinfo
  {volume} {1}},\ \bibinfo {pages} {62} (\bibinfo {year} {2010})}\BibitemShut
  {NoStop}%
\bibitem [{\citenamefont {Kitsak}\ \emph {et~al.}(2017)\citenamefont {Kitsak},
  \citenamefont {Papadopoulos},\ and\ \citenamefont
  {Krioukov}}]{Kitsak2017latent}%
  \BibitemOpen
  \bibfield  {author} {\bibinfo {author} {\bibfnamefont {M.}~\bibnamefont
  {Kitsak}}, \bibinfo {author} {\bibfnamefont {F.}~\bibnamefont
  {Papadopoulos}}, \ and\ \bibinfo {author} {\bibfnamefont {D.}~\bibnamefont
  {Krioukov}},\ }\bibfield  {title} {\emph {\bibinfo {title} {{Latent geometry
  of bipartite networks}},\ }}\href {\doibase 10.1103/PhysRevE.95.032309}
  {\bibfield  {journal} {\bibinfo  {journal} {Phys. Rev. E}\ }\textbf {\bibinfo
  {volume} {95}},\ \bibinfo {pages} {032309} (\bibinfo {year}
  {2017})}\BibitemShut {NoStop}%
\bibitem [{\citenamefont {Aldecoa}\ \emph {et~al.}(2015)\citenamefont
  {Aldecoa}, \citenamefont {Orsini},\ and\ \citenamefont
  {Krioukov}}]{Aldecoa2015}%
  \BibitemOpen
  \bibfield  {author} {\bibinfo {author} {\bibfnamefont {R.}~\bibnamefont
  {Aldecoa}}, \bibinfo {author} {\bibfnamefont {C.}~\bibnamefont {Orsini}}, \
  and\ \bibinfo {author} {\bibfnamefont {D.}~\bibnamefont {Krioukov}},\
  }\bibfield  {title} {\emph {\bibinfo {title} {{Hyperbolic graph generator}},\
  }}\href {\doibase 10.1016/j.cpc.2015.05.028} {\bibfield  {journal} {\bibinfo
  {journal} {Comput. Phys. Commun.}\ }\textbf {\bibinfo {volume} {196}},\
  \bibinfo {pages} {492} (\bibinfo {year} {2015})}\BibitemShut {NoStop}%
\bibitem [{\citenamefont {Garc{\'{i}}a-P{\'{e}}rez}\ \emph
  {et~al.}(2018)\citenamefont {Garc{\'{i}}a-P{\'{e}}rez}, \citenamefont
  {Bogu{\~{n}}{\'{a}}},\ and\ \citenamefont {Serrano}}]{Garcia2018multiscale}%
  \BibitemOpen
  \bibfield  {author} {\bibinfo {author} {\bibfnamefont {G.}~\bibnamefont
  {Garc{\'{i}}a-P{\'{e}}rez}}, \bibinfo {author} {\bibfnamefont
  {M.}~\bibnamefont {Bogu{\~{n}}{\'{a}}}}, \ and\ \bibinfo {author}
  {\bibfnamefont {M.~{\'{A}}.}\ \bibnamefont {Serrano}},\ }\bibfield  {title}
  {\emph {\bibinfo {title} {{Multiscale unfolding of real networks by geometric
  renormalization}},\ }}\href {\doibase 10.1038/s41567-018-0072-5} {\bibfield
  {journal} {\bibinfo  {journal} {Nat. Phys.}\ }\textbf {\bibinfo {volume}
  {14}},\ \bibinfo {pages} {1} (\bibinfo {year} {2018})}\BibitemShut {NoStop}%
\bibitem [{\citenamefont {Muscoloni}\ and\ \citenamefont
  {Cannistraci}(2018{\natexlab{c}})}]{muscoloni2018nonuniform}%
  \BibitemOpen
  \bibfield  {author} {\bibinfo {author} {\bibfnamefont {A.}~\bibnamefont
  {Muscoloni}}\ and\ \bibinfo {author} {\bibfnamefont {C.~V.}\ \bibnamefont
  {Cannistraci}},\ }\bibfield  {title} {\emph {\bibinfo {title} {{A nonuniform
  popularity-similarity optimization (nPSO) model to efficiently generate
  realistic complex networks with communities}},\ }}\href {\doibase
  10.1088/1367-2630/aac06f} {\bibfield  {journal} {\bibinfo  {journal} {New J.
  Phys.}\ }\textbf {\bibinfo {volume} {20}},\ \bibinfo {pages} {052002}
  (\bibinfo {year} {2018}{\natexlab{c}})}\BibitemShut {NoStop}%
\bibitem [{\citenamefont {Garc{\'\i}a-P{\'e}rez}\ \emph
  {et~al.}(2019)\citenamefont {Garc{\'\i}a-P{\'e}rez}, \citenamefont {Allard},
  \citenamefont {Serrano},\ and\ \citenamefont
  {Bogu{\~n}{\'a}}}]{Perez2019mercator}%
  \BibitemOpen
  \bibfield  {author} {\bibinfo {author} {\bibfnamefont {G.}~\bibnamefont
  {Garc{\'\i}a-P{\'e}rez}}, \bibinfo {author} {\bibfnamefont {A.}~\bibnamefont
  {Allard}}, \bibinfo {author} {\bibfnamefont {M.~{\'A}.}\ \bibnamefont
  {Serrano}}, \ and\ \bibinfo {author} {\bibfnamefont {M.}~\bibnamefont
  {Bogu{\~n}{\'a}}},\ }\bibfield  {title} {\emph {\bibinfo {title} {Mercator:
  uncovering faithful hyperbolic embeddings of complex networks},\ }}\href
  {\doibase 10.1088/1367-2630/ab57d2} {\bibfield  {journal} {\bibinfo
  {journal} {New Journal of Physics}\ }\textbf {\bibinfo {volume} {21}},\
  \bibinfo {pages} {123033} (\bibinfo {year} {2019})}\BibitemShut {NoStop}%
\bibitem [{\citenamefont {Davis}\ and\ \citenamefont
  {Goadrich}(2006)}]{Davis2006relationship}%
  \BibitemOpen
  \bibfield  {author} {\bibinfo {author} {\bibfnamefont {J.}~\bibnamefont
  {Davis}}\ and\ \bibinfo {author} {\bibfnamefont {M.}~\bibnamefont
  {Goadrich}},\ }in\ \href {\doibase 10.1145/1143844.1143874} {\emph {\bibinfo
  {booktitle} {Proc. 23rd Int. Conf. Mach. Learn. - ICML '06}}}\ (\bibinfo
  {year} {2006})\ pp.\ \bibinfo {pages} {233--240}\BibitemShut {NoStop}%
\bibitem [{\citenamefont {Alanis-Lobato}\ and\ \citenamefont
  {Andrade-Navarro}(2016)}]{Alanis-Lobato2016distance}%
  \BibitemOpen
  \bibfield  {author} {\bibinfo {author} {\bibfnamefont {G.}~\bibnamefont
  {Alanis-Lobato}}\ and\ \bibinfo {author} {\bibfnamefont {M.~A.}\ \bibnamefont
  {Andrade-Navarro}},\ }\bibfield  {title} {\emph {\bibinfo {title} {{Distance
  distribution between complex network nodes in hyperbolic space}},\ }}\href
  {\doibase 10.25088/ComplexSystems.25.3.223} {\bibfield  {journal} {\bibinfo
  {journal} {Complex Syst.}\ }\textbf {\bibinfo {volume} {25}},\ \bibinfo
  {pages} {223} (\bibinfo {year} {2016})}\BibitemShut {NoStop}%
\bibitem [{\citenamefont {Ma}\ and\ \citenamefont
  {Zeng}(2003)}]{ma2003reconstruction}%
  \BibitemOpen
  \bibfield  {author} {\bibinfo {author} {\bibfnamefont {H.}~\bibnamefont
  {Ma}}\ and\ \bibinfo {author} {\bibfnamefont {A.-P.}\ \bibnamefont {Zeng}},\
  }\bibfield  {title} {\emph {\bibinfo {title} {{Reconstruction of metabolic
  networks from genome data and analysis of their global structure for various
  organisms}},\ }}\href {\doibase 10.1093/bioinformatics/19.2.270} {\bibfield
  {journal} {\bibinfo  {journal} {Bioinformatics}\ }\textbf {\bibinfo {volume}
  {19}},\ \bibinfo {pages} {270} (\bibinfo {year} {2003})}\BibitemShut
  {NoStop}%
\bibitem [{rou()}]{routeviews}%
  \BibitemOpen
  \href@noop {} {\bibfield  {title} {\emph {\bibinfo {title} {{University of
  Oregon Route Views Project }},\ }}}\bibinfo {note}
  {\url{http://www.routeviews.org/routeviews/}}\BibitemShut {NoStop}%
\bibitem [{ope()}]{openpgp}%
  \BibitemOpen
  \href@noop {} {\bibfield  {title} {\emph {\bibinfo {title} {{The Open PGP
  Alliance}},\ }}}\bibinfo {note} {\url{http://www.openpgp.org/}}\BibitemShut
  {NoStop}%
\bibitem [{\citenamefont {L{\"{u}}}\ \emph {et~al.}(2015)\citenamefont
  {L{\"{u}}}, \citenamefont {Pan}, \citenamefont {Zhou}, \citenamefont
  {Zhang},\ and\ \citenamefont {Stanley}}]{Lu2015toward}%
  \BibitemOpen
  \bibfield  {author} {\bibinfo {author} {\bibfnamefont {L.}~\bibnamefont
  {L{\"{u}}}}, \bibinfo {author} {\bibfnamefont {L.}~\bibnamefont {Pan}},
  \bibinfo {author} {\bibfnamefont {T.}~\bibnamefont {Zhou}}, \bibinfo {author}
  {\bibfnamefont {Y.-C.}\ \bibnamefont {Zhang}}, \ and\ \bibinfo {author}
  {\bibfnamefont {H.~E.}\ \bibnamefont {Stanley}},\ }\bibfield  {title} {\emph
  {\bibinfo {title} {{Toward link predictability of complex networks}},\
  }}\href {\doibase 10.1073/pnas.1424644112} {\bibfield  {journal} {\bibinfo
  {journal} {Proc. Natl. Acad. Sci.}\ }\textbf {\bibinfo {volume} {112}},\
  \bibinfo {pages} {2325} (\bibinfo {year} {2015})}\BibitemShut {NoStop}%
\bibitem [{\citenamefont {Bl{\"{a}}sius}\ \emph {et~al.}(2016)\citenamefont
  {Bl{\"{a}}sius}, \citenamefont {Friedrich}, \citenamefont {Krohmer},\ and\
  \citenamefont {Laue}}]{Blasius2016efficient}%
  \BibitemOpen
  \bibfield  {author} {\bibinfo {author} {\bibfnamefont {T.}~\bibnamefont
  {Bl{\"{a}}sius}}, \bibinfo {author} {\bibfnamefont {T.}~\bibnamefont
  {Friedrich}}, \bibinfo {author} {\bibfnamefont {A.}~\bibnamefont {Krohmer}},
  \ and\ \bibinfo {author} {\bibfnamefont {S.}~\bibnamefont {Laue}},\ }in\
  \href {\doibase http://dx.doi.org/10.4230/LIPIcs.ESA.2016.16} {\emph
  {\bibinfo {booktitle} {24th Annu. Eur. Symp. Algorithms (ESA 2016)}}},\
  Vol.~\bibinfo {volume} {57}\ (\bibinfo {year} {2016})\ pp.\ \bibinfo {pages}
  {16:1--16:18}\BibitemShut {NoStop}%
\bibitem [{\citenamefont {Wang}\ \emph {et~al.}(2016)\citenamefont {Wang},
  \citenamefont {Wu}, \citenamefont {Li}, \citenamefont {Jin},\ and\
  \citenamefont {Xiong}}]{Wang2016link}%
  \BibitemOpen
  \bibfield  {author} {\bibinfo {author} {\bibfnamefont {Z.}~\bibnamefont
  {Wang}}, \bibinfo {author} {\bibfnamefont {Y.}~\bibnamefont {Wu}}, \bibinfo
  {author} {\bibfnamefont {Q.}~\bibnamefont {Li}}, \bibinfo {author}
  {\bibfnamefont {F.}~\bibnamefont {Jin}}, \ and\ \bibinfo {author}
  {\bibfnamefont {W.}~\bibnamefont {Xiong}},\ }\bibfield  {title} {\emph
  {\bibinfo {title} {{Link prediction based on hyperbolic mapping with
  community structure for complex networks}},\ }}\href {\doibase
  10.1016/j.physa.2016.01.010} {\bibfield  {journal} {\bibinfo  {journal}
  {Phys. A Stat. Mech. Appl.}\ }\textbf {\bibinfo {volume} {450}},\ \bibinfo
  {pages} {609} (\bibinfo {year} {2016})}\BibitemShut {NoStop}%
\bibitem [{\citenamefont {Alanis-Lobato}\ \emph
  {et~al.}(2016{\natexlab{a}})\citenamefont {Alanis-Lobato}, \citenamefont
  {Mier},\ and\ \citenamefont {Andrade-Navarro}}]{Alanis-Lobato2016efficient}%
  \BibitemOpen
  \bibfield  {author} {\bibinfo {author} {\bibfnamefont {G.}~\bibnamefont
  {Alanis-Lobato}}, \bibinfo {author} {\bibfnamefont {P.}~\bibnamefont {Mier}},
  \ and\ \bibinfo {author} {\bibfnamefont {M.~A.}\ \bibnamefont
  {Andrade-Navarro}},\ }\bibfield  {title} {\emph {\bibinfo {title} {{Efficient
  embedding of complex networks to hyperbolic space via their Laplacian}},\
  }}\href {\doibase 10.1038/srep30108} {\bibfield  {journal} {\bibinfo
  {journal} {Sci. Rep.}\ }\textbf {\bibinfo {volume} {6}},\ \bibinfo {pages}
  {30108} (\bibinfo {year} {2016}{\natexlab{a}})}\BibitemShut {NoStop}%
\bibitem [{\citenamefont {Alanis-Lobato}\ \emph
  {et~al.}(2016{\natexlab{b}})\citenamefont {Alanis-Lobato}, \citenamefont
  {Mier},\ and\ \citenamefont {Andrade-Navarro}}]{Alanis-Lobato2016manifold}%
  \BibitemOpen
  \bibfield  {author} {\bibinfo {author} {\bibfnamefont {G.}~\bibnamefont
  {Alanis-Lobato}}, \bibinfo {author} {\bibfnamefont {P.}~\bibnamefont {Mier}},
  \ and\ \bibinfo {author} {\bibfnamefont {M.~A.}\ \bibnamefont
  {Andrade-Navarro}},\ }\bibfield  {title} {\emph {\bibinfo {title} {{Manifold
  learning and maximum likelihood estimation for hyperbolic network
  embedding}},\ }}\href {\doibase 10.1007/s41109-016-0013-0} {\bibfield
  {journal} {\bibinfo  {journal} {Appl. Netw. Sci.}\ }\textbf {\bibinfo
  {volume} {1}},\ \bibinfo {pages} {10} (\bibinfo {year}
  {2016}{\natexlab{b}})}\BibitemShut {NoStop}%
\bibitem [{\citenamefont {Muscoloni}\ \emph {et~al.}(2017)\citenamefont
  {Muscoloni}, \citenamefont {Thomas}, \citenamefont {Ciucci}, \citenamefont
  {Bianconi},\ and\ \citenamefont {Cannistraci}}]{Muscoloni2017machine}%
  \BibitemOpen
  \bibfield  {author} {\bibinfo {author} {\bibfnamefont {A.}~\bibnamefont
  {Muscoloni}}, \bibinfo {author} {\bibfnamefont {J.~M.}\ \bibnamefont
  {Thomas}}, \bibinfo {author} {\bibfnamefont {S.}~\bibnamefont {Ciucci}},
  \bibinfo {author} {\bibfnamefont {G.}~\bibnamefont {Bianconi}}, \ and\
  \bibinfo {author} {\bibfnamefont {C.~V.}\ \bibnamefont {Cannistraci}},\
  }\bibfield  {title} {\emph {\bibinfo {title} {{Machine learning meets complex
  networks via coalescent embedding in the hyperbolic space}},\ }}\href
  {\doibase 10.1038/s41467-017-01825-5} {\bibfield  {journal} {\bibinfo
  {journal} {Nat. Commun.}\ }\textbf {\bibinfo {volume} {8}},\ \bibinfo {pages}
  {1615} (\bibinfo {year} {2017})}\BibitemShut {NoStop}%
\bibitem [{\citenamefont {Faqeeh}\ \emph {et~al.}(2018)\citenamefont {Faqeeh},
  \citenamefont {Osat},\ and\ \citenamefont
  {Radicchi}}]{faqeeh2018scharacterizing}%
  \BibitemOpen
  \bibfield  {author} {\bibinfo {author} {\bibfnamefont {A.}~\bibnamefont
  {Faqeeh}}, \bibinfo {author} {\bibfnamefont {S.}~\bibnamefont {Osat}}, \ and\
  \bibinfo {author} {\bibfnamefont {F.}~\bibnamefont {Radicchi}},\ }\bibfield
  {title} {\emph {\bibinfo {title} {{Characterizing the Analogy Between
  Hyperbolic Embedding and Community Structure of Complex Networks}},\ }}\href
  {\doibase 10.1103/PhysRevLett.121.098301} {\bibfield  {journal} {\bibinfo
  {journal} {Phys. Rev. Lett.}\ }\textbf {\bibinfo {volume} {121}},\ \bibinfo
  {pages} {098301} (\bibinfo {year} {2018})}\BibitemShut {NoStop}%
\bibitem [{\citenamefont {Colomer-de Sim{\'{o}}n}\ \emph
  {et~al.}(2013)\citenamefont {Colomer-de Sim{\'{o}}n}, \citenamefont
  {Serrano}, \citenamefont {Beir{\'{o}}}, \citenamefont {Alvarez-Hamelin},\
  and\ \citenamefont {Bogu{\~{n}}{\'{a}}}}]{colomer2013clustering}%
  \BibitemOpen
  \bibfield  {author} {\bibinfo {author} {\bibfnamefont {P.}~\bibnamefont
  {Colomer-de Sim{\'{o}}n}}, \bibinfo {author} {\bibfnamefont {M.~{\'{A}}.}\
  \bibnamefont {Serrano}}, \bibinfo {author} {\bibfnamefont {M.~G.}\
  \bibnamefont {Beir{\'{o}}}}, \bibinfo {author} {\bibfnamefont {J.~I.}\
  \bibnamefont {Alvarez-Hamelin}}, \ and\ \bibinfo {author} {\bibfnamefont
  {M.}~\bibnamefont {Bogu{\~{n}}{\'{a}}}},\ }\bibfield  {title} {\emph
  {\bibinfo {title} {{Deciphering the global organization of clustering in real
  complex networks}},\ }}\href {\doibase 10.1038/srep02517} {\bibfield
  {journal} {\bibinfo  {journal} {Sci Rep}\ }\textbf {\bibinfo {volume} {3}},\
  \bibinfo {pages} {2517} (\bibinfo {year} {2013})}\BibitemShut {NoStop}%
\bibitem [{met()}]{metabolicdetails}%
  \BibitemOpen
  \href@noop {} {}\bibinfo {note} {Further details on the dataset can be found
  at \url{http://snap.stanford.edu/data/as.html}}\BibitemShut {NoStop}%
\bibitem [{ced()}]{cederlof}%
  \BibitemOpen
  \href@noop {} {\bibfield  {title} {\emph {\bibinfo {title} {{OpenPGP web of
  trust database}},\ }}}\bibinfo {note}
  {\url{http://www.lysator.liu.se/~jc/wotsap/wots2/}}\BibitemShut {NoStop}%
\bibitem [{\citenamefont {Voitalov}\ \emph {et~al.}(2019)\citenamefont
  {Voitalov}, \citenamefont {van~der Hoorn}, \citenamefont {van~der Hofstad},\
  and\ \citenamefont {Krioukov}}]{voitalov2019scale}%
  \BibitemOpen
  \bibfield  {author} {\bibinfo {author} {\bibfnamefont {I.}~\bibnamefont
  {Voitalov}}, \bibinfo {author} {\bibfnamefont {P.}~\bibnamefont {van~der
  Hoorn}}, \bibinfo {author} {\bibfnamefont {R.}~\bibnamefont {van~der
  Hofstad}}, \ and\ \bibinfo {author} {\bibfnamefont {D.}~\bibnamefont
  {Krioukov}},\ }\bibfield  {title} {\emph {\bibinfo {title} {{Scale-free
  networks well done}},\ }}\href {\doibase 10.1103/PhysRevResearch.1.033034}
  {\bibfield  {journal} {\bibinfo  {journal} {Phys. Rev. Res.}\ }\textbf
  {\bibinfo {volume} {1}},\ \bibinfo {pages} {033034} (\bibinfo {year}
  {2019})},\ \Eprint {http://arxiv.org/abs/1811.02071} {arXiv:1811.02071}
  \BibitemShut {NoStop}%
\bibitem [{\citenamefont {Liben-Nowell}\ and\ \citenamefont
  {Kleinberg}(2003)}]{Liben2003link}%
  \BibitemOpen
  \bibfield  {author} {\bibinfo {author} {\bibfnamefont {D.}~\bibnamefont
  {Liben-Nowell}}\ and\ \bibinfo {author} {\bibfnamefont {J.}~\bibnamefont
  {Kleinberg}},\ }\bibfield  {title} {\emph {\bibinfo {title} {{The Link
  Prediction Problem for Social Networks}},\ }}\href {\doibase
  10.1002/asi.v58:7} {\bibfield  {journal} {\bibinfo  {journal} {Proc. Twelfth
  Annu. ACM Int. Conf. Inf. Knowl. Manag.}\ ,\ \bibinfo {pages} {556}}
  (\bibinfo {year} {2003})}\BibitemShut {NoStop}%
\bibitem [{\citenamefont {Zhou}\ \emph {et~al.}(2009)\citenamefont {Zhou},
  \citenamefont {L{\"{u}}},\ and\ \citenamefont {Zhang}}]{Zhou2009predicting}%
  \BibitemOpen
  \bibfield  {author} {\bibinfo {author} {\bibfnamefont {T.}~\bibnamefont
  {Zhou}}, \bibinfo {author} {\bibfnamefont {L.}~\bibnamefont {L{\"{u}}}}, \
  and\ \bibinfo {author} {\bibfnamefont {Y.~C.}\ \bibnamefont {Zhang}},\
  }\bibfield  {title} {\emph {\bibinfo {title} {{Predicting missing links via
  local information}},\ }}\href {\doibase 10.1140/epjb/e2009-00335-8}
  {\bibfield  {journal} {\bibinfo  {journal} {Eur. Phys. J. B}\ }\textbf
  {\bibinfo {volume} {71}},\ \bibinfo {pages} {623} (\bibinfo {year}
  {2009})}\BibitemShut {NoStop}%
\bibitem [{\citenamefont {Cannistraci}\ \emph {et~al.}(2013)\citenamefont
  {Cannistraci}, \citenamefont {Alanis-Lobato},\ and\ \citenamefont
  {Ravasi}}]{Cannistraci2013b}%
  \BibitemOpen
  \bibfield  {author} {\bibinfo {author} {\bibfnamefont {C.~V.}\ \bibnamefont
  {Cannistraci}}, \bibinfo {author} {\bibfnamefont {G.}~\bibnamefont
  {Alanis-Lobato}}, \ and\ \bibinfo {author} {\bibfnamefont {T.}~\bibnamefont
  {Ravasi}},\ }\bibfield  {title} {\emph {\bibinfo {title} {{From
  link-prediction in brain connectomes and protein interactomes to the
  local-community-paradigm in complex networks}},\ }}\href {\doibase
  10.1038/srep01613} {\bibfield  {journal} {\bibinfo  {journal} {Sci. Rep.}\
  }\textbf {\bibinfo {volume} {3}},\ \bibinfo {pages} {1613} (\bibinfo {year}
  {2013})}\BibitemShut {NoStop}%
\bibitem [{\citenamefont {Jaccard}(1901)}]{Jaccard1901}%
  \BibitemOpen
  \bibfield  {author} {\bibinfo {author} {\bibfnamefont {P.}~\bibnamefont
  {Jaccard}},\ }\bibfield  {title} {\emph {\bibinfo {title} {{{\'{E}}tude
  comparative de la distribution florale dans une portion des Alpes et des
  Jura}},\ }}\href {\doibase 10.5169/seals-266450} {\bibfield  {journal}
  {\bibinfo  {journal} {Bull. del la Soci{\'{e}}t{\'{e}} Vaudoise des Sci.
  Nat.}\ }\textbf {\bibinfo {volume} {37}},\ \bibinfo {pages} {547} (\bibinfo
  {year} {1901})}\BibitemShut {NoStop}%
\bibitem [{\citenamefont {Peixoto}(2017)}]{peixoto2017nonparametric}%
  \BibitemOpen
  \bibfield  {author} {\bibinfo {author} {\bibfnamefont {T.~P.}\ \bibnamefont
  {Peixoto}},\ }\bibfield  {title} {\emph {\bibinfo {title} {{Nonparametric
  Bayesian inference of the microcanonical stochastic block model}},\ }}\href
  {\doibase 10.1103/PhysRevE.95.012317} {\bibfield  {journal} {\bibinfo
  {journal} {Phys. Rev. E}\ }\textbf {\bibinfo {volume} {95}},\ \bibinfo
  {pages} {012317} (\bibinfo {year} {2017})},\ \Eprint
  {http://arxiv.org/abs/1610.02703} {arXiv:1610.02703} \BibitemShut {NoStop}%
\bibitem [{\citenamefont {Peixoto}(2014{\natexlab{a}})}]{Peixoto2014}%
  \BibitemOpen
  \bibfield  {author} {\bibinfo {author} {\bibfnamefont {T.~P.}\ \bibnamefont
  {Peixoto}},\ }\bibfield  {title} {\emph {\bibinfo {title} {{Hierarchical
  block structures and high-resolution model selection in large networks}},\
  }}\href {\doibase 10.1103/PhysRevX.4.011047} {\bibfield  {journal} {\bibinfo
  {journal} {Phys. Rev. X}\ }\textbf {\bibinfo {volume} {4}} \bibinfo {pages} {011047} (\bibinfo {year}
  {2014}{\natexlab{a}})}\BibitemShut {NoStop}%
\bibitem [{\citenamefont
  {Peixoto}(2014{\natexlab{b}})}]{peixoto_graph-tool_2014}%
  \BibitemOpen
  \bibfield  {author} {\bibinfo {author} {\bibfnamefont {T.~P.}\ \bibnamefont
  {Peixoto}},\ }\bibfield  {title} {\emph {\bibinfo {title} {The graph-tool
  python library},\ }}\href {\doibase 10.6084/m9.figshare.1164194} {\bibfield
  {journal} {\bibinfo  {journal} {figshare}\ } (\bibinfo {year}
  {2014}{\natexlab{b}})}\BibitemShut {NoStop}%
\bibitem [{\citenamefont {Peixoto}(2014{\natexlab{c}})}]{peixoto2014efficient}%
  \BibitemOpen
  \bibfield  {author} {\bibinfo {author} {\bibfnamefont {T.~P.}\ \bibnamefont
  {Peixoto}},\ }\bibfield  {title} {\emph {\bibinfo {title} {{Efficient Monte
  Carlo and greedy heuristic for the inference of stochastic block models}},\
  }}\href {\doibase 10.1103/PhysRevE.89.012804} {\bibfield  {journal} {\bibinfo
   {journal} {Phys. Rev. E}\ }\textbf {\bibinfo {volume} {89}},\ \bibinfo
  {pages} {012804} (\bibinfo {year} {2014}{\natexlab{c}})},\ \Eprint
  {http://arxiv.org/abs/1310.4378} {1310.4378} \BibitemShut {NoStop}%
\bibitem [{\citenamefont {Bogu{\~{n}}{\'{a}}}\ and\ \citenamefont
  {Pastor-Satorras}(2003)}]{boguna2003class}%
  \BibitemOpen
  \bibfield  {author} {\bibinfo {author} {\bibfnamefont {M.}~\bibnamefont
  {Bogu{\~{n}}{\'{a}}}}\ and\ \bibinfo {author} {\bibfnamefont
  {R.}~\bibnamefont {Pastor-Satorras}},\ }\bibfield  {title} {\emph {\bibinfo
  {title} {{Class of correlated random networks with hidden variables}},\
  }}\href {\doibase 10.1103/PhysRevE.68.036112} {\bibfield  {journal} {\bibinfo
   {journal} {Phys. Rev. E}\ }\textbf {\bibinfo {volume} {68}},\ \bibinfo
  {pages} {036112} (\bibinfo {year} {2003})}\BibitemShut {NoStop}%
\bibitem [{\citenamefont {Fisher}\ and\ \citenamefont
  {Lee}(1981)}]{fisher1981nonparametric}%
  \BibitemOpen
  \bibfield  {author} {\bibinfo {author} {\bibfnamefont {N.~I.}\ \bibnamefont
  {Fisher}}\ and\ \bibinfo {author} {\bibfnamefont {A.~J.}\ \bibnamefont
  {Lee}},\ }\bibfield  {title} {\emph {\bibinfo {title} {{Nonparametric
  measures of angular-linear association}},\ }}\href {\doibase
  10.1093/biomet/68.3.629} {\bibfield  {journal} {\bibinfo  {journal}
  {Biometrika}\ }\textbf {\bibinfo {volume} {68}},\ \bibinfo {pages} {629}
  (\bibinfo {year} {1981})}\BibitemShut {NoStop}%
\end{thebibliography}
\end{document}